\documentclass[12pt,preprint]{aastex}
\usepackage{lscape}
\usepackage{color}

\shorttitle{Characterization of AGN and their hosts.}
\shortauthors{Ramos Almeida et al.}

\begin{document}

\title{Characterization of AGN and their hosts in the Extended Groth Strip: a multiwavelength analysis.}

\author{C. Ramos Almeida\altaffilmark{1}, J.M. Rodr\'\i guez Espinosa\altaffilmark{1},
G. Barro\altaffilmark{2}, J. Gallego\altaffilmark{2}, and P. G. P\'{e}rez-Gonz\'{a}lez\altaffilmark{2,3}}

\altaffiltext{1}{Instituto de Astrof\'\i sica de Canarias (IAC), 
              C/V\'\i a L\'{a}ctea, s/n, E-38205, La Laguna, Tenerife, Spain.
cra@iac.es and jre@iac.es}

\altaffiltext{2}{Departamento de Astrof\'\i sica  y Ciencias de la Atm\'{o}sfera, Facultad de Ciencias 
F\'\i sicas, Universidad Complutense de Madrid, E-28040 Madrid, Spain.
gbc@astrax.fis.ucm.es, jgm@astrax.fis.ucm.es, and pgperez@astrax.fis.ucm.es}

\altaffiltext{3}{Associate Astronomer at Steward Observatory, University of Arizona, Tucson, AZ 85721.}

\begin{abstract}

We have employed a reliable technique of classification of Active Galactic Nuclei (AGN) based on 
the fit of well-sampled spectral energy distributions (SEDs) with a complete set of AGN and 
starburst galaxy templates. We have compiled ultraviolet, optical, and infrared data
for a sample of 116 AGN originally selected for their X-ray and mid-infrared emissions 
(96 with single detections and 20 with double optical counterparts).  
This is the most complete compilation of multiwavelength data for such a big sample of AGN in the Extended
Groth Strip (EGS).
Through these SEDs, we are able to obtain highly reliable photometric redshifts and to 
distinguish between pure and host-dominated AGN. 
For the objects with unique detection we find that they can be separated into 
five main groups, namely:
{\it Starburst-dominated AGN} (24 \% of the sample), {\it Starburst-contaminated AGN} (7 \%), 
{\it Type-1 AGN} (21 \%), {\it Type-2 AGN} (24 \%), and {\it Normal galaxy hosting AGN} (24 \%). 
We find these groups concentrated at different redshifts: {\it Type-2 AGN} and 
{\it Normal galaxy hosting AGN} are concentrated at low redshifts, whereas {\it Starburst-dominated AGN} 
and {\it Type-1 AGN} show a larger span.
Correlations between hard/soft X-ray and ultraviolet, optical and infrared luminosities, respectively, 
are reported for the first time for such a sample of AGN spanning a wide range of redshifts.
For the 20 objects with double detection the percentage of {\it Starburst-dominated AGN} increases up to 48\%.

\end{abstract}

\keywords{galaxies:active - galaxies:nuclei - galaxies:starburst - ultraviolet:galaxies - 
infrared:galaxies - X-rays:galaxies}

\section{Introduction}

The role of AGN in the formation and evolution of galaxies is still not well
established. It is not clear whether AGN represent episodic phenomena in the life of galaxies, are random
processes (given that the Supermassive Black Hole is already there), or are more fundamental. 
Some authors claim that AGN are
key in quenching the star formation bursts in their host galaxies \citep{Granato04,Springel05}.
It has also been shown that the mass dependence
of the peak star-formation epoch appears to mirror the mass dependence of Black Hole (BH) activity, 
as recently seen in redshift surveys of both radio-and X-ray-selected active 
galactic nuclei \citep{Waddington01,Hasinger03b}.
For these reasons, searching for signatures of AGN feedback in 
the properties of AGN host galaxies is one of the most promising ways of testing the role of AGN in galaxy evolution. 

One way of finding variations in the AGN population with redshift is to compare their SEDs 
defined over a broad wavelength range. The SED of an AGN can reveal the presence of the underlying 
central engine, together with the luminosity of the host galaxy, the reddening, and
the role of the star formation in the various frequency regimes. SED determination in samples of
AGN at different redshifts is an efficient method to search for evolutionary trends.
Accuracy in the photometry and a filterset spanning a broad wavelength range are required to characterize correctly  
different types of AGN.

Multiwavelength surveys are fundamental in the study of active galactic nuclei, since these appear 
considerably
different depending on the wavelength range of consideration. The hard X-ray selection of AGN using 
deep observations is one of the most reliable methods of finding AGN \citep{Mushotzky04}, although a percentage of them  
remains undetected using this technique \citep{Peterson06}, specially the most highly obscured objects.
For this reason, it is important to characterize AGN at different wavelength ranges, 
in order to be capable of identifying them by more than one selection technique, and to distinguish between
the different groups of active nuclei, including those that could be contaminated, or even hidden, by starbursts. 
Mid-infrared surveys have been very successful in finding X-ray undetected AGN in large numbers, 
but in this case it 
is crucial to distinguish the AGN from the non-active star-forming galaxies. This can be achieved  
using typical mid-infrared colors of AGN \citep{Lacy04,Stern05,Alonso06,Donley08} or by combining mid-infrared 
and radio detections \citep{Donley05,Alonso06,Martinez-Sansigre05,Martinez-Sansigre07,Lacy07,Park08}.

The Extended Groth Strip ($\alpha$ = 14$^{h}$ 17$^{m}$, $\delta$ = 52$^{o}$ 30') enlarges the Hubble Space
Telescope Groth-Westphal strip \citep{Groth94} up to 2$^{o}$x15', having the advantage of
being a low extinction area in the northern sky, with low galactic and zodiacal infrared emission, and
good schedulability by space observatories. For these reasons, there is a huge amount of public data at
different wavelength ranges that only require to be compiled and cross-correlated in a consistent way.
The overall majority of the observational work in the EGS
have been coordinated by the AEGIS proyect\footnote{The AEGIS project is a collaborative effort to obtain both 
deep imaging covering all major wavebands from X-ray to radio and optical spectroscopy over a large area 
of sky. http://aegis.ucolick.org/index.html} \citep{Davis07}.

With the huge amount of data available for this region of the sky, we have constructed a robust AGN sample, 
detected in the X-rays and in the mid-infrared, intermediate in depth and area in comparison with other surveys
\citep{Jannuzi99,Dickinson01,Lonsdale03,Eisenhardt04,Franceschini05}.
The photometry has been performed over the publicy available images, in several bands, in order to compile
as best-sampled SEDs as possible.
The biggest advantage of our AGN sample, compared with other works, is the robustness of the photometry, 
performed in a consistent way among the different bands, and its multiwavelength nature: it is the most complete 
compilation of data for such a big sample of AGN in the EGS. This allows us to determine accurate photometric redshifts, 
and to distinguish clearly between the different groups of AGN.
We have used a comprenhensive set of AGN plus starburst templates 
from \citet{Polletta07}, to fit the SEDs of the galaxies in the sample, 
and to separate them in five different main groups. Through this classification, we can study the properties 
of the different types of AGN in this sample.
Section 2 describes the sample and the cross-matching of the multiwavelength data, Section 3 explains the  
technique of classification of AGN based on the fit of their spectral energy distributions, in Section 4 the 
results of this paper are discussed, and finally in Section 5 the main conclusions are summarized.
Throughout this paper  we assume an 
H$_{0}$=75 km s$^{-1}$ Mpc$^{-1}$ and a $\Lambda$CDM cosmology with $\Omega_{m}$=0.3 and 
$\Omega_{\Lambda}$=0.7.

\section{Sample and multi-wavelength data}
\label{sample}

	The sample studied in this work, that comprises 116 AGN candidates, was built from the 
previously published X-ray catalogs found in \citet{Waskett03}, \citet{Nandra05}, and \citet{Barmby06}. 
These objects were originally selected by \citet{Barmby06}, both in the X-ray (Chandra and XMM-Newton) 
and in the mid-infrared (Spitzer). 
The X-ray and mid-infrared observations in the EGS are intermediate in depth and area between GOODS
\citep{Dickinson01}, the shallower NOAO Deep-Wide Field \citep{Jannuzi99,Eisenhardt04}, and SWIRE
\citep{Lonsdale03} surveys. Therefore, this region provides a valuable test of AGN properties at
intermediate fluxes. 
In addition to this, we have compiled ultraviolet, optical, and near-infrared archival data for these AGN,  
in order to increase the definition of their SEDs. 
These well-sampled SEDs allow us, first, to classify the objects in different galaxy population
groups and calculate their photometric redshifts, and secondly, to study the physical 
properties of this representative sample of AGN.

The Chandra data were
taken with ACIS-I in 2002 August \citep{Nandra05}, consisting of a 200 ks exposure with a limiting
full-band flux (0.5-10 keV) of 3.5x10$^{-16}$ erg cm$^{-2}$ s$^{-1}$. The XMM-Newton data were obtained
in 2000 July with a 56 ks exposure and with a limiting 0.5-10 keV flux of 2x10$^{-15}$ 
erg cm$^{-2}$ s$^{-1}$ \citep{Waskett03}. \citet{Barmby06} combined both catalogues producing a list of
152 sources within the limits of the Spitzer mid-infrared observations. 

The Spitzer data (IRAC and MIPS) are part of the Infrared Array Camera Deep Survey,
taken during 2003 December and 2004 
June-July with 2.7 hr exposure per pointing. In the case of the MIPS 
data, the observations were done in 2004 January and June with a depth of 1200 s per pointing.
The 5$\sigma$ limiting AB magnitudes are 24.0, 24.0, 21.9, and 22.0 for the IRAC bands, and 19.1 in the
case of MIPS. \citet{Barmby06} finally selected 138 objects with secure detections in all four IRAC bands,
out of the 152 X-ray emitters. 
The detection of these objects in both the X-rays and the mid-infrared gives confidence in their 
classification as AGN. Besides, we have also checked that the values of the hard X-ray and 24 \micron~fluxes 
lie inside the AGN-characteristic region (see Figure 1 of \citet{Alonso04}).

In addition to the previous, we have compiled near and far-ultraviolet
images from the GALEX GR2/GR3 data release (3$\sigma$ limiting AB magnitudes = 25 in both Far- and Near-UV filters);  
optical images from the CFHT Legacy Survey, T0003 worldwide release (Gwyn et al., in preparation),
taken with the MegaCam imager on the 4 m Canada-France Hawaii Telescope \citep{Boulade03} (5$\sigma$ limiting 
AB magnitudes = 26.3, 27.0, 26.5, 26.0, and 25.0 in u,g,r,i, and z bands);
and J and K$_{S}$ data from the version 3.3 of the Palomar-WIRC K-selected catalog of \citet{Bundy06}, 
(5$\sigma$ limiting Vega magnitudes = 23 and 20.6 in the J and K$_{S}$ bands). 

The fluxes employed in this work have been measured in a
compilation of publicly available imaging data, which is outlined briefly
in \citet{Villar08} and will be described in detail in Barro et al., in preparation   
(see also \citet{Perez08b}). Photometry
in consistent apertures was measured in all bands with available
imaging data following the procedure described in \citet{Perez05,Perez08}.
In the near-infrared, no deep J- and K$_{S}$-band imaging data were
available and we have used the photometric catalogs published by \citet{Bundy06}.
The same happens with the X-ray data, that have been drawn from the catalogs \citep{Waskett03,Nandra05,Barmby06}.

We have performed the cross-matching of the 138 sources between the X-ray and Spitzer data, adding  
ultraviolet, optical, and near-infrared data points,  
avoiding the false matches that \citet{Barmby06} expected in their sample. We identify these 
objects through their IRAC positions in our merged photometric catalog \citep{Perez05,Perez08}. 
The source coordinates
on the IRAC 3.6~\micron~images are then cross-correlated with each one of the ultraviolet, optical and
near-infrared catalogs using a search radius of 2.5\arcsec~, starting with the deepest images. When a source is
identified in one of these images, the \citet{Kron80} elliptical aperture from this reference image is taken and 
overlaid onto each of the other bands. The aperture employed is large enough to enclose the PSF in all the 
ultraviolet, optical and
near-infrared images (the seeing being less than 1.5\arcsec). For IRAC and MIPS, because of their large PSFs,
integrated magnitudes measured in small apertures (applying aperture corrections) are employed. 
The hard and soft X-ray fluxes are obtained by cross-correlating the IRAC positions with the X-ray catalogs, 
using a search radius of 2\arcsec~in this case. 
Uncertainties of each measured flux are obtained from the sky pixel-to-pixel variations, 
detector readout noise, Poisson noise in the measured fluxes, errors in the World Coordinate 
System, and errors in the absolute photometric calibration. 

In some cases,  for a single IRAC source,  there are several 
counterparts in the ground-based images within the 2.5\arcsec~search radius. For these objects, the ground-based
optical/near-infrared reference image is used to determine the positions of each source separately. 
The IRAC images are then deconvolved using the IRAC PSFs.
Although the IRAC PSFs have
FWHMs of aproximately 2\arcsec, determination of the central position of each IRAC source can be performed more 
accurately, and sources can be resolved for separations $\sim$1\arcsec~from each other.   
IRAC fluxes are then remeasured by fixing the positions of the objects in each pair, and by scaling the
flux of each object in an aperture of 0.9\arcsec.  
For a more detailed description of the cross-matching and aperture photometry see \citet{Perez05,Perez08}. 

Out of the 138 sources that comprise the final sample chosen by \citet{Barmby06}, we find 96 sources
that have unique detections in all bands, plus other 20 objects with double detection in the ground-based 
images. We discard the remaining 22 objects because 21 of them show multiple (more than two) 
detections in the optical/near-infrared images, leading to possible source confusion, plus another object 
that shows a star-like SED.
The analysis of the data will be done first for the 96 objects that are definitely free from contamination 
from other sources. Nevertheless, in Section 4.5, we
analyse the images and photometric redshifts of those additional 20 objects with double detection.

\section{Spectral energy distributions and photometric redshifts of objects with unique detection.}
\label{spectral}

In order to classify the 96 spectral energy distributions and to estimate their photometric redshifts, 
we combine optical (u,g,r,i,z), near-infrared (J,K), and mid-infrared data 
(IRAC 3.6, 4.5, 5.8, 8 \micron~and MIPS 24 \micron) to build well-sampled SEDs that we then fit with 
the library of starburst, AGN and galaxy templates 
taken from \citet{Polletta07}. We make use of the photometric redshift code
HyperZ \citep{Bolzonella00} to perform the fits. This code determines the best photometric redshifts
(z$_{phot}$) by minimization of the $\chi^{2}$ derived from a comparison between the photometric 
SEDs and the set of template spectra, leaving the redshift as a variable. 
The code also takes into account the effects of dust extinction according to the selected reddening law 
\citep{Calzetti00}. Choosing a wide range of reddening values seems to be essential to reproduce the SEDs of 
high redshift galaxies \citep{Bolzonella00}.
According to \citet{Steidel99}, the typical E(B-V) for galaxies up
to z~$\sim$~4 is 0.15 mag, thus A$_{V}$~$\sim$~0.6 mag when using the Calzetti's law. 
The maximum A$_{V}$ allowed in our calculations is about 2 times this value, thus A$_{V}$ ranges 
from 0.0 to 1.2, with a step between them of 0.3. Similar values of A$_{V}$ are typically chosen 
in the literature \citep{Bolzonella00,Babbedge04}.

The chosen set of templates contains 23 SED-types, that we have arranged into the following five 
main groups: 
{\it Starburst-dominated AGN} (which includes four Starbursts and 
Starburst/ULIRGs templates), {\it Starburst-contaminated AGN} (three templates, namely: 
Starburst/ULIRG/Seyfert 1, Starburst/Seyfert 2, and Starburst/ULIRG/Seyfert 2), 
{\it Type-1 AGN} (three Type-1 QSO templates), {\it Type-2 AGN} 
(Type-2 QSO, Torus-QSO, Seyfert 1.8, and Seyfert 2 templates), and finally, {\it Normal galaxy hosting AGN}
(nine templates including 2, 5, and 13 Gyr ellipticals plus S0, Sa, Sb, Sc, Sd, Sdm type spirals). 
These templates span the range in wavelength between 0.1 and 1000
\micron. For a detailed explanation of their synthesis see \citet{Polletta07}. Our main 
interest is to classify all of our sources into these five main groups and to determine the 
distribution of the sources into each of these groups.
Notice that although all of the sources are AGN, the {\it Starburst-dominated AGN} have their SEDs 
dominated by the starburst emission from the optical to the mid-infrared; the 
{\it Normal galaxy hosting AGN} would be low-luminosity AGN embeded in an otherwise normal galaxy
emission; and
in the case of the {\it Starburst-contaminated AGN}, the emission of both the starburst 
or the AGN dominate depending on the wavelength we are looking at. Indeed, some of these objects 
show noticeably the AGN power-law beyond the near-infrared.

We fit data from the optical $u$ band up to the MIPS 24 \micron~band. We avoid the use of GALEX data because 
few galaxies have these, and because their use introduce big errors in the fits.
As explained in \citet{Polletta07}, including
mid-infrared data improves considerably the photometric redshift calculations, since some spectral types suffer
degeneration that is broken by the non-extinguished longer wavelengths, even if the errors in the magnitudes are 
larger in the mid-infrared than in the optical and near-infrared bands. 

\begin{figure}[!h]
\centering
\includegraphics[width=11cm,angle=90]{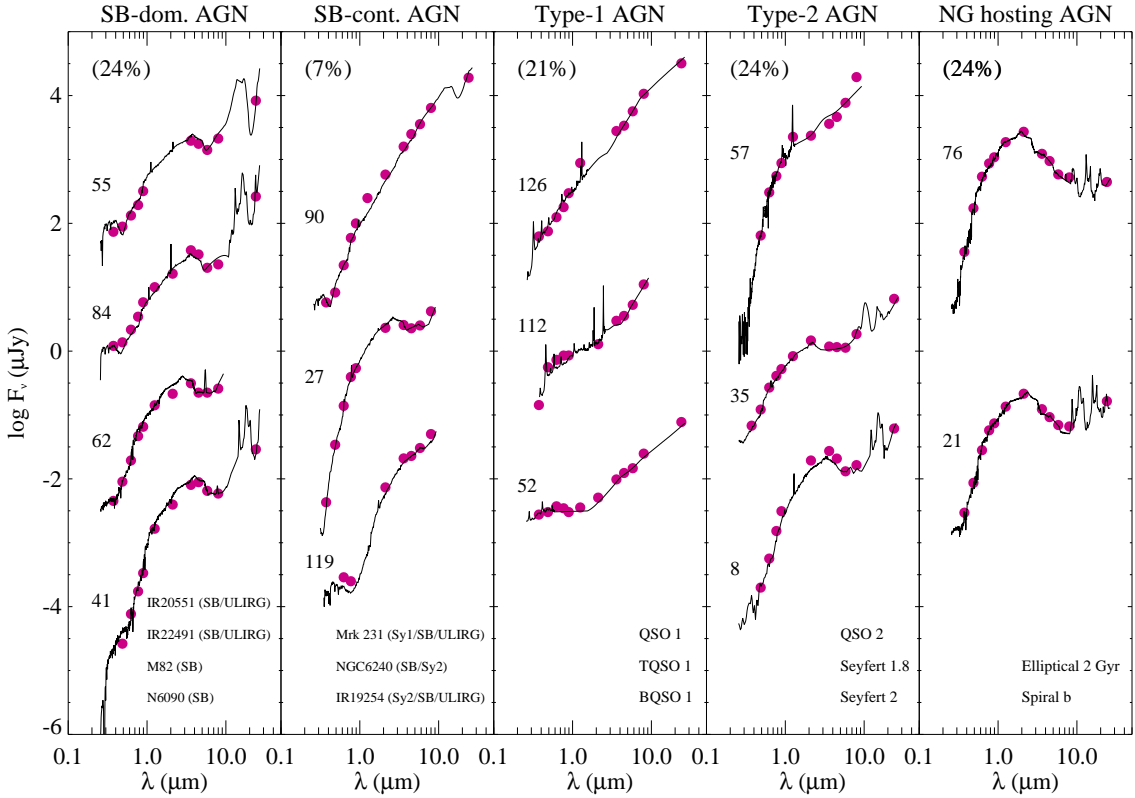}
\caption{\footnotesize{Examples of SEDs in our sample (dots) fitted with different templates from
\citet{Polletta07} for the five main groups considered. The legends in the bottom right of each pannel 
refer to the actual template within the groups from \citet{Polletta07}.
The data have been scaled for clarity. The X-axis corresponds to observed wavelength.
The fifteen galaxies represented here are labelled with the ID from Table \ref{photoz}. The 
percentages of objects enclosed in each group are typed in the upper left corner of each pannel.}
\label{fits}}
\end{figure}

Examples of HyperZ fits for each of the employed templates are shown in Figure \ref{fits}. 
In the {\it Type-2 AGN} pannel,
only three templates are shown because none of the 96 galaxies were fitted with the Torus-QSO template. 
The {\it Normal galaxy hosting AGN} pannel contains only one example of elliptical template 
(the 2 Gyr elliptical) and one
example of spiral (Sb). See Table \ref{photoz} to check the SED types and their corresponding group.
Photometric redshifts derived from the fits 
are reported in Table \ref{photoz}, together with the $\chi_{\nu}^{2}$ and probabilities given by 
HyperZ, the A$_{V}$, and the template used for the fit of each galaxy.
In the cases where spectroscopic redshifts are available, these are also given in Table \ref{photoz}.

\begin{figure}[!h]
\centering
\includegraphics[width=13cm,height=11cm]{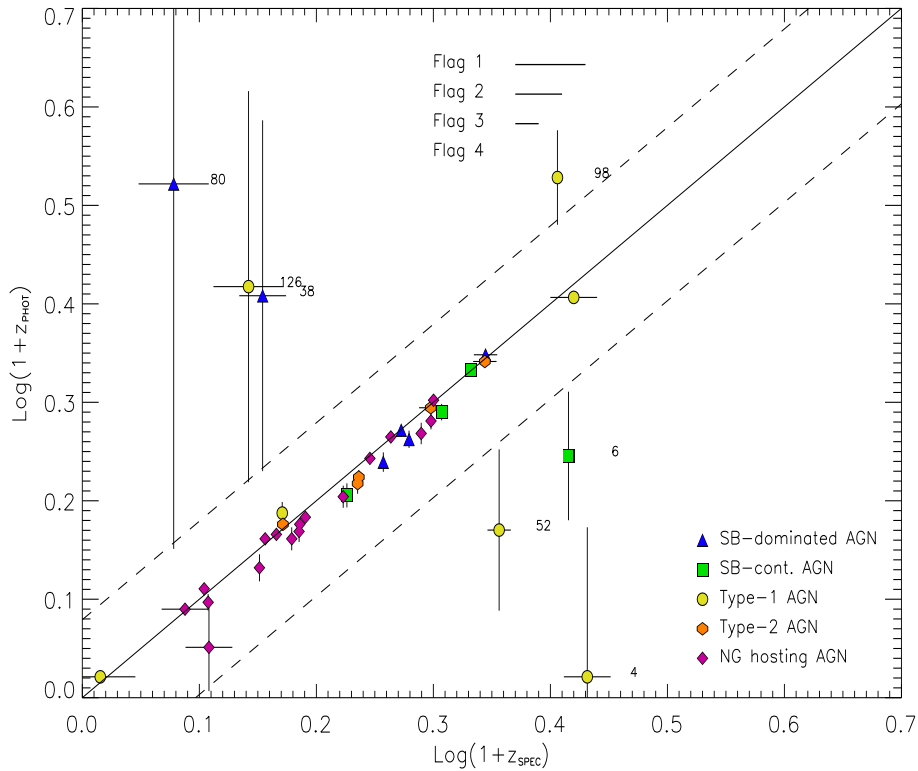}
\caption{\footnotesize{Comparison between photometric and spectrospic redshifts for the 39 sources with publicly
available $z_{spec}$ from the DEEP data archive \citep{Weiner05}. The solid line corresponds to 
z$_{spec}$=z$_{phot}$.
Dashed lines represent 20\% boundaries in (1+$z$). Horizontal error bars 
indicating the reliability of the $z_{spec}$ (flag=1 for the less reliable and flag=4 for the most reliable) 
are represented. Vertical error bars ilustrate the discrepancies between the $z_{spec}$ 
and $z_{phot}$. The seven outliers are labelled (4, 6, 38, 52, 80, 98 and 126). 
Note that there are only three outliers when only the 31 galaxies with reliable $z_{spec}$ are considered.}
\label{z}}
\end{figure}

A comparison between the photometric and spectroscopic redshifts for the 39 sources with publicly
available $z_{spec}$ from the DEEP data archive \citep{Weiner05} is shown in Figure \ref{z}. Notice that 
only 31 out of these 39 galaxies have reliable spectroscopic redshifts (flag = 3 or 4 in the DEEP data archive). 
Horizontal error bars 
indicating the reliability of the $z_{spec}$ are represented in Figure \ref{z}, together with vertical error bars
that indicate the discrepancies between the $z_{spec}$ and $z_{phot}$.
The dashed lines represent 20\%
agreement in (1+$z$). The fractional error $\Delta$z=${z_{phot}-z_{spec} \over 1+ z_{spec}}$ quantifies the 
number of catastrophic outliers, which are those with $\mid\Delta z\mid$ $>$ 0.2. Our measured mean
$\Delta$z for the 39 sources with spectroscopic redshifts is 0.05, with a $\sigma_{z}$ = 0.37, and an outlier 
fraction of $\sim$18\%, corresponding to seven
discordant objects, labelled in Figure \ref{z}. However, if we consider only the 31 objects with reliable 
$z_{spec}$ (flags = 3 or 4), $\Delta$z = -0.03, and $\sigma_{z}$ = 0.11, with three outliers (8\%). 
These results point to the goodness of our fits, and thus we rather trust our photometric redshifts 
better than the spectroscopic ones for the outliers indicated (all of them with $z_{spec}$ with 
flags = 1 or 2 in the DEEP database).
We nevertheless note a slight underestimation of our photometric redshifts (see Figure \ref{z}) 
in comparison with 
the spectroscopic ones ($\Delta$z = -0.03). Although this effect is negligible,
we are aware of it, and we assume that all the calculated z$_{phot}$'s might be affected by this slight underestimation.


Based on the good agreement between spectroscopic and photometric redshifts in this fairly large 
subsample of sources (the results shown are better than
those typically obtained for AGN samples \citep{Babbedge04,Kitsionas05,Bundy08} and with practically the same 
$\sigma_{z}$ and outlier fraction than those reported by \citet{Polletta07}), we can
confidently extrapolate the results to the rest of the sample. This, together with the SED classification into the 
five groups described above, allows us to perform a reliable 
statistical analysis of the different AGN populations. 

It is worth to mention that the distribution of object type in the subsample
with spectroscopic redshifts is completely different from the total sample. 
Considering only the 31 objects with highly reliable z$_{spec}$, there are 4 {\it Starburst-dominated AGN}, 
4 {\it Starburst-contaminated AGN}, 3 {\it Type-1 AGN}, 5 {\it Type-2 AGN}, and 15 {\it Normal galaxy hosting AGN}.
Thus, it is very difficult to check the redshift failure rate for the different groups. Only for the 
{\it Normal galaxy hosting AGN} we can confirm the success in the redshift determination with this set of
templates, since $\sim$50\% of the total number of objects fitted with elliptical or spiral templates have 
z$_{spec}$ to compare with. Due to the flat and featureless SED typical of Type-1 QSOs, 
the {\it Type-1 AGN} group of templates could produce the less reliable photometric redshifts of the sample
\citep{Franceschini05}. 
We can not discard then that any subset of templates produces higher redshift failure rates
than others, but looking at the distribution of the objects belonging to the different groups of AGN in the various 
diagnostic diagrams in the following sections, and at the correlations displayed by them, we are confident 
that our SED classification and redshift determination are as good as for the {\it Normal galaxy hosting AGN} 
for the rest of the groups.

\section{Discussion}
\subsection{Classification by SEDs and photometric redshift distribution of the sample.}

Together with the photometric redshift calculations reported in the previous section, we obtain spectral
energy distribution fits, that allow us to distinguish between different types of AGN populations, i.e., whether
they are pure AGN, AGN hosted by starburst-dominated galaxies, or AGN in otherwise normal 
galaxies.

For the five main groups described before we obtain the following distribution:
{\it Starburst-dominated AGN} (24 \% of the sample), {\it Starburst-contaminated AGN} (7 \%),
{\it Type-1 AGN} (21 \%), {\it Type-2 AGN} (24 \%), and {\it Normal galaxy hosting AGN} (24 \%).

We consider the {\it Type-1 AGN},  {\it Type-2 AGN} and {\it Starburst-contaminated AGN} as 
representative groups of AGN-dominated galaxies (since their SEDs are AGN-like at all or almost all 
wavelength ranges). The {\it Starburst-dominated AGN} and {\it Normal galaxy hosting AGN} are likewise considered 
AGN somehow masked by their host emission. With this simple classification, we find that 52\% of the 
sample is AGN-dominated while
48\% is host galaxy-dominated; i.e., half of the objects in the EGS sample of AGN
show AGN-like SEDs while the other half show host-dominated SEDs.
This is consistent with the finding that between 40\% and 60\% of the Chandra-selected galaxies in the 
Hawaii Deep Survey Field SSA13 and in the Chandra Deep Field North (\citet{Barger01} and \citet{Hornschemeier01}, 
respectively) have optical spectra with no-signs of nuclear activity.

Also \citet{Barmby06}, based on the IRAC slopes ($\alpha <$ 0 for the red power-law IRAC SEDs, 
and $\alpha >$ 0 for the blue ones) divided their 
sample in sources where the central engine dominates the IRAC SEDs and stellar-dominated galaxies.
They found that 40\% of the sources have red power-law SEDs, another 40\% have blue host-dominated 
mid-infrared SEDs, and the remaining 20\% could not be fitted with a power-law.

The method employed in this paper constitutes a powerful technique of classification of high redshift 
AGN provided we are able to procure well-sampled SEDs. This is important, for instance, for 
multi-band deep surveys of galaxies for which spectroscopic data will be necessarily scarce. 
Having SEDs over the largest wavelength range as possible is mandatory to identify the entire 
AGN population \citep{Dye08}. Otherwise, depending on the observed wavelength, the galaxies could be
missclasified. This is crucial, for example, for our {\it Starburst-contaminated AGN}, that in the optical
range look like starburst galaxies, and towards redder wavelengths appear as Type-1 or Type-2 AGN.
\citet{Dye08} finds also that the results of the SED fitting show little difference between two
filtersets that span the same wavelength range, despite the number of filters used. Nevertheless, from our work, 
we find that including a large number of filters can reveal 
details in the SED shape that help the code choosing between different templates. This is crucial 
to distinguish among the different templates of a given group, for which little differences in the SED
determine the type of object, or its age \citep{Polletta07}.

\begin{figure}[!h]
\centering
\includegraphics[width=14cm]{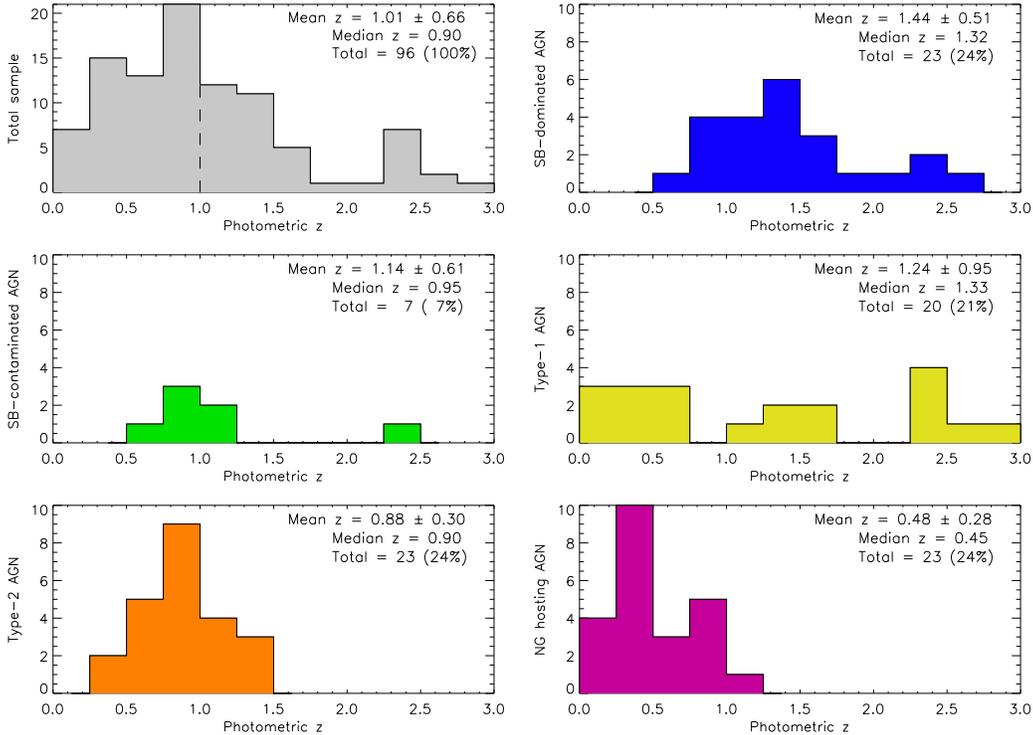}
\caption{\footnotesize{Distribution of photometric redshifts for the 96 objects with unique detection in our sample and 
for all the five main groups considered in this work. 
Mean photometric redshift for each group and corresponding standard deviation, together with the 
median redshift and the number of objects are labelled in each pannel.}
\label{hyperz}}
\end{figure}

We use now the classification of the galaxies obtained to investigate the properties of the different 
AGN groups. The distribution of redshifts for all the 96 objects with unique detection in our sample is shown 
in the top-left pannel of Figure \ref{hyperz}. 58\% of the sample have z$<$1, with the rest of the sources distributed in a
decreasing tail up to z=3. This is expected for X-ray selected samples with similar or even deeper flux
limits \citep{Hasinger03,Barger05}. Figure \ref{hyperz} also shows histograms for the photometric redshift 
distributions of the {\it Starburst-dominated AGN}, {\it Starburst-contaminated AGN}, {\it Type-1 AGN}, 
{\it Type-2 AGN} , and the {\it Normal galaxy hosting AGN} groups. 

{\it Type-2 AGN} and {\it Normal galaxy hosting AGN} are concentrated at lower redshifts, whereas the 
{\it Starburst-dominated AGN} show a high concentration around z$_{phot}$=1.3. The {\it Type-1 AGN} group has 
the largest spread in redshift, its mean value being z$_{phot}$=1.24$\pm$0.95. 
The {\it Starburst-contaminated AGN} group contains only seven objects, six of them within the redshift range [0.6,1.2], 
the other having a z$_{phot}$=2.45

This indicates that the {\it Starburst-dominated AGN} constitute the high-redshift population of AGN
masked by powerful host emission, whilst the {\it Normal galaxy hosting AGN} group represents the low-redshift 
population of low-luminosity AGN also masked by their host galaxies. Previous studies suggest that most
low luminosity AGN are found in massive, mostly spheroidal galaxies \citep{Dunlop03,Kauffmann03,Grogin05,Pierce07}.
Something similar happens with the AGN-dominated group: 
the {\it Type-1 AGN} span a large redshift distribution, the {\it Starburst-contaminated AGN} 
are located at intermediate values of redshift, and
finally, the {\it Type-2 AGN} are the low-z objects in this subsample.

\citet{Alonso04} found that $\sim$25\% of their X-ray and 24 \micron~selected sources in both the EGS and the
Lockman Hole (45 in total) show pure type-1 AGN SEDs, while more than half of the sample have stellar 
emission-dominated 
or obscured SEDs. \citet{Franceschini05} detected 99 AGN in the X-rays and mid-infrared with Spitzer
in the SWIRE survey \citep{Lonsdale03}, sorting them in three main groups:  Type-1 AGN (39\%), Type-2 AGN (23\%),
and normal and starburst galaxies (38\%).
By adding \citet{Piccinotti82} and \citet{Kuraszkiewicz03} samples, there are 32 AGN 
with z$\le$0.12, also selected both in the
hard X-rays and mid-infrared, with more than half of these sources being type-1 AGN according to their SEDs. 
Ours and other works (e.g., \citet{Alonso04,Franceschini05}) performed with SED classification of X-ray and 
mid-infrared selected AGN in a wide range of redshift (up to z $\sim$ 2-3),
when compared with the results obtained for local samples of AGN selected in the same bands, 
seem to indicate that the percentage of type-1 objects decreases with redshift, 
while the number of obscured AGN at high redshift increases.

Although the data used in this paper do not allow a deep study of the AGN feedback phenomenon, 
it is worth noticing that a redshift sequence can be readily seen in Figure \ref{hyperz}. 
Indeed, the {\it Starburst-dominated AGN} would have the highest redshifts in a decreasing 
sequence that goes through the {\it Type-1 AGN}, {\it Starburst-contaminated AGN}, and {\it Type-2 AGN},
ending with the {\it Normal galaxy hosting AGN} group, that shows the lowest redshifts. This evolutionary
sequence has been noticed for early-type galaxies by \citet{Schawinski07}. According to this recent work, 
the starbursts would start and be the dominant player after its onset. Subsequently, as the BH accretes 
enough mass, the AGN feedback reveals itself as the BH competes for the gas reservoir with the starbursts 
eventually quenching the star formation. The starburst phase thus declines, the AGN becoming dominant.
The {\it Starburst-contaminated AGN} phase would be the transition phase mentioned by \citet{Schawinski07}.
This process continue through lower ionization phases and it will end with the more quiescent 
{\it Normal galaxy hosting AGN} phases at lower redshifts.


\subsection{Correlations}

\subsubsection{Correlations for the whole sample}

One of the main advantages of the sample we are discussing is the multiwavelength coverage of the data, 
which allows us to study for the first time various correlations between ultraviolet/optical/infrared 
luminosities and X-ray luminosities for such a big AGN sample and within this range of redshift. 
The aim is to understand the behaviour of these sources in the different wavelength ranges. 

Absolute magnitudes (M$_{ABS}$) computed by HyperZ in each filter using the photometric redshifts 
and the chosen cosmological parameters, 
are used here to derive luminosities for the 96 objects with unique detection.
The HyperZ code provides the M$_{ABS}$ (including the K correction) in the
ultraviolet, optical, near- and mid-infrared filters considered. 
Regarding the X-ray data, the observed rest-frame hard and soft
X-ray luminosities are obtained from the equation $L_X = \frac{4 \pi d_{L}^{2} f_X}{(z+1)^{2-\Gamma}}$, where 
$d_{L}$ is the luminosity distance (cm), f$_{X}$ is the X-ray flux (ergs cm$^{-2}$ s$^{-1}$), and $\Gamma$ is the
photon index. In this
case, the K correction vanishes since we assume a photon index $\Gamma$ = 2 \citep{Krumpe07,Alexander03,Mainieri02},
which is the canonical value for unobscured AGN \citep{George00}. Obscured active nuclei have
considerably flatter effective X-ray spectral slopes, due to the energy-dependent photoelectric absorption of the
X-ray emission \citep{Risaliti99}. However, \citet{Mainieri02} find the same intrinsic 
slope of the X-ray spectrum for both type-1 and type-2 AGN whatever their absorption levels, with $\Gamma \sim$~2  
for an X-ray selected sample in the Lockman Hole. We therefore assume a photon index $\Gamma$ = 2 for either obscured
and unobscured AGN, and consequently no K correction is needed for the X-ray luminosities.

\begin{figure}[!h]
\includegraphics[width=12cm,angle=90]{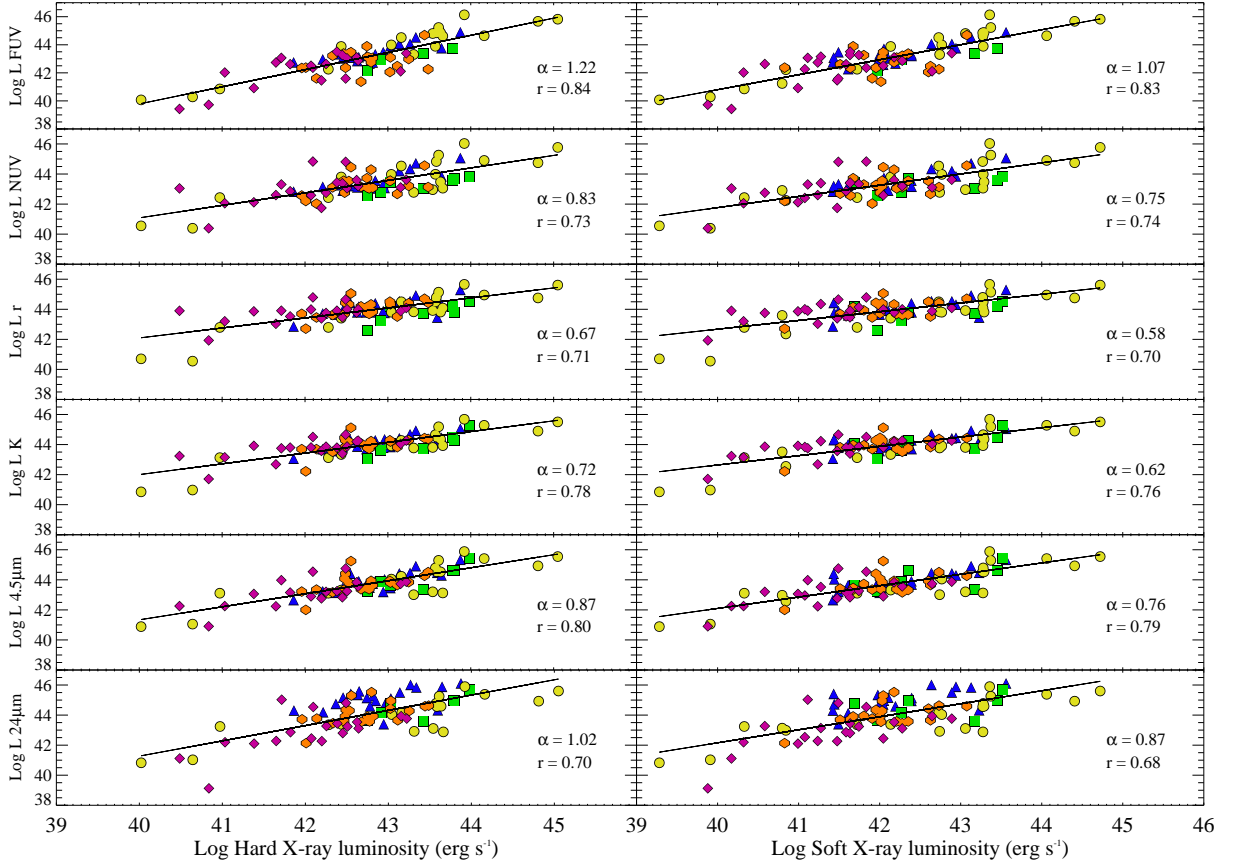}
\caption{\footnotesize{Examples of luminosity-luminosity scatter diagrams for all the objects in our sample with unique
detections and published data in the considered bands. Logarithms of $\nu$L$_{\nu}$ in the Far-UV, near-UV, r, K, 
IRAC 4.5 \micron, and MIPS 24 \micron~bands (erg s$^{-1}$) are represented versus their hard and soft X-ray
counterparts. Symbols are the same as in Figure \ref{z}, indicating the template fitting classification.}
\label{correlations}}
\end{figure}

The first row of Table \ref{types} shows the fitting slopes and correlation coefficients ($r$) 
of each scatter diagram between the far-UV/near-UV/ugriz/JK/IRAC/MIPS luminosities and the hard/soft
X-ray luminosities for the fits including all the objects with unique detection. 
In all cases the Spearman's rank correlation test has been performed, 
confirming that all the correlations are significant (p$<$0.01). 
Examples of these correlations for the far-UV/near-UV/r/K/IRAC 4.5 \micron\ /MIPS 24 \micron~luminosities 
and the hard/soft X-ray luminosities are shown in Figure \ref{correlations}. 

The expected slopes for AGN-dominated 
objects should be close to unity, since if the active nucleus is the dominant emitting source 
at all wavelengths, tight linear correlations should be drawn. Reality is different, and AGN 
are actually hosted by different types of galaxies. As it has been seen in previous sections, 
these host galaxies contaminate or even mask the AGN emission, thus deviating correlations from linear and
worsening them. Both the X-ray and mid-infrared emissions are mostly dominated by the active nuclei, 
whereas the optical and, to a lesser extent, the near-infrared bands are more affected by extinction, 
by stellar emission from the host galaxy, or by both.
This is clearly reflected in the slopes and correlation coefficients (hereafter $\alpha$ and $r$) 
of the global fits (see first row of Table \ref{types}).
Although correlations are all significant, with both the slopes and correlation coefficients close 
to unity, they begin getting slightly blurred as wavelength increases from the bluest optical bands 
up to the K band, improving again in the mid-infrared. 
The blurring is more noticeable when soft instead of hard X-rays are considered, 
due to the higher obscuration that affects the lower energies. 

Correlations between ultraviolet and X-ray luminosities are also good. 
The slopes are $\alpha \sim$~1.2 and 1.1 for the far-UV versus both the hard and soft X-rays 
luminosities, respectively, in good agreement with early X-ray studies of AGN that find
correlations between X-ray and ultraviolet monocromatic luminosities with similar slopes: L$_{X}$ $\propto$
L$_{UV}^{\beta}$, with $\beta \sim 0.7-0.8$, thus $\alpha \sim$ 1.4-1.2 \citep{Wilkes94,Vignali03,Strateva05,Steffen06}.
Nevertheless, this range of $\alpha$ was determined by using 2 keV and 2500 \AA~luminosities, which correspond to soft
X-rays and Near-UV, respectively. The slopes measured by us for the near-UV versus both the hard and soft X-rays 
luminosities are $\alpha \sim$~0.8 and 0.7, that are lower than expected. Nevertheless, \citet{LaFranca95}
found a correlation consistent with $\alpha$ = 1 using a generalized orthogonal regression that is in better agreement
with our values.

\subsubsection{Correlations for the main AGN groups}

We also report slopes and correlation coefficients for the five main groups considered 
in the last sections separately in Table \ref{types}. 
Spearman's rank correlation tests have been performed for all scatter diagrams. Thus,
the values reported in Table \ref{types} correspond only to objects showing significant correlations (p$<$0.01).

Looking at the {\it Type-1 AGN} in Table \ref{types} we conclude that they are the
less contaminated active nuclei of the sample. Since we are seeing a direct view of the central engine, 
the emission is dominated by the AGN at all wavelengths. Therefore, these objects draw
the most tight correlations between each photometric band and either the hard 
or the soft X-ray luminosities\footnote{Note that in order to check that the good correlation displayed by {\it Type-1 AGN} luminosities is not due to a
distance effect (this group of galaxies presents the largest spread in redshift, as shown in 
Figure \ref{hyperz}), we have also analysed the corresponding fluxes, instead of luminosities, 
for all the groups considered in
this section. This way, the distance effect is eliminated from the fits. 
We find the same linear and significant correlations
for {\it Type-1 AGN} fluxes, confirming that the correlations displayed for this group of AGN are only due to their
intrinsic properties. }. 

If we look at the {\it Starburst-contaminated AGN}, we find significant correlations between 
the hard X-rays and the ultraviolet and infrared bands, dissapearing  
for the soft X-rays, due to the higher obscuration in this wavelength range.

For the {\it Starburst-dominated AGN}, correlations including the hard X-ray data are better and 
more robust than those with the soft X-ray ones, for which both the slope and $r$ values are far from unity. 
This is certainly due to the higher obscuration affecting the soft X-ray emission in these objects.  
The host galaxy emission and the dust are indeed masking the AGN.  
The same, but more dramatically, happens with the {\it Type-2 AGN} group, for which all the correlations 
involving the soft X-ray emission are not significant.

It is worth to mention the worsening of the fits for the {\it Starburst-dominated AGN} when 
the hard X-ray and either the IRAC~8 \micron~or MIPS 24 \micron~emission are considered. The slopes and 
correlation coefficients of both fits move away from unity, something that is interpreted as due to 
the increasing importance of the starburst emission at these longer wavelengths.
If we look at the MIPS 24 \micron~luminosity-luminosity scatter diagrams (bottom of Figure \ref{correlations}), the
overall majority of the {\it Starburst-dominated AGN} are located above the fit line. This indicates that there 
is an excess of mid-infrared emission, as compared with the X-ray luminosity coming principally from 
the AGN. This mid-infrared excess comes from warm dust heated by the intense
star formation bursts taking place in the galaxy (in addition to the dust heated by the AGN), hence deviating the 
{\it Starburst-dominated AGN} group from the linear fit, and making the correlation non-significant when 
the MIPS 24 \micron~luminosity is considered.

The behaviour of {\it Normal galaxy hosting AGN} is completely different: correlations when either the soft or 
hard X-rays are considered are quite similar, improving towards longer wavelengths, 
where the AGN resurfaces. This group of galaxies include low-luminosity AGN hosted in normal 
galaxies that dominates 
the optical and near-infrared bands, but not the mid-infrared emission. This explain why in some fits 
performed with this subset of templates, the IRAC 8 and MIPS 24 \micron~are not completely well 
reproduced by the fit.


\subsection{X-ray properties}

Looking at the hard and soft luminosity ranges (see Table \ref{luminosity}) for each of the five main 
groups described above, we find that 
{\it Type-1 AGN} present the largest spread in luminosity, together with the highest luminosity values 
in both bands (L$_{Hard}=10^{40-45}$ and L$_{Soft}=10^{39-45}~erg~s^{-1}$, not corrected for absorption).
\citet{Alonso06} found that the majority of galaxies in their sample of X-ray detected sources in the
CDF-S fitted with Broad-line AGN (BLAGN) QSO templates showed hard X-ray luminosities in the range 
10$^{43-44}erg~s^{-1}$ (also not corrected for absorption).
The same has been found when a spectroscopic classification of the objects has been possible
\citep{Zheng04,Szokoly04,Barger05}. Our hard X-ray luminosity range for {\it Type-1 AGN} agrees with
the literature in the sense that the most luminous X-ray sources are enclosed in that range, whilst 
five sources show L$_{Hard} < 10^{43}~erg~s^{-1}$, and only three have 
L$_{Hard} < 10^{42}~erg~s^{-1}$, namely irac068644, irac027980, and irac018192, all of them with 
z$_{phot} < 0.2$. The most X-ray luminous AGN in our sample is irac040934, with a 
L$_{Hard} = 10^{45}~erg~s^{-1}$ and z$_{phot}$ = 2.42.

The behaviour of the {\it Starburst-contaminated AGN} is very similar to that of the majority
of the {\it Type-1 AGN} and exactly coincides with the hard X-ray luminosity range found by
\citet{Alonso06} for BLAGN-fitted objects. This indicate that, despite the starburst appearance of the
SEDs of these objects at longer wavelengths, in regard to their X-ray emission their AGN nature dominates.

The {\it Starburst-dominated AGN} are contained in a narrower interval of X-ray luminosities 
(L$_{Hard}=10^{42-44}~erg~s^{-1}$ and L$_{Soft}=10^{41-44}~erg~s^{-1}$), 
although reaching high values,  indicating that these 
galaxies are not only starbursts, but also masked-AGN that show strong in their X-ray emission. 
Indeed, very few {\it bona fide} starburst galaxies have L$_{X} > 10^{42}$ erg s$^{-1}$, even including
luminous sources at moderate redshifts \citep{Zezas01}. 
Only for warm ultraluminous infrared galaxies (ULIRGs)
luminosities of up to 10$^{42}$ erg s$^{-1}$ are expected \citep{Franceschini03}.
{\it Type-2 AGN} display hard X-ray luminosities ranging from 10$^{42}$ to 10$^{43}$ erg s$^{-1}$, 
staying in a much narrower range and with lower values than those of {\it Type-1 AGN}. 
The values of hard X-ray luminosities that we find for {\it Starburst-dominated AGN} and {\it Type-2 AGN}
coincide with those found by \citet{Alonso06} for their galaxies fitted with Narrow-line AGN 
(NLAGN)+ULIRG templates.

Finally, the {\it Normal galaxy hosting AGN} group shows the lowest luminosity range of any of the
groups (L$_{Hard} = L_{Soft} = 10^{40-43}~erg~s^{-1}$), which is consistent with 
the fact that they are hosting low-luminosity AGN \citep{Dunlop03,Kauffmann03,Grogin05,Pierce07}. 
The hard X-ray luminosity range 
of this group of objects coincides with typical luminosities (L$_{Hard} <$ 2 x 10$^{42}~erg~s^{-1}$) 
of the local cool ULIRGs population, except for four sources, namely irac045337, irac019616, irac016716, and
irac049420.

These results, together with the mean redshift of each group reported in Section 4.1.,   
point out that the evolution of AGN is luminosity-dependent, with low-luminosity AGN 
peaking at lower redshifts than luminous active nuclei 
\citep{Hasinger03,Hasinger05,Fiore03,Ueda03,LaFranca05,Brandt05,Bongiorno07}.

\subsection{Infrared and optical properties}

The IRAC mid-infrared colors have been used as a diagnostic tool to separate AGN from non-active 
galaxies and stars in
different samples \citep{Lacy04,Hatziminaoglou05,Stern05,Alonso06,Barmby06,Donley07}. Particularly, 
\citet{Stern05} show an IRAC color-color diagram for the AGES sample, with all their objects 
spectroscopically classified. They found that BLAGN are clearly separated from Galactic 
stars and ordinary galaxies in their diagram, with the NLAGN located both inside 
and outside of the active galaxies area. 

\begin{figure}[!h]
\includegraphics[width=10cm,angle=90]{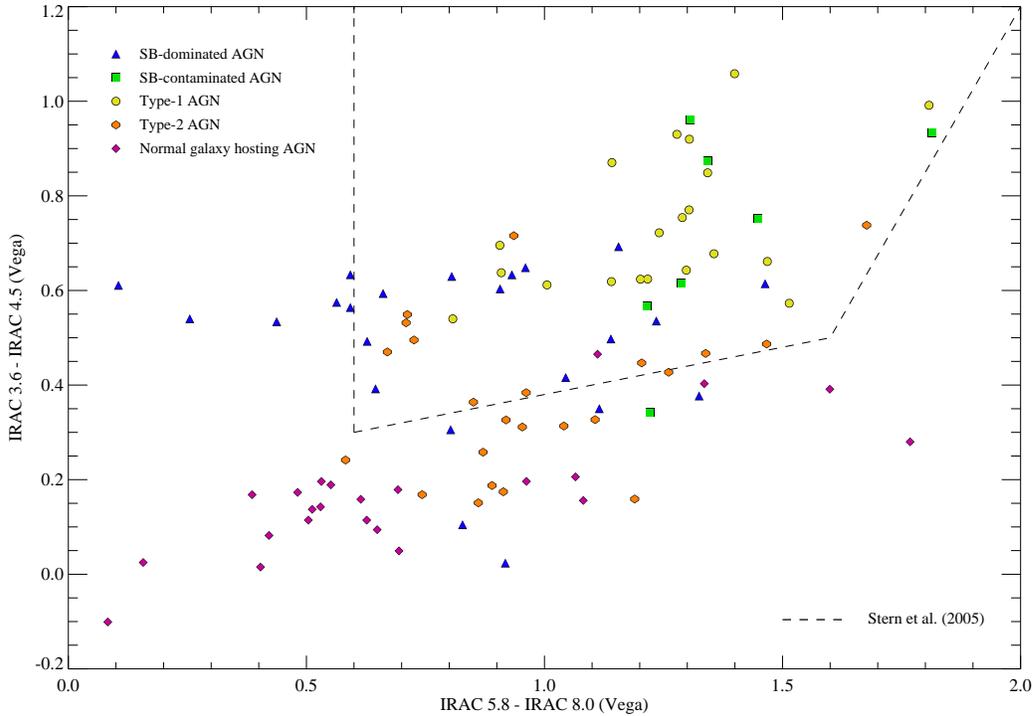}
\caption{\footnotesize{IRAC color-color plot for the 96 sources in our sample. Symbols are the same as in 
Figure \ref{z}, indicating the template fitting classification. The dashed line 
corresponds to the \citet{Stern05} empirical separation of AGN in their sample.}
\label{color}}
\end{figure}

An IRAC colour-colour diagram for our sample is represented in Figure \ref{color}. The different symbols 
indicate the template fitting classification. The dashed line in Figure 
\ref{color} corresponds to the \citet{Stern05} empirical separation of AGN in their sample. 
In our case, this region includes all the {\it Type-1 AGN}, and all but one of the 
{\it Starburst-contaminated AGN}. This is expected, since five of the seven galaxies belonging to 
that group were fitted with the Sy1/SB/ULIRG template (SED type = 4, see Table \ref{photoz}),
while the one located outside the AGN region was fitted with the Sy2/SB template (SED type = 6). 

The only-galaxy classified as {\it Starburst-contaminated AGN} fitted with a Sy2/SB/ULIRG template 
(SED type = 13) that is contained in the \citet{Stern05} AGN region is irac046309, its
photometric redshift being $z$=2.45. 
The redshift of this source is mentioned here because, as 
\citet{Barmby06} discuss and ilustrate in their Figure 6, the AGN-dominated
templates have red mid-infrared colors and thus, lie inside the \citet{Stern05} region at all redshifts, whereas
the star-forming galaxy templates begin to move into this area as the redshift increases.
This explains why all {\it Type-1 AGN} are located inside the AGN region marked by the
dashed line, as well as the six {\it Starburst-contaminated AGN}: five are fitted with the 
Sy1/SB/ULIRG template, and the galaxy irac046309 is a high redshift Sy2/SB/ULIRG.
The {\it Normal galaxy hosting AGN}-fitted objects 
(that have the bluest IRAC colors of the sample) are excluded of this region (except for one of them).

As shown in \citet{Stern05}, the active galaxy region is contaminated with Galactic stars and normal
galaxies, with the NLAGN located both inside and outside of this area. The same happens in our
Figure \ref{color}: 
{\it Starburst-dominated AGN} and {\it Type-2 AGN} are partly contained in this area, and partly not. We have estimated
the mean redshifts of both groups of galaxies for the in- and out-objects, finding that 
the {\it Starburst-dominated AGN} 
lying outside the pure-AGN region have a mean $z$ = 1.35$\pm$0.54, while those inside have a mean 
$z$ = 1.52$\pm$0.50. Following the same trend, the {Type-2 AGN} mean redshift is $z$ = 0.79$\pm$0.26 for the
outside objects, and $z$ = 0.97$\pm$0.32 for the galaxies included in the \citet{Stern05} region. This is again consistent
with the evolution of mid-infrared colors with redshift for star-forming galaxies \citep{Barmby06,Donley08}. However, 
these mean redshifts for {\it Starburst-dominated AGN} and {Type-2 AGN} lying inside and outside the \citet{Stern05} region 
are only orientative, since the differences between them are not statistically significant.
 
The reliability of these type of diagram (mid-infrared color selection) in selectioning AGN have been questioned in 
the literature \citep{Cardamone08,Donley08}. It seems that they fail to identify a large number of X-ray selected AGN, 
finding only the most luminous. In our work, the \citet{Stern05} region wraps all the Type-1 objects, all but one
of the {\it Starburst-contaminated AGN}, and half of the {\it Starburst-dominated AGN} and {Type-2 AGN}. 52\% of our 
sample is included in this area, but the low-luminosity AGN (most of them {\it Normal galaxy hosting AGN} and several 
{\it Starburst-dominated AGN} and {\it Type-2 AGN}) are excluded. \citet{Cardamone08} find that 76\% of their
spectroscopically-selected BLAGNs fall inside this region, but only 40\% of the X-ray selected objects are included. 
Summarizing, although the diagram in Figure \ref{color} only includes half of our sample in the \citet{Stern05} region, 
it seems very effective segregating the different AGN groups.


\begin{figure}[!h]
\includegraphics[width=11cm,angle=90]{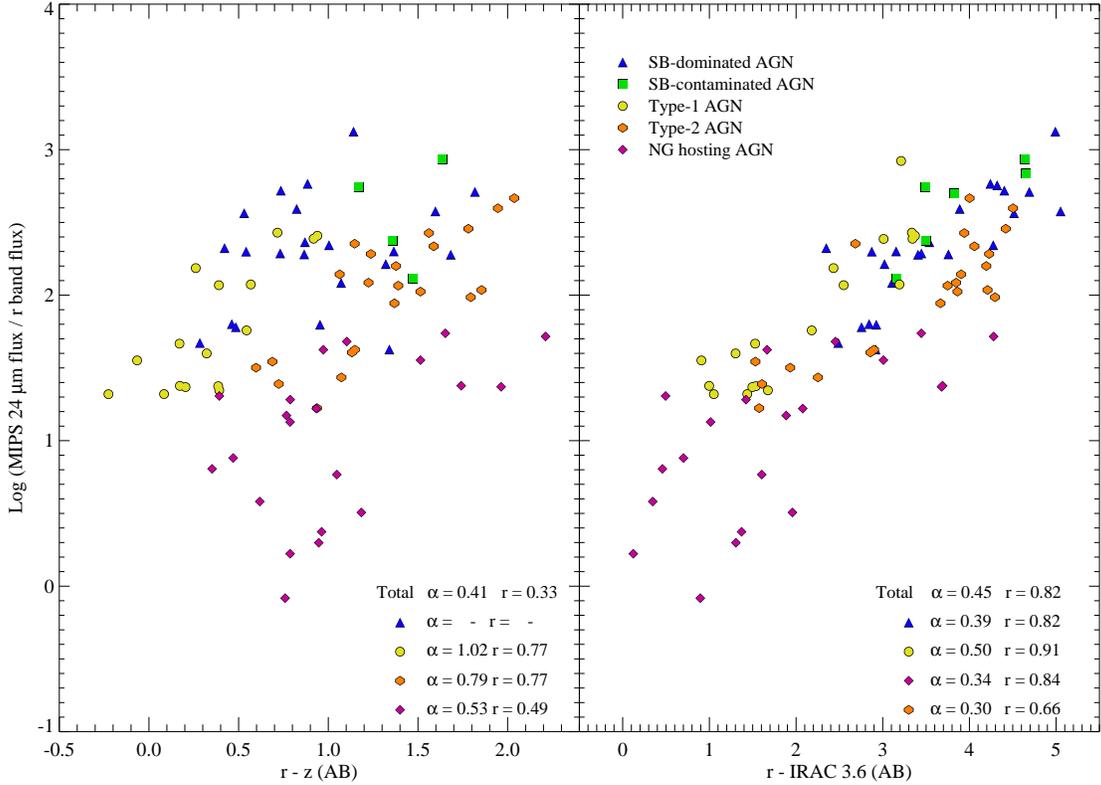}
\caption{\footnotesize{Log (MIPS 24 \micron~/ r band) flux ratio versus the r-z (AB) color for objects 
with unique detection in our
sample and available r and z magnitudes (left pannel) and the same ratio versus the r-IRAC 3.6 \micron~
color (right pannel). The slopes and correlation coefficients are labeled for the global fit and for each of the 
four groups.}
\label{obscured1}}
\end{figure}

	The left pannel of Figure \ref{obscured1} shows the mid-infrared 24~\micron~to optical 
(r band) flux ratio versus the (r-z)$_{AB}$ color for galaxies with unique detection. 
The 24~\micron~to optical flux ratio is a rough estimator of obscured activity in galaxies, 
since the 24~\micron~sources with with faint optical counterparts should be luminous AGN obscured
by dust and/or gas in the optical range \citep{Fiore08}.   
The (r-z)$_{AB}$ color depends on the obscuration present in the galaxy.

As expected for pure AGN, we find a significant correlation between the 24~\micron~to $r$
flux ratio and (r-z)$_{AB}$ for the {\it Type-1 AGN} and {\it Type-2 AGN}, because the nuclear
emission dominates both in the optical and mid-infrared wavelengths \citep{Fiore08}. However, the 
correlation is not significant for the {\it Starburst-dominated AGN} group, since they have an excess
in their mid-infrared emission, coming from the dust heated by the starbursts, in addition to the 
dust heated by the AGN. {\it Normal galaxy hosting AGN} display also a correlation between the two quantities, 
but with a different slope and lower correlation coefficient than the pure AGN objects. 
The corresponding slopes and correlation coefficients are indicated in
Figure \ref{obscured1}, except for the  {\it Starburst-contaminated AGN} group, 
due to the low number of objects fitted with this set of templates.

A segregation between the different groups is noticeable in the plot: the {\it Starburst-dominated AGN} and 
{\it Starburst-contaminated AGN} 
are shifted towards the highest values of the mid-infrared to optical ratio 
Log (24 \micron / r band flux) $>$ 1.6 , {\it Type-1 AGN} and 
{\it Type-2 AGN} are located at intermediate values, and the 
{\it Normal galaxy hosting AGN} have the lowest values of this ratio (Log [24 \micron / r band flux] $<$ 1.8).
Obscured AGN should be located towards the top right of Figure \ref{obscured1} (left pannel), 
since they have the reddest optical colors and the highest 24 \micron/r band flux ratios. 
{\it Starburst-dominated AGN}, 
{\it Starburst-contaminated AGN}, and {\it Type-2 AGN}-fitted objects 
are the most obscured galaxies in our sample, according to this diagram, although they are not as obscured as 
those in \citet{Fiore08}. We have chosen the (r-z)$_{AB}$ color instead of the most common 
(r-K)$_{AB}$ due to the lower number of objects that have available K magnitudes.

In the right pannel of Figure \ref{obscured1} the same mid-infrared 24~\micron~/ r band flux ratio is shown
against the (r-IRAC 3.6 \micron)$_{AB}$ color for galaxies with unique detection.
As much as the (r-z)$_{AB}$ color is contaminated  by the host galaxy contribution, the 
(r-IRAC 3.6 \micron)$_{AB}$ color
is dominated by the hot dust emission heated by the AGN and/or intense star formation \citep{Brusa05}. 
In this case,  all the individual groups of objects as well as the whole sample show significant
and tight correlations. The segregation between
the different groups mentioned before is clear again in this graph. 
The {\it Starburst-dominated AGN}  and {\it Starburst-contaminated AGN} clearly show the reddest colors of the
sample (r-IRAC 3.6 \micron~$>$ 2.3), while the {\it Normal galaxy hosting AGN} display the bluest, concentrated around 
r-IRAC 3.6 \micron~$\sim$ 1.6. These objects occupy the left bottom corner of the right pannel of Figure 
\ref{obscured1} because the host galaxy outshines the AGN emission at all wavelengths (except in the X-rays). 


\subsection{Objects with double detection in the optical bands.}
\label{double}

\begin{figure}[!h]
\centering
{\par
\includegraphics[width=4cm,height=4cm,angle=90]{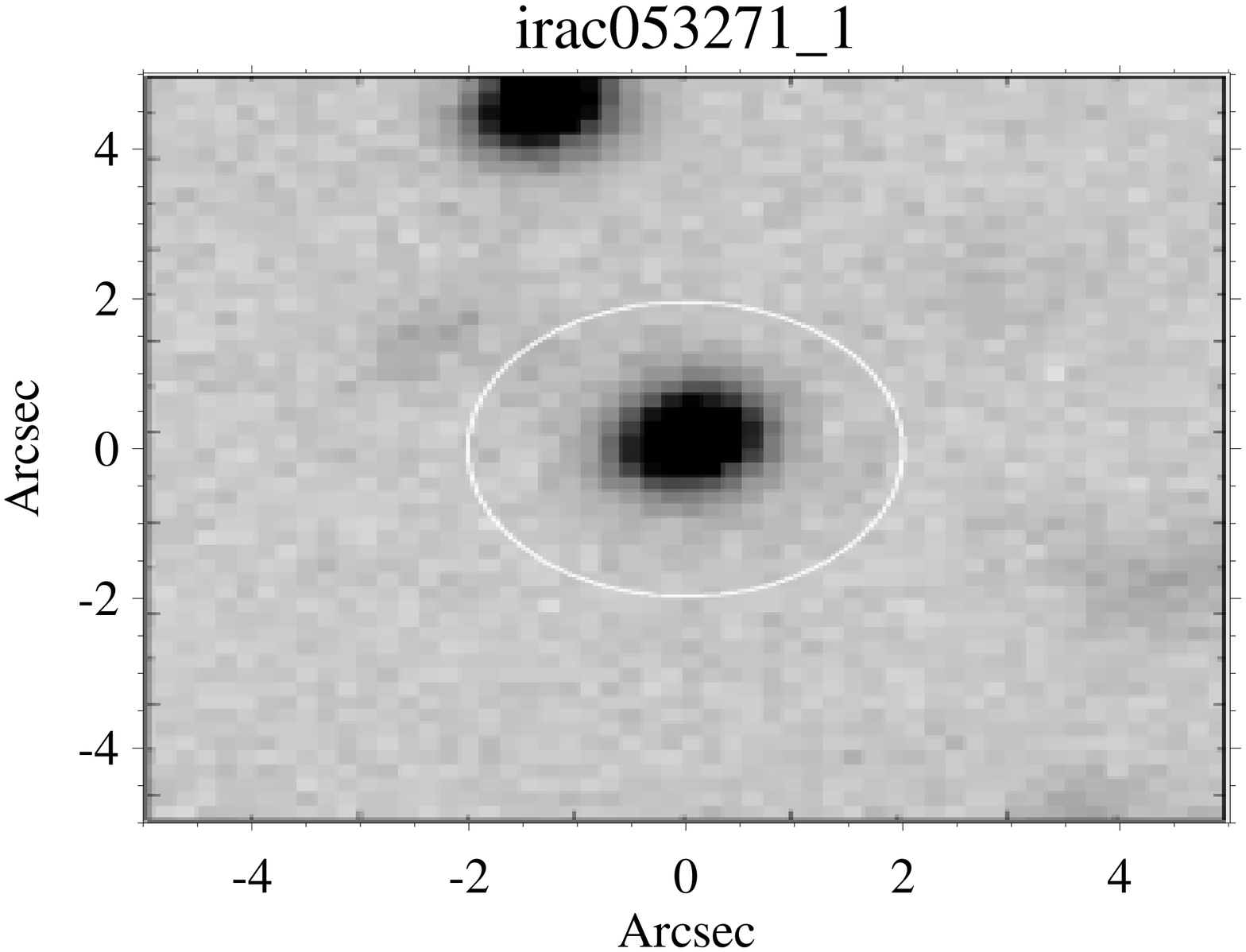}
\includegraphics[width=4cm,height=4cm,angle=90]{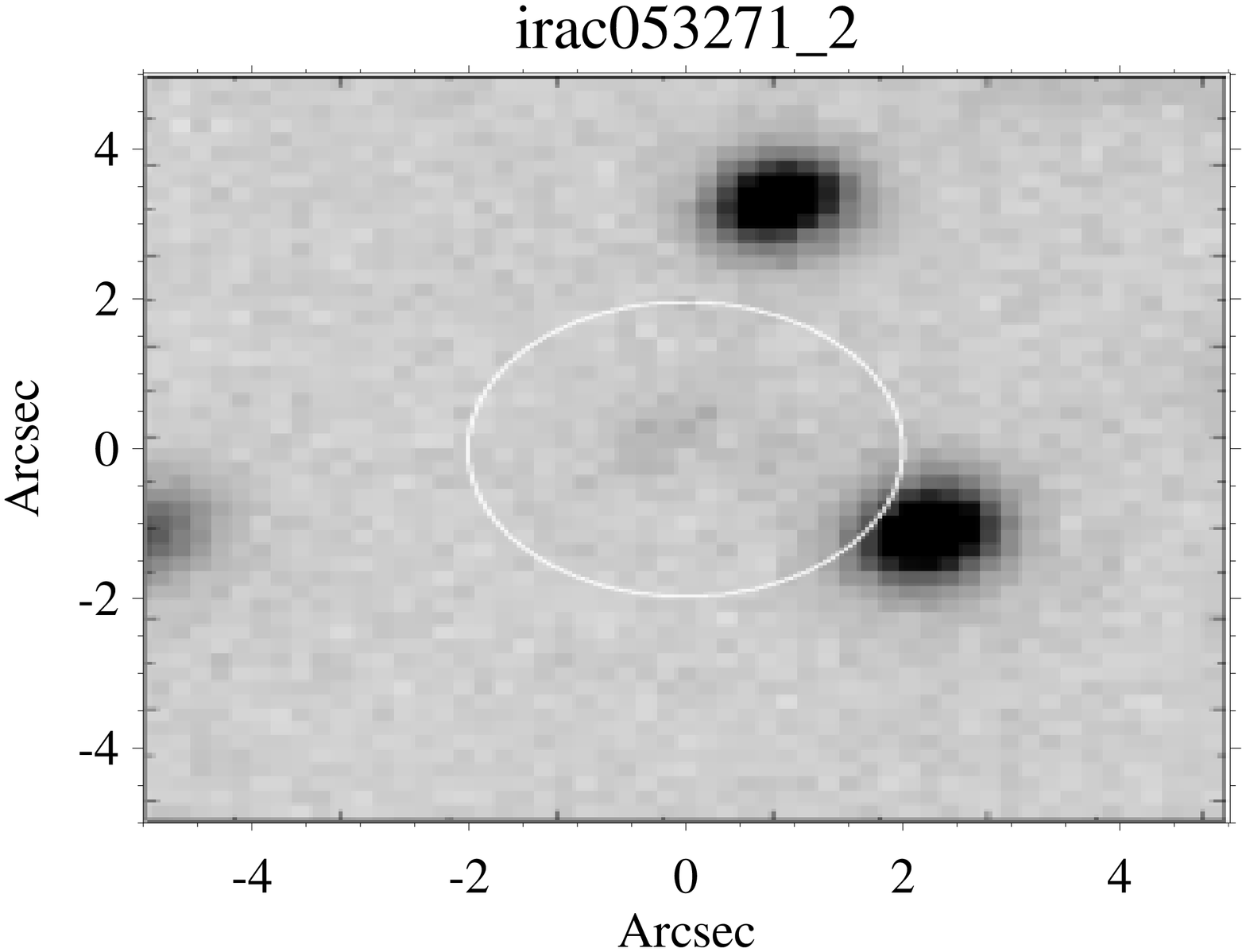}
\includegraphics[width=4cm,height=4cm,angle=90]{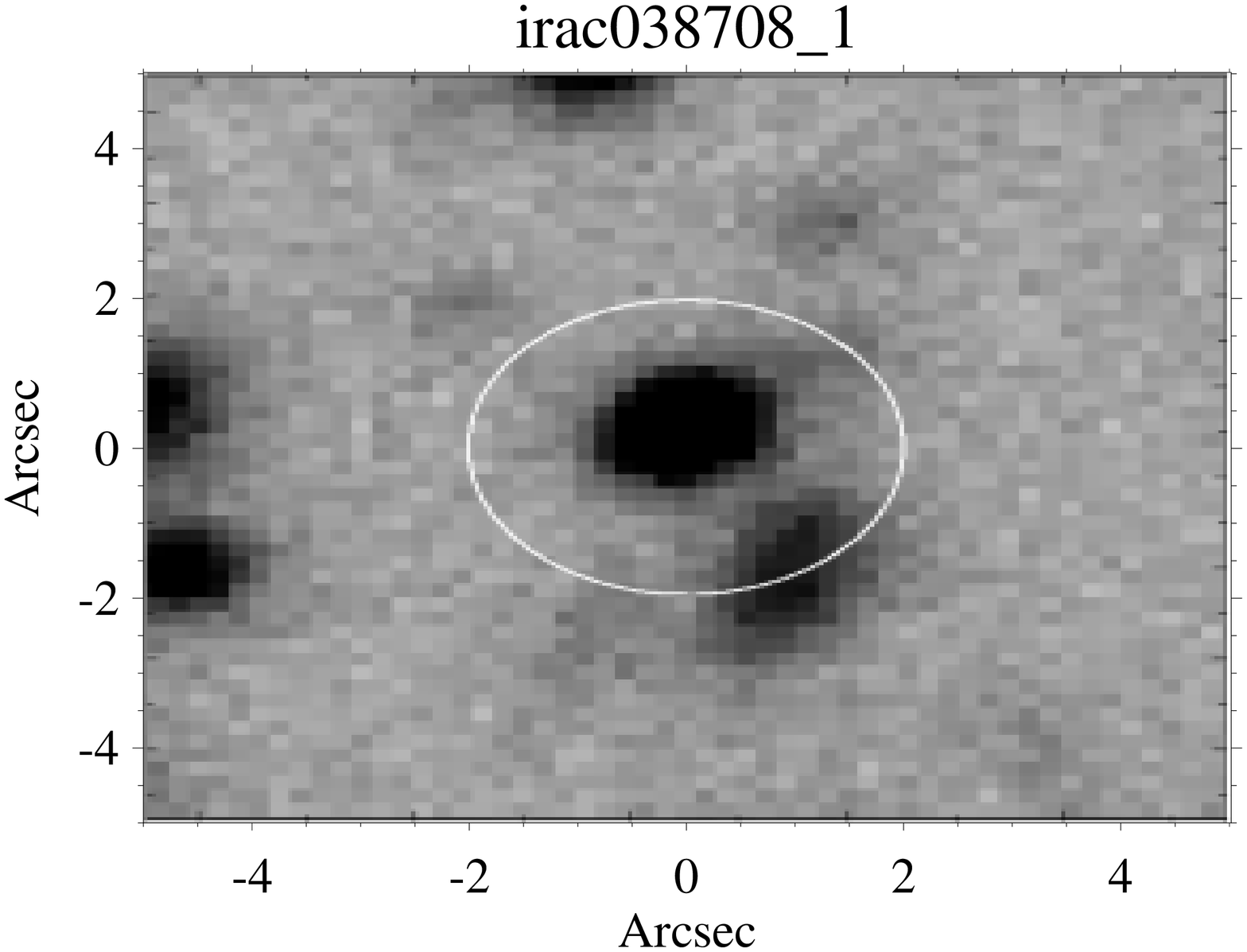}
\includegraphics[width=4cm,height=4cm,angle=90]{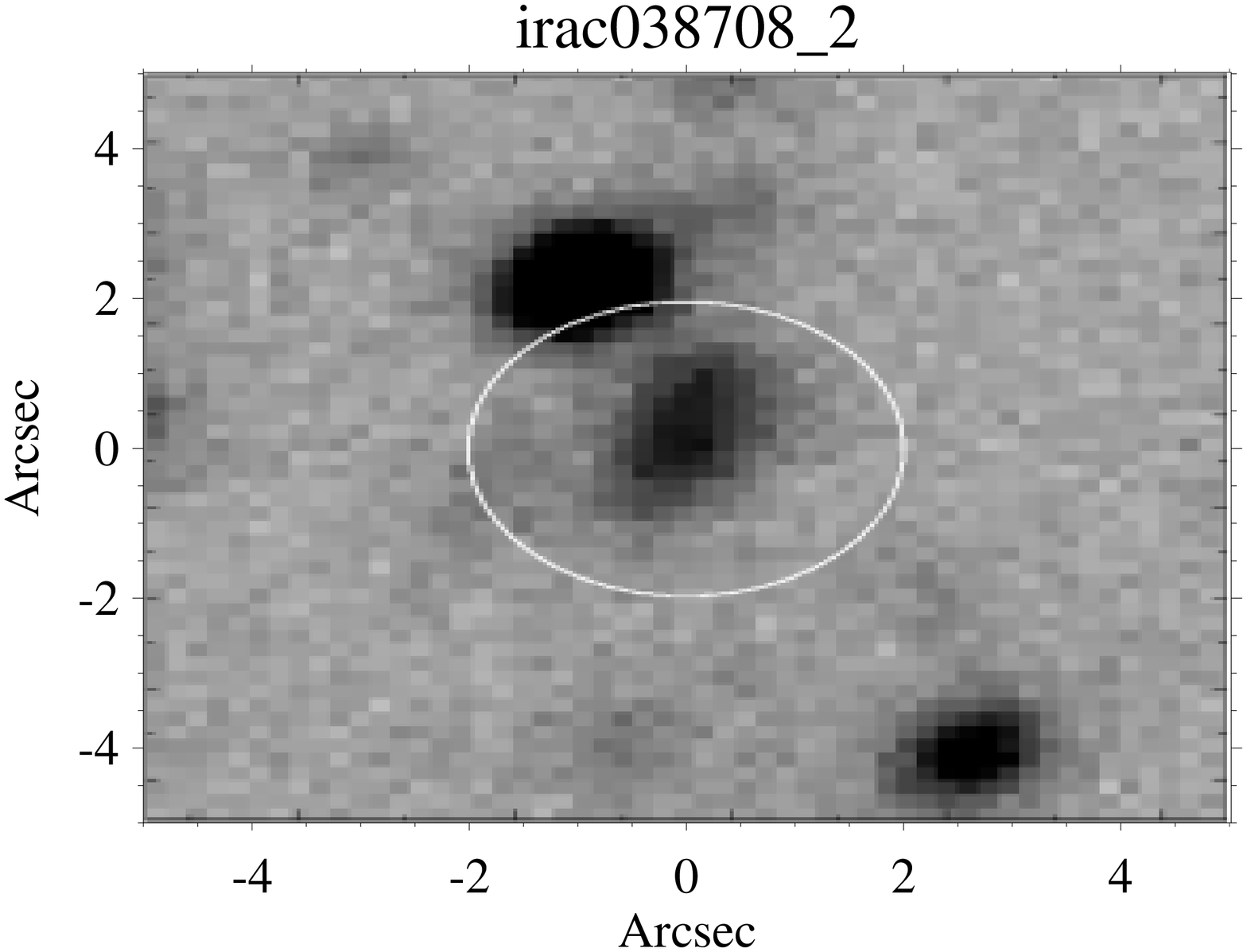}\par}
{\par
\includegraphics[width=4cm,height=4cm,angle=90]{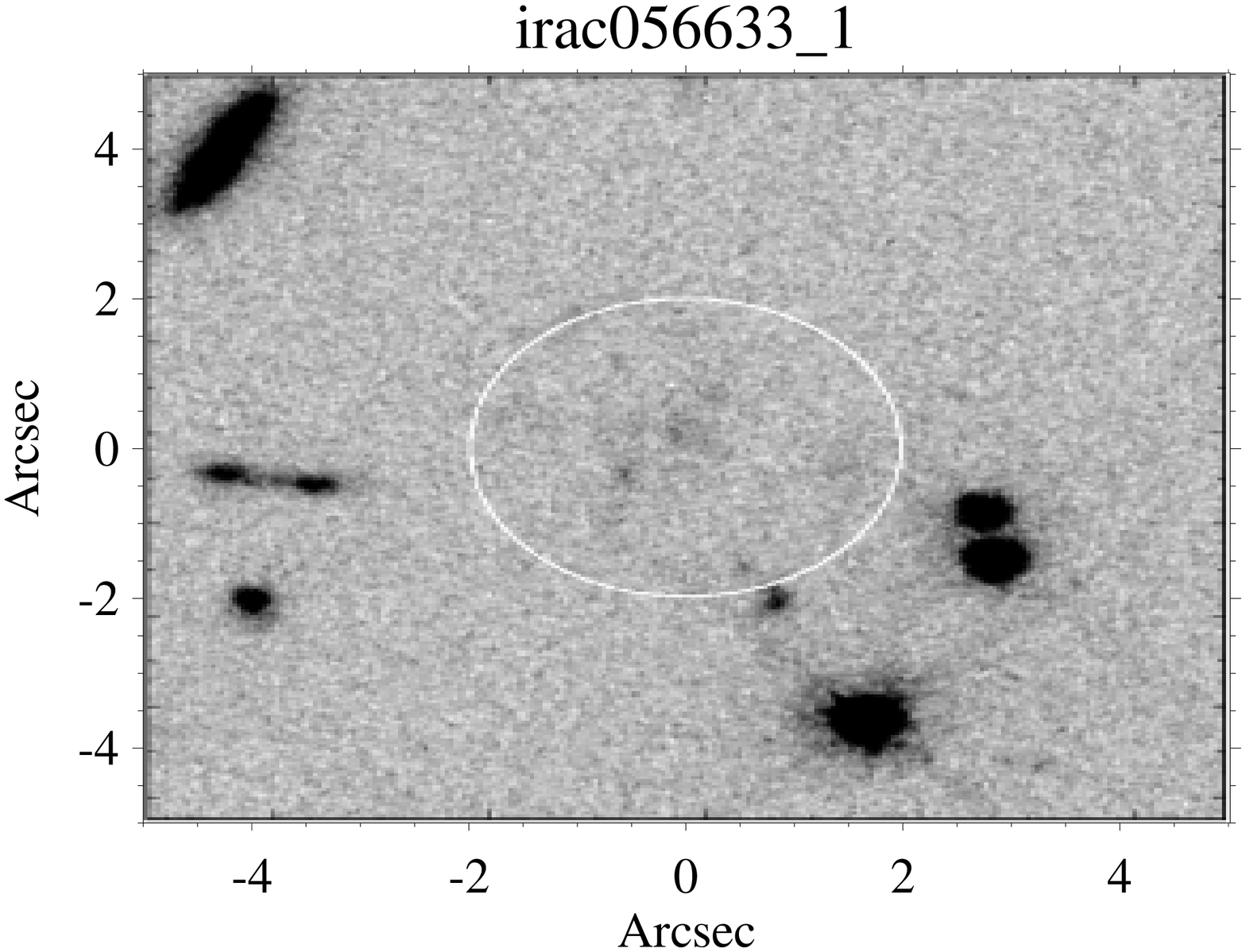}
\includegraphics[width=4cm,height=4cm,angle=90]{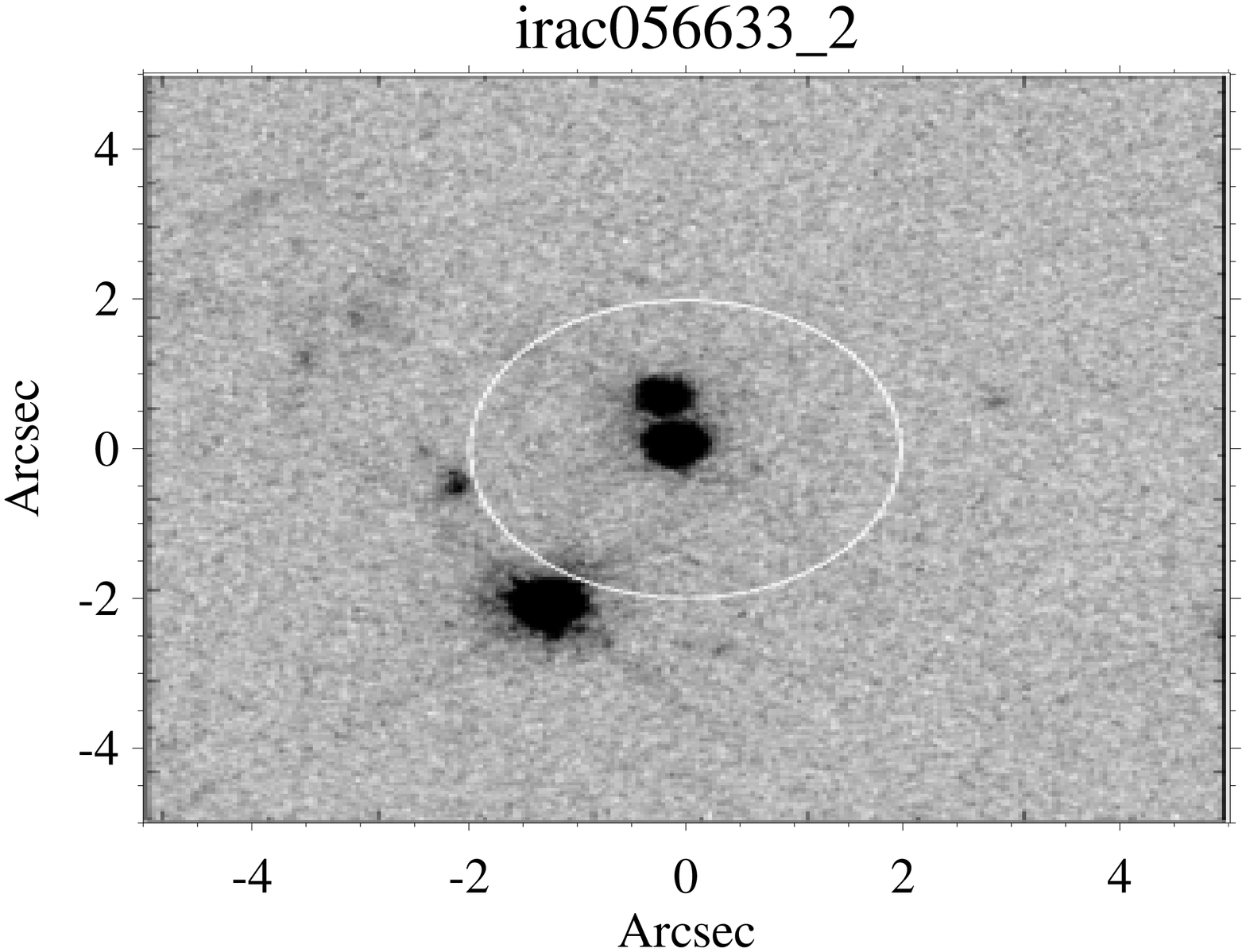}
\includegraphics[width=4cm,height=4cm,angle=90]{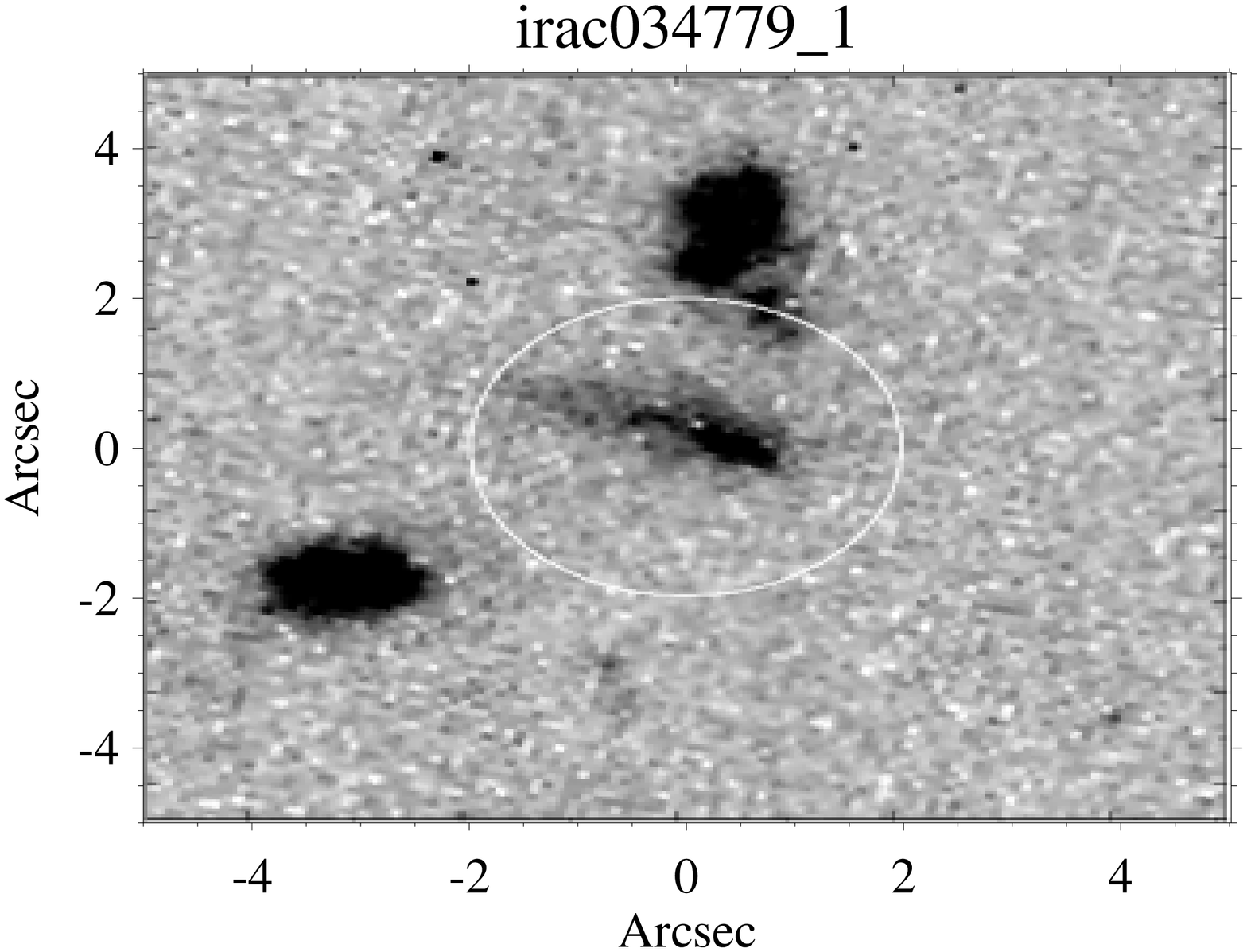}
\includegraphics[width=4cm,height=4cm,angle=90]{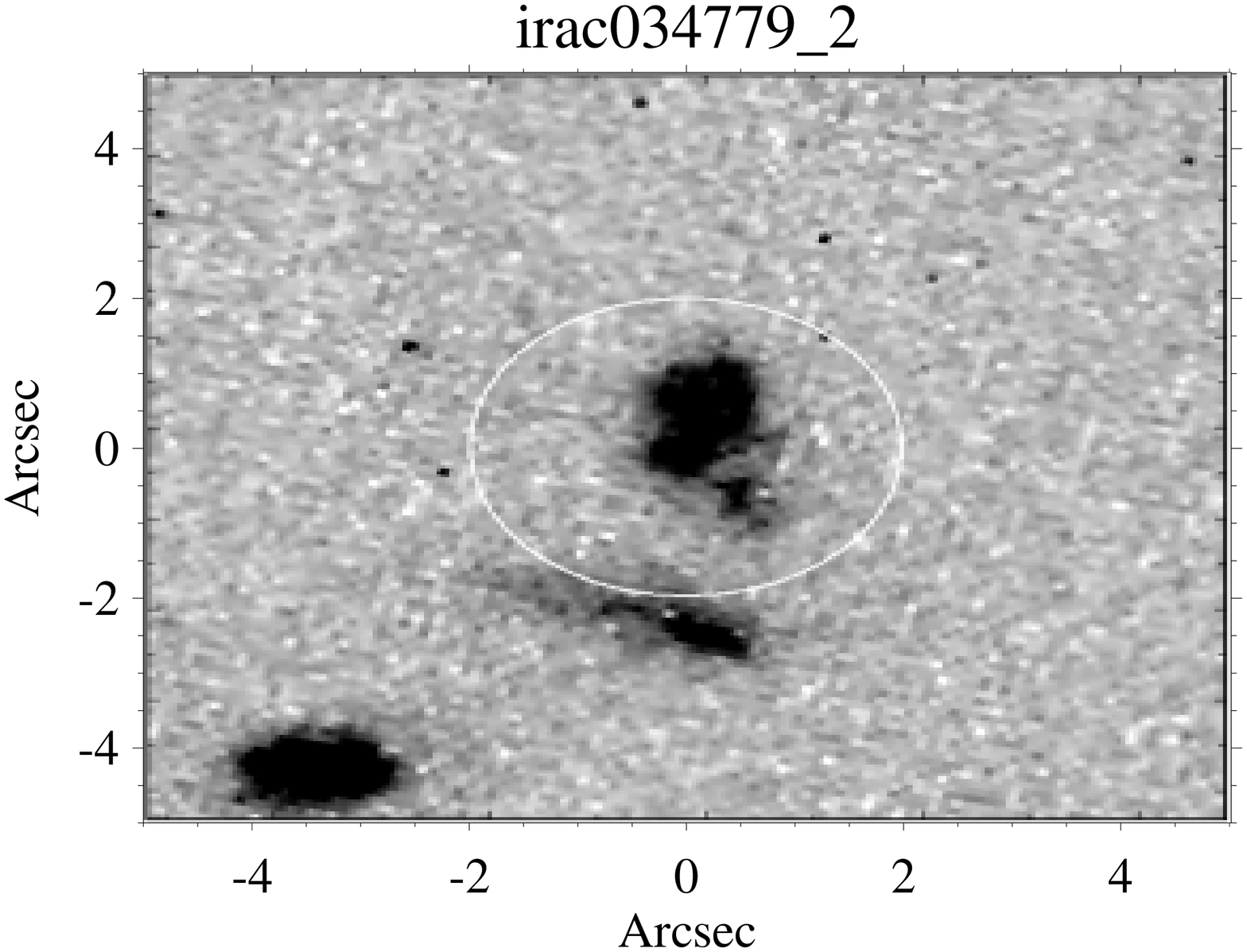}\par}
{\par
\includegraphics[width=4cm,height=4cm,angle=90]{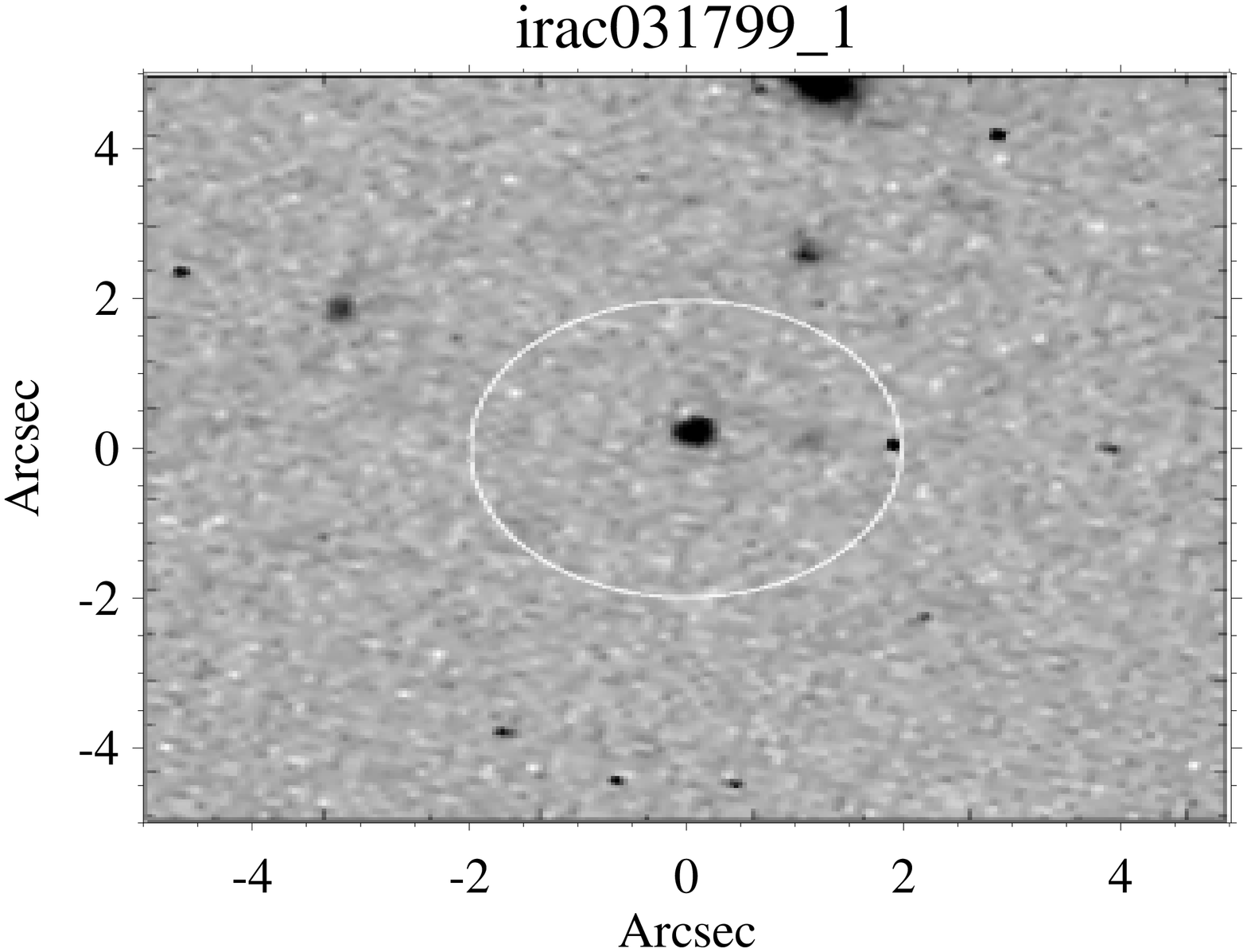}
\includegraphics[width=4cm,height=4cm,angle=90]{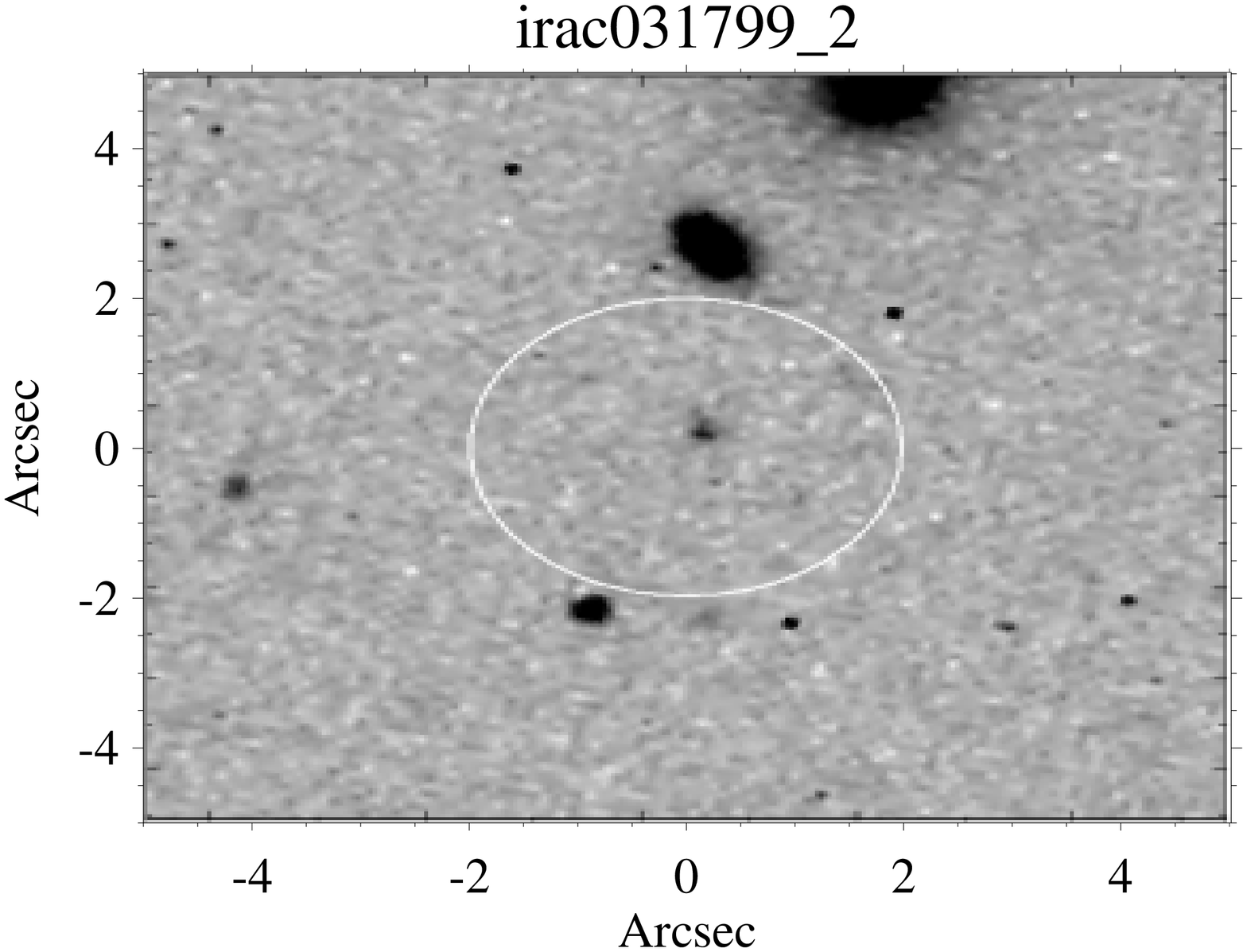}
\includegraphics[width=4cm,height=4cm,angle=90]{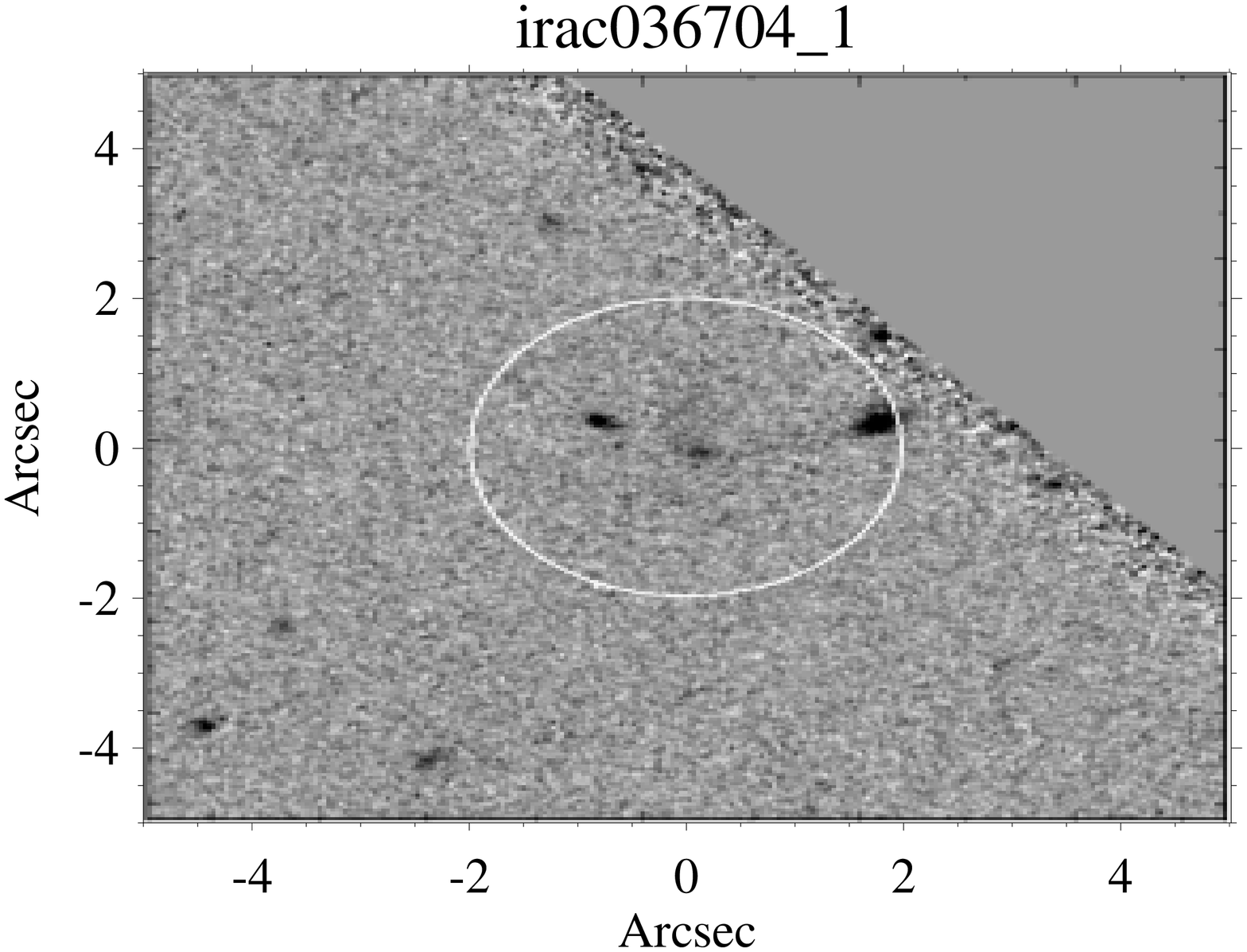}
\includegraphics[width=4cm,height=4cm,angle=90]{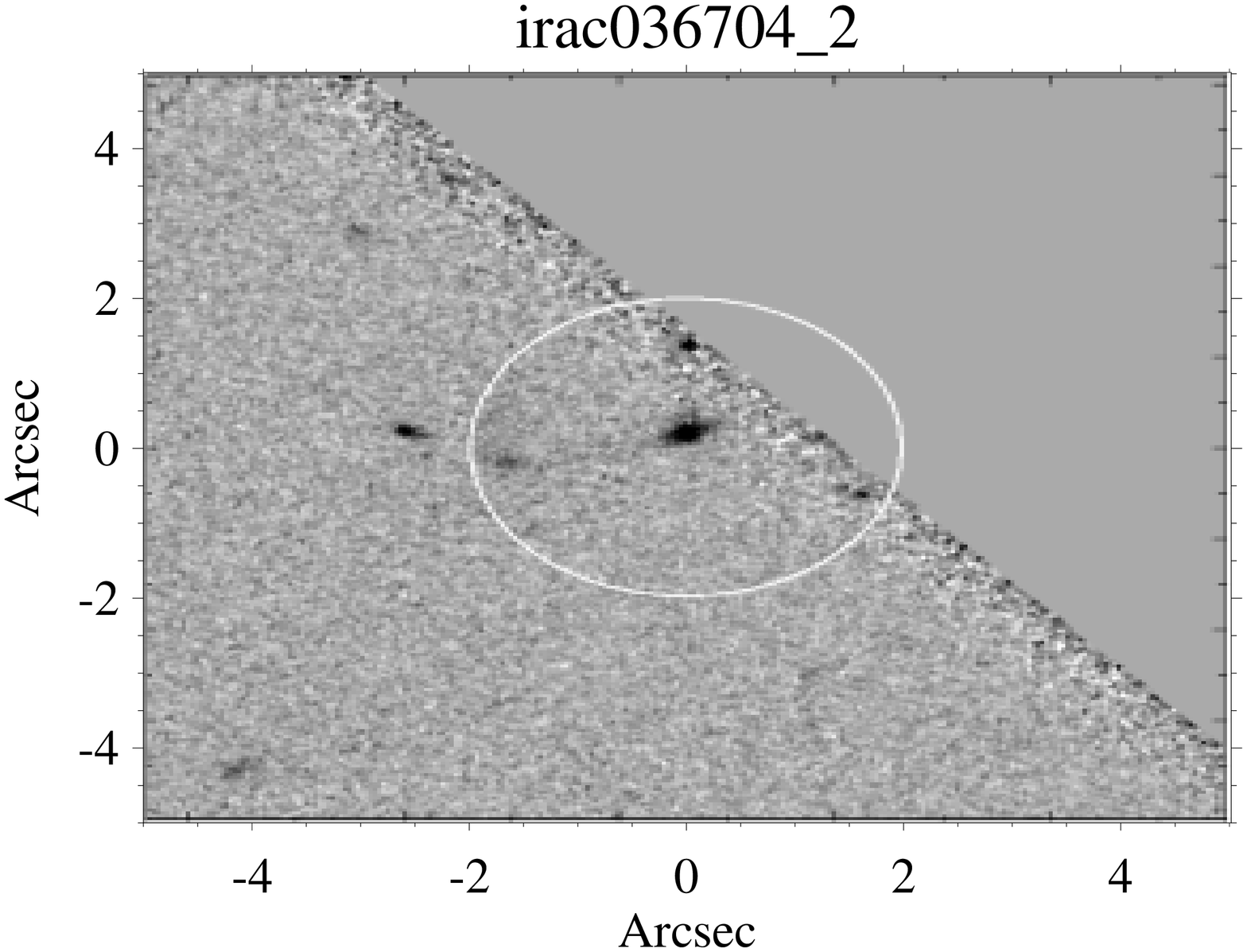}\par}
{\par
\includegraphics[width=4cm,height=4cm,angle=90]{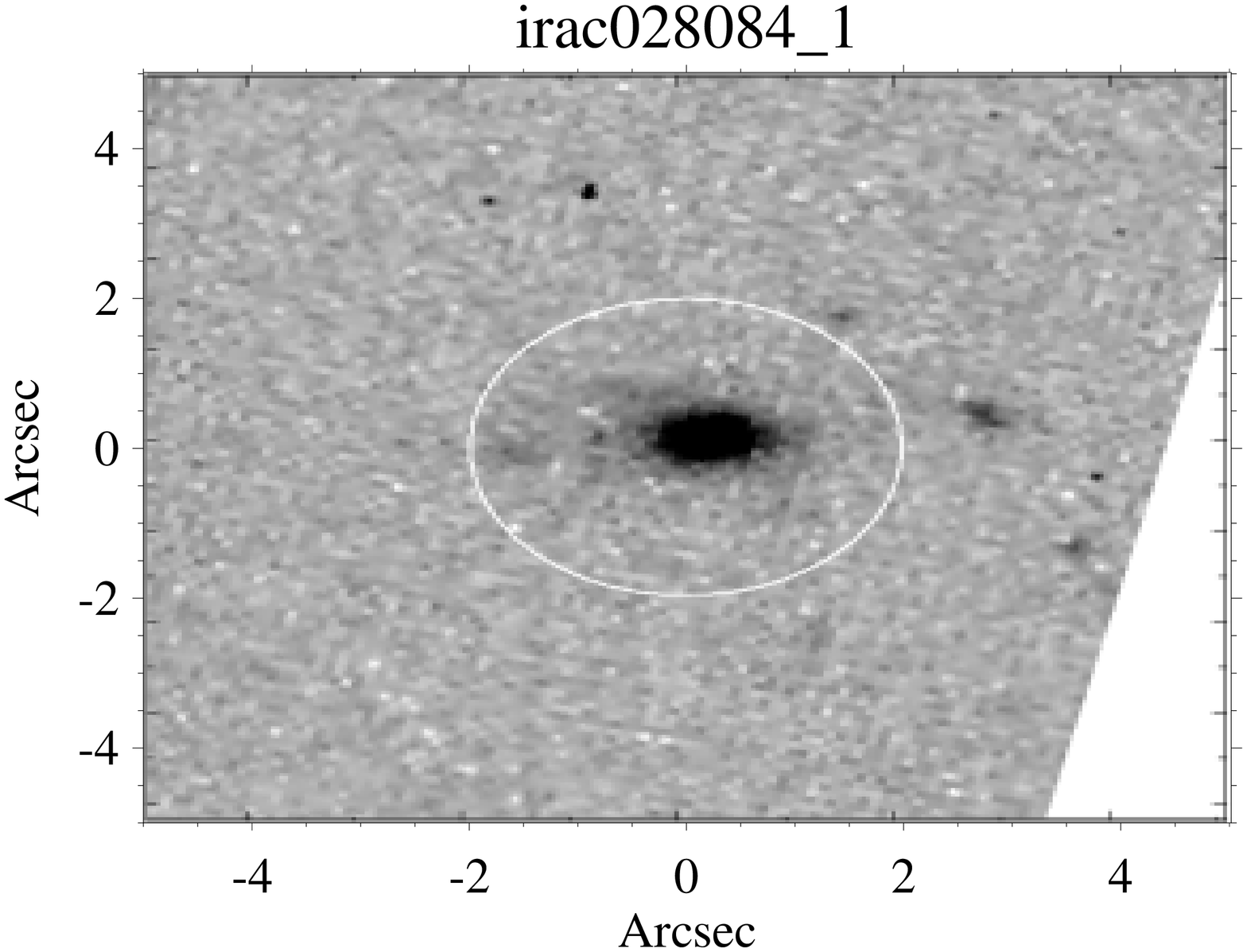}
\includegraphics[width=4cm,height=4cm,angle=90]{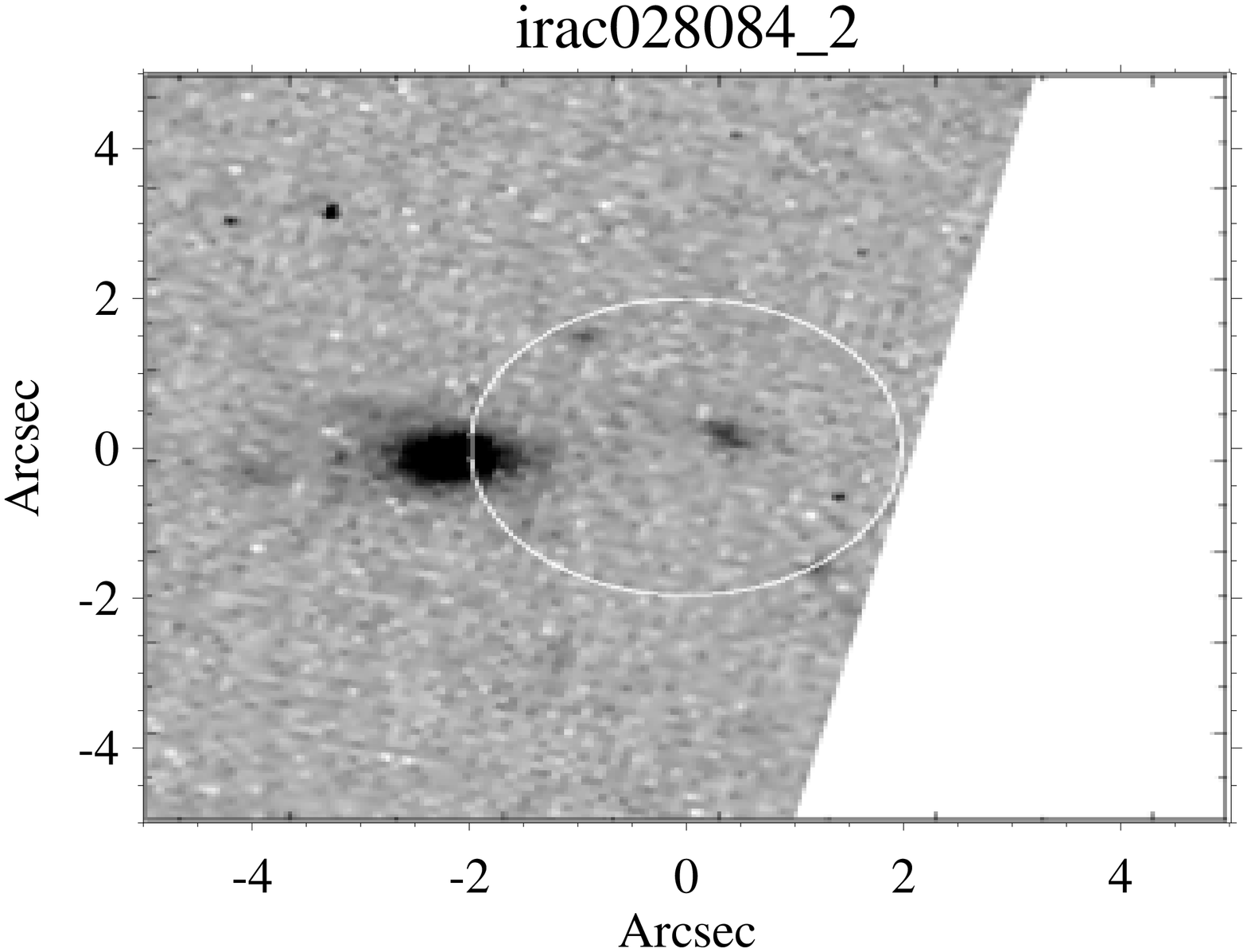}
\includegraphics[width=4cm,height=4cm,angle=90]{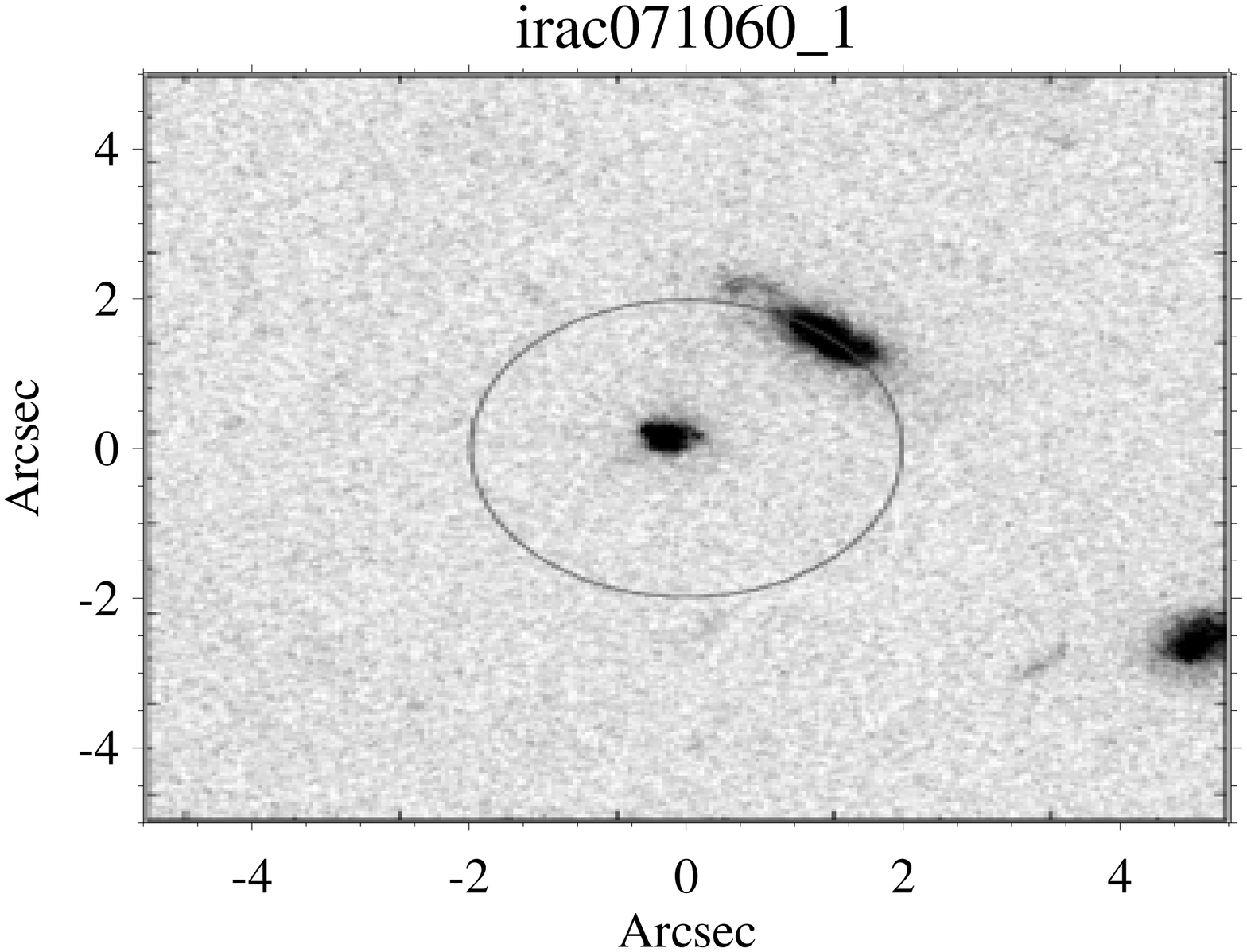}
\includegraphics[width=4cm,height=4cm,angle=90]{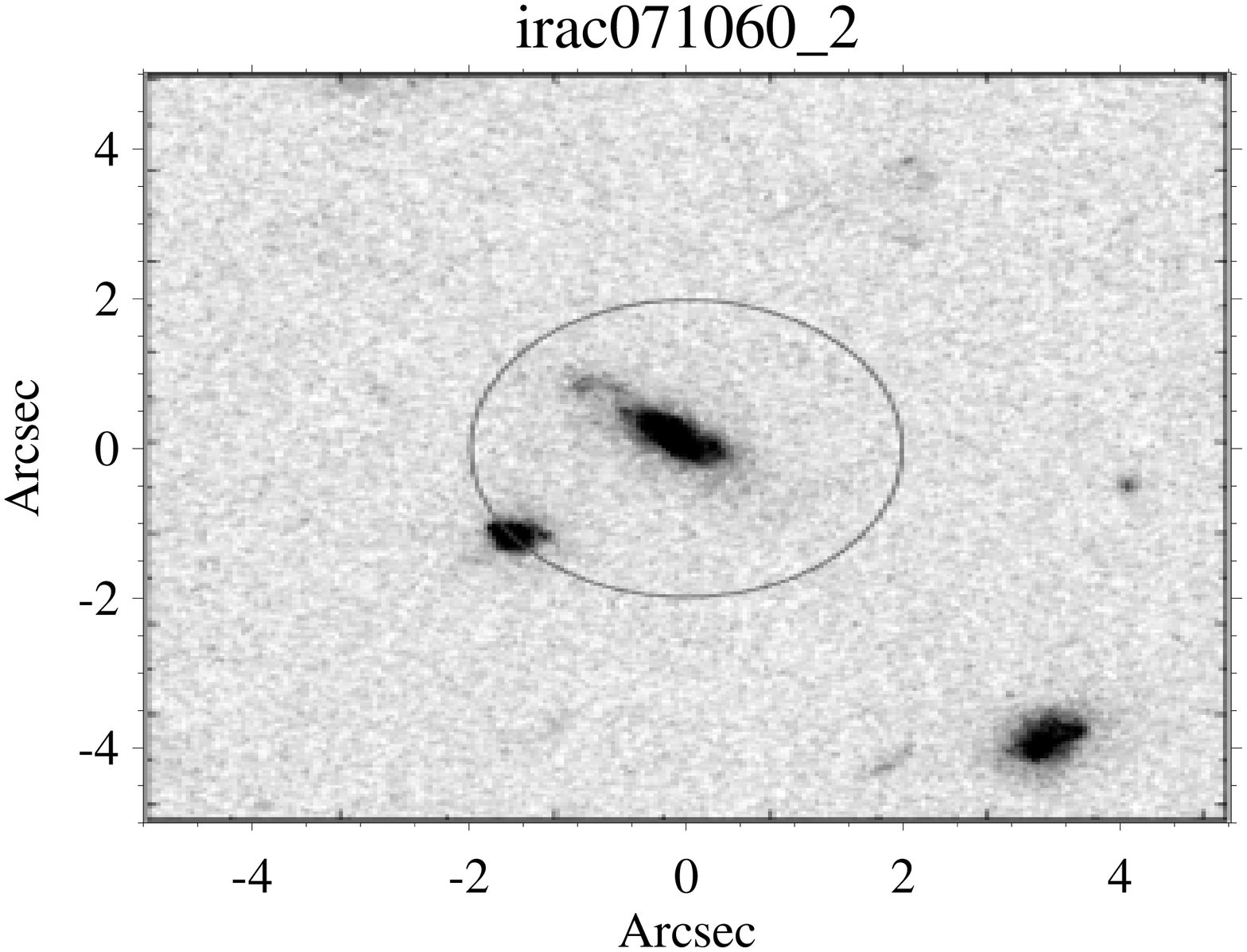}\par}
{\par
\includegraphics[width=4cm,height=4cm,angle=90]{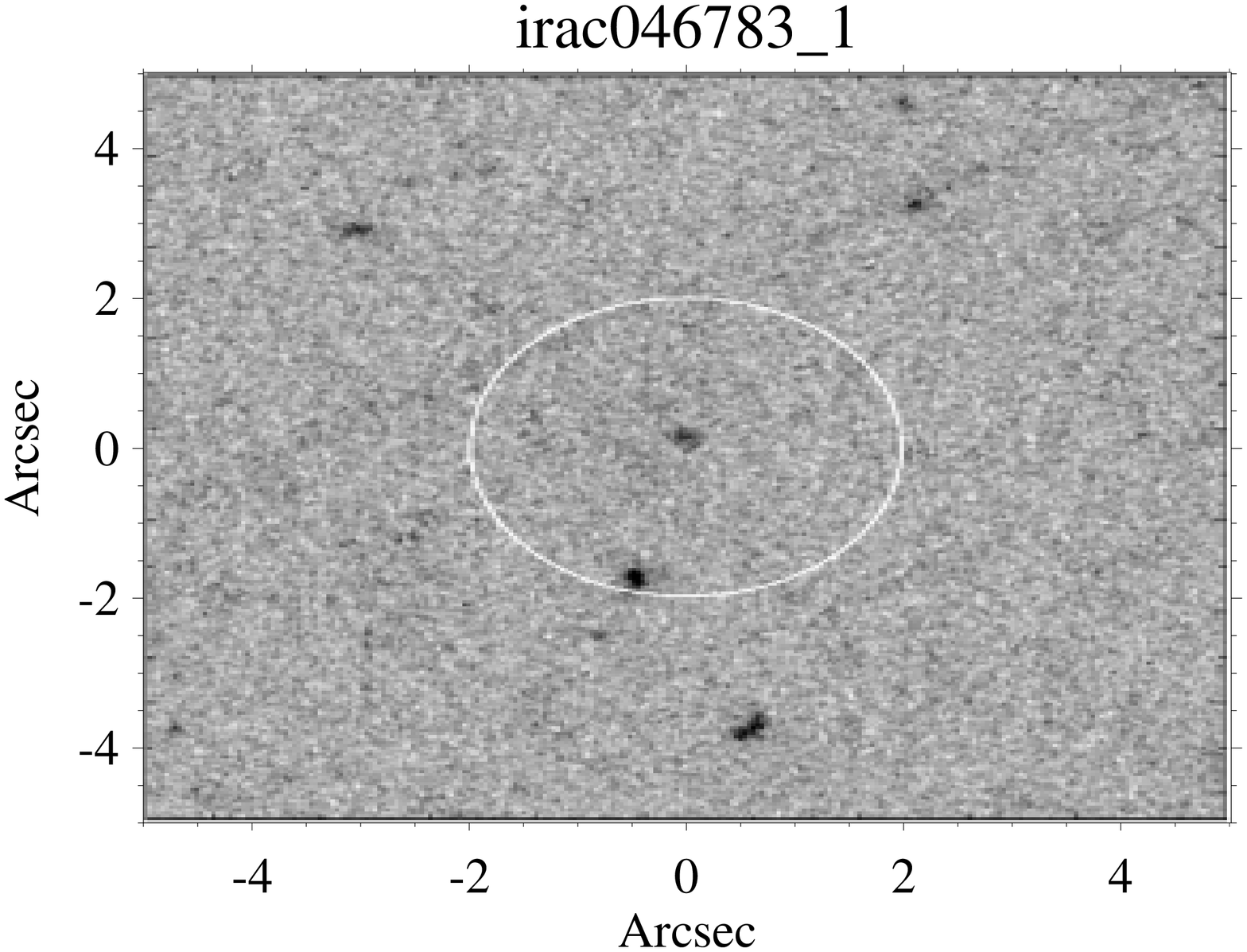}
\includegraphics[width=4cm,height=4cm,angle=90]{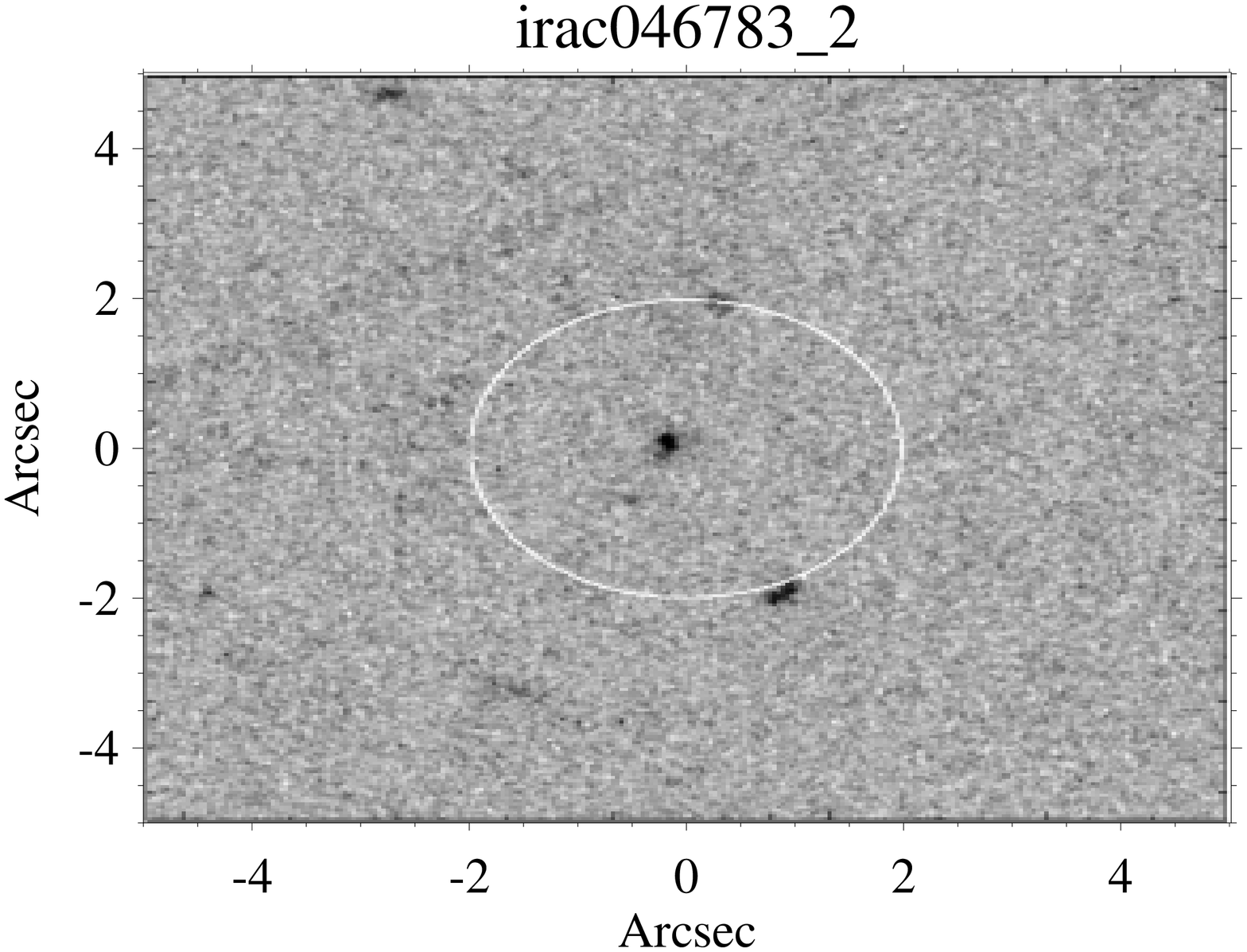}
\includegraphics[width=4cm,height=4cm,angle=90]{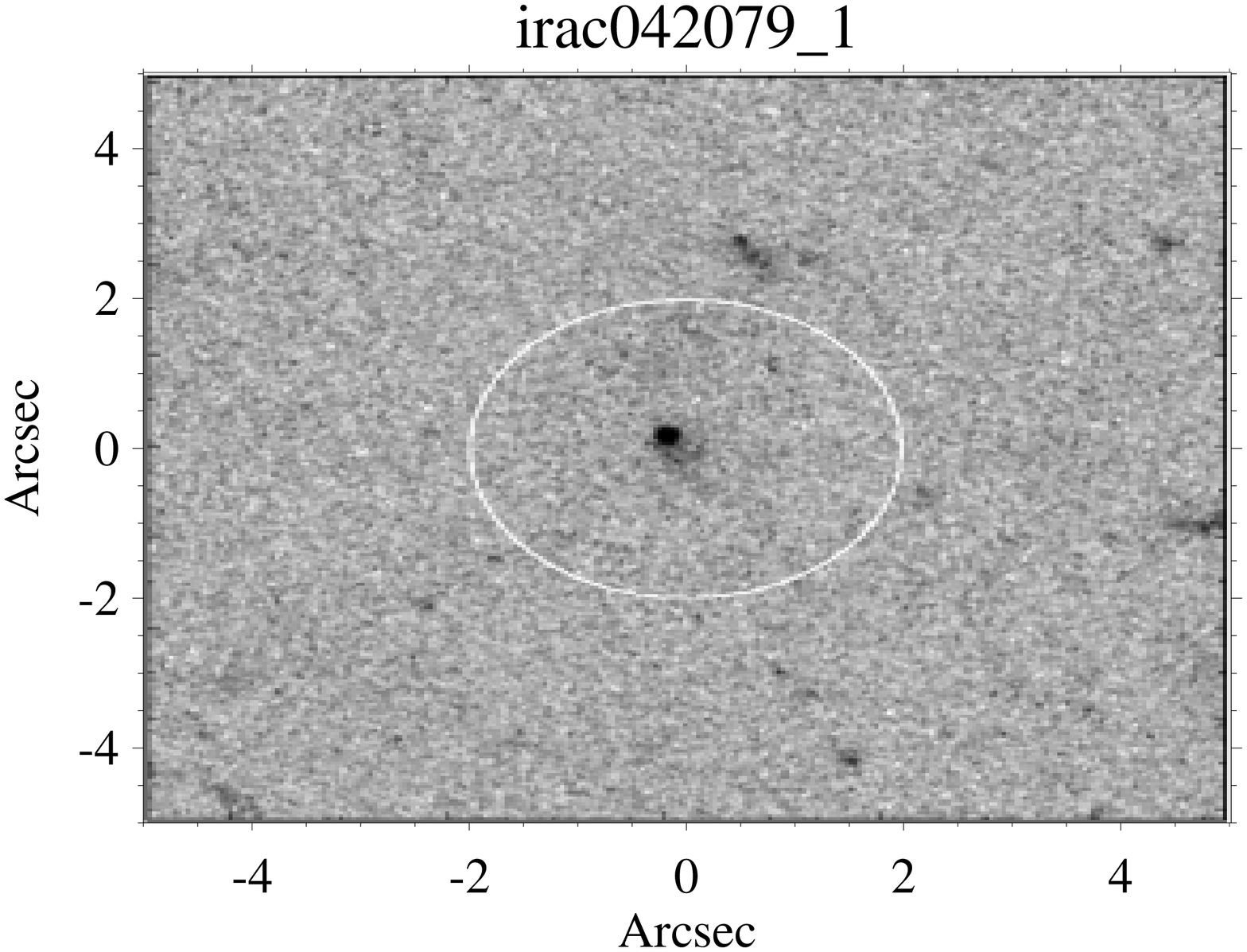}
\includegraphics[width=4cm,height=4cm,angle=90]{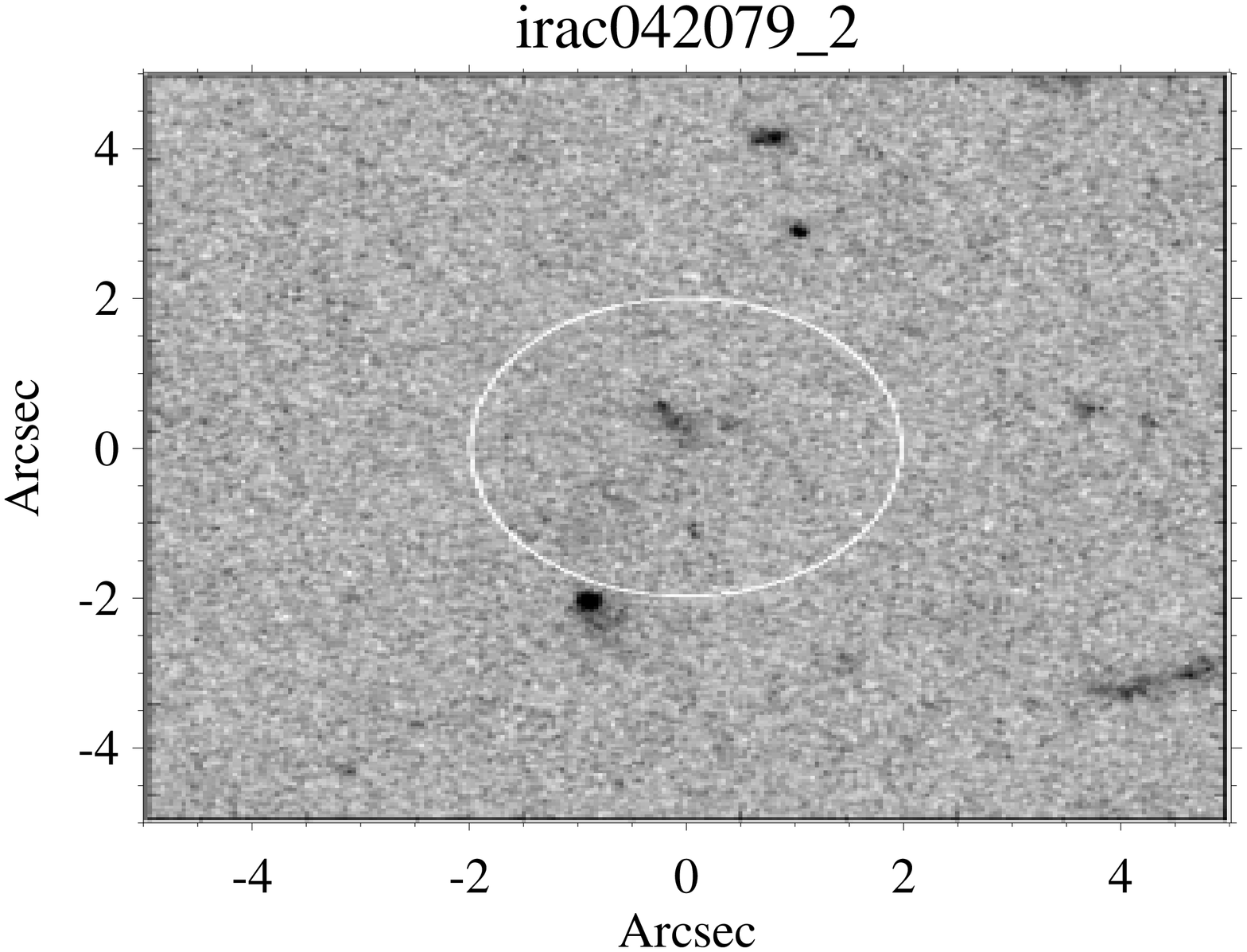}\par}
\figcaption{\footnotesize{Optical images (10\arcsec~x10\arcsec) of the sources with double detection
in our sample. 
For objects irac053271 and irac038708, R-band images from CFHTLS are shown. The rest are ACS V-band images:
irac056633, irac034779, irac031799, irac036704, irac028084, irac071060, irac046783, and irac042079.
Two stamps are shown for each pair of galaxies, the circle indicating the position of each source candidate.}
\label{cromo1}}
\end{figure}

\begin{figure}[!h]
\centering
\centering
{\par
\includegraphics[width=4cm,height=4cm,angle=90]{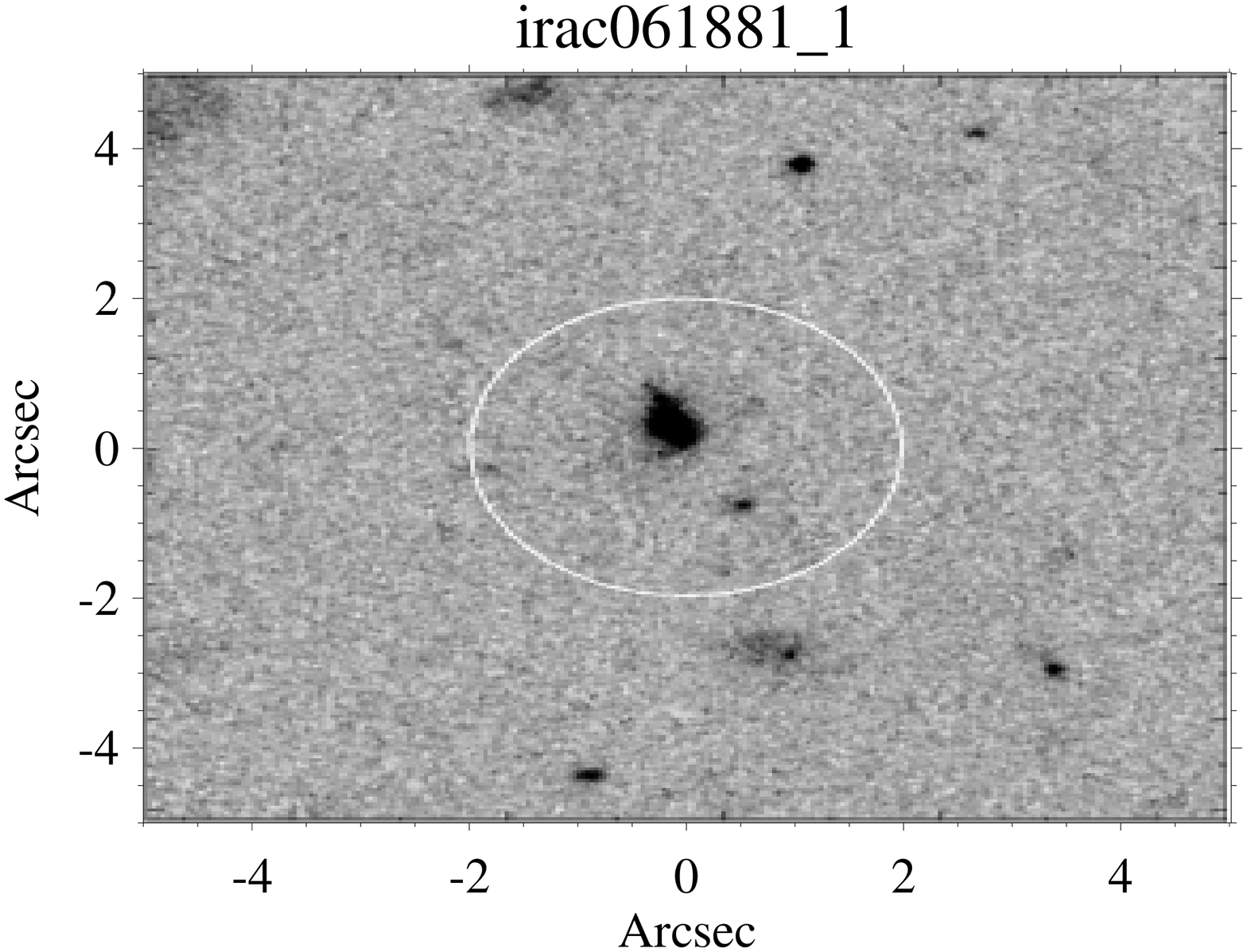}
\includegraphics[width=4cm,height=4cm,angle=90]{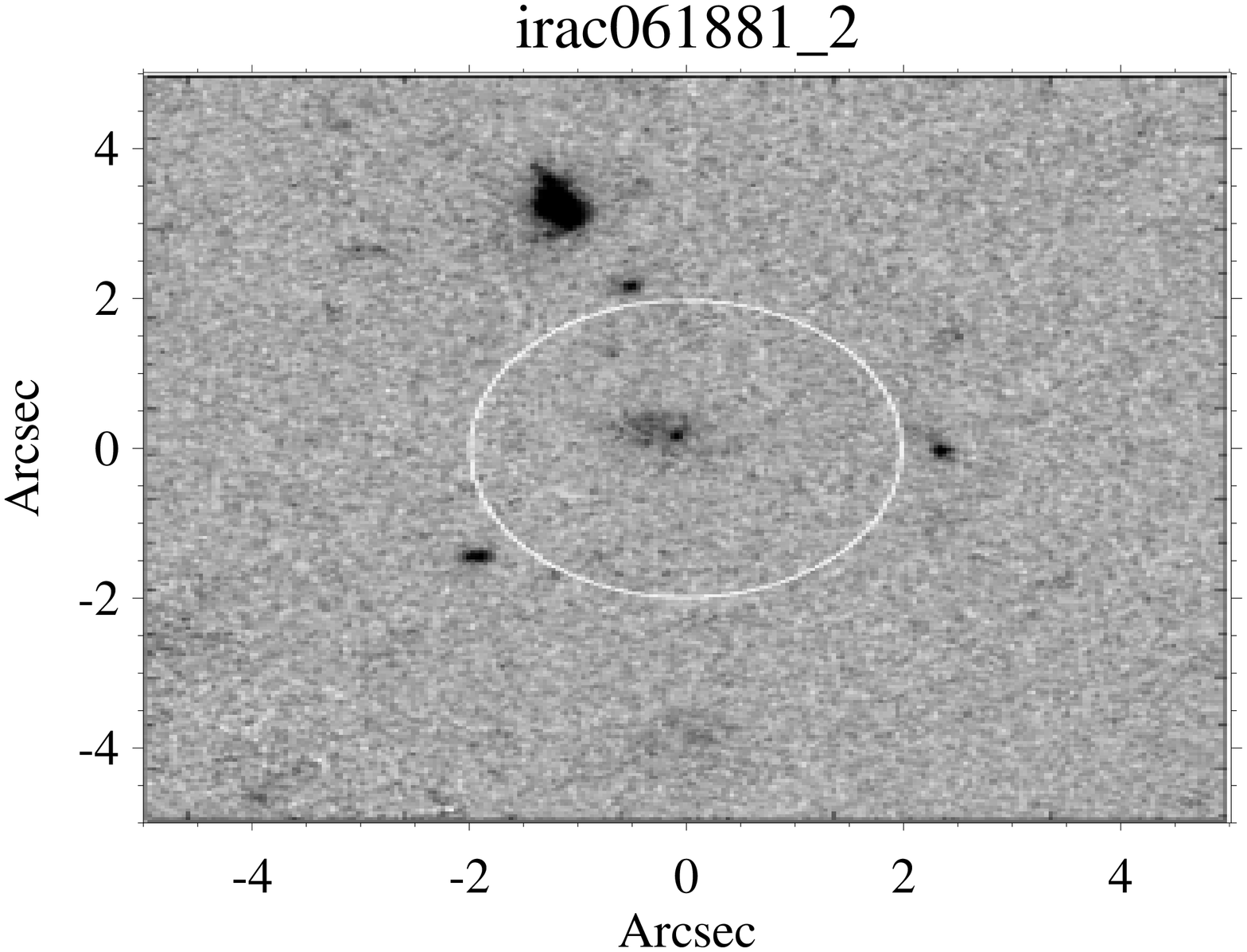}
\includegraphics[width=4cm,height=4cm,angle=90]{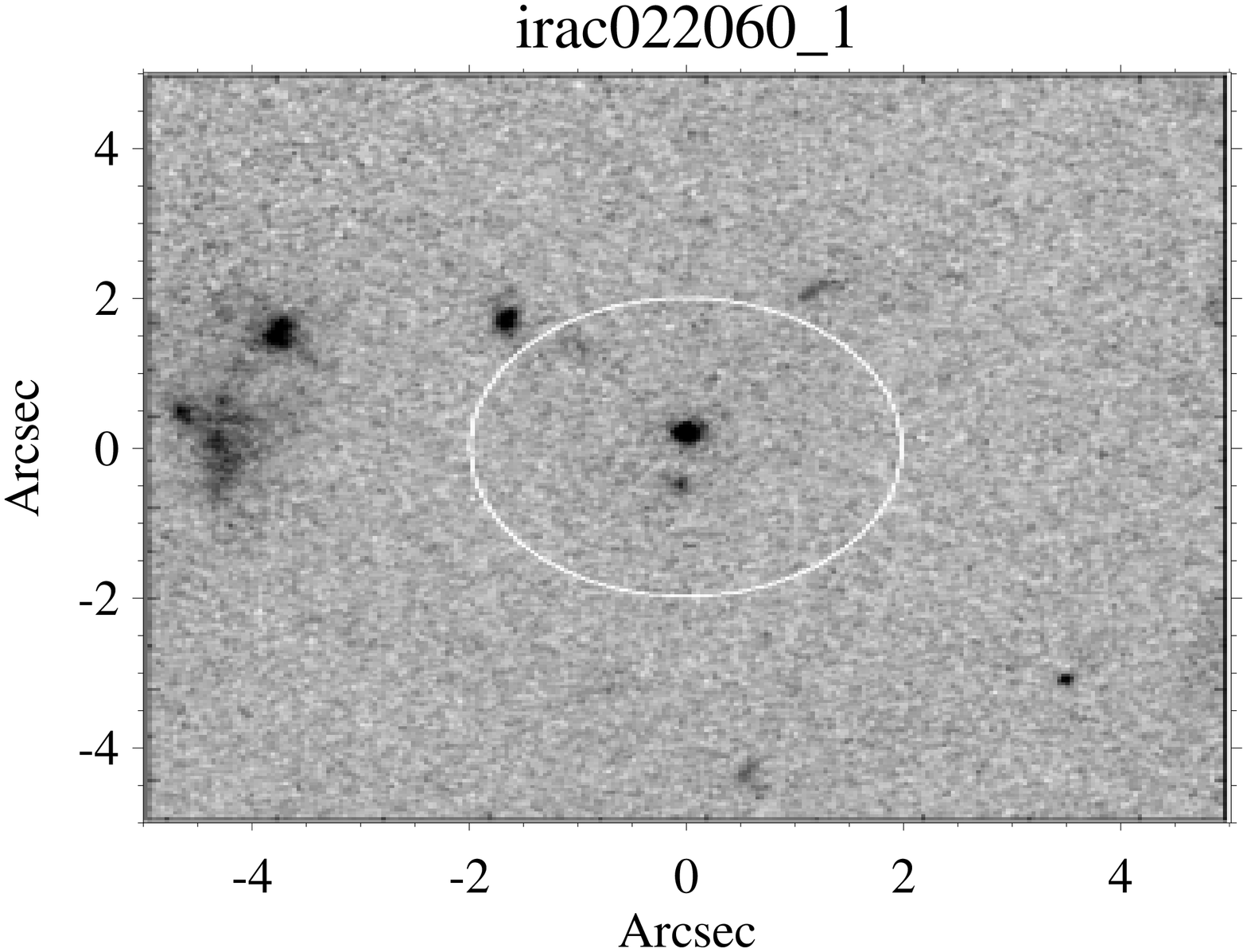}
\includegraphics[width=4cm,height=4cm,angle=90]{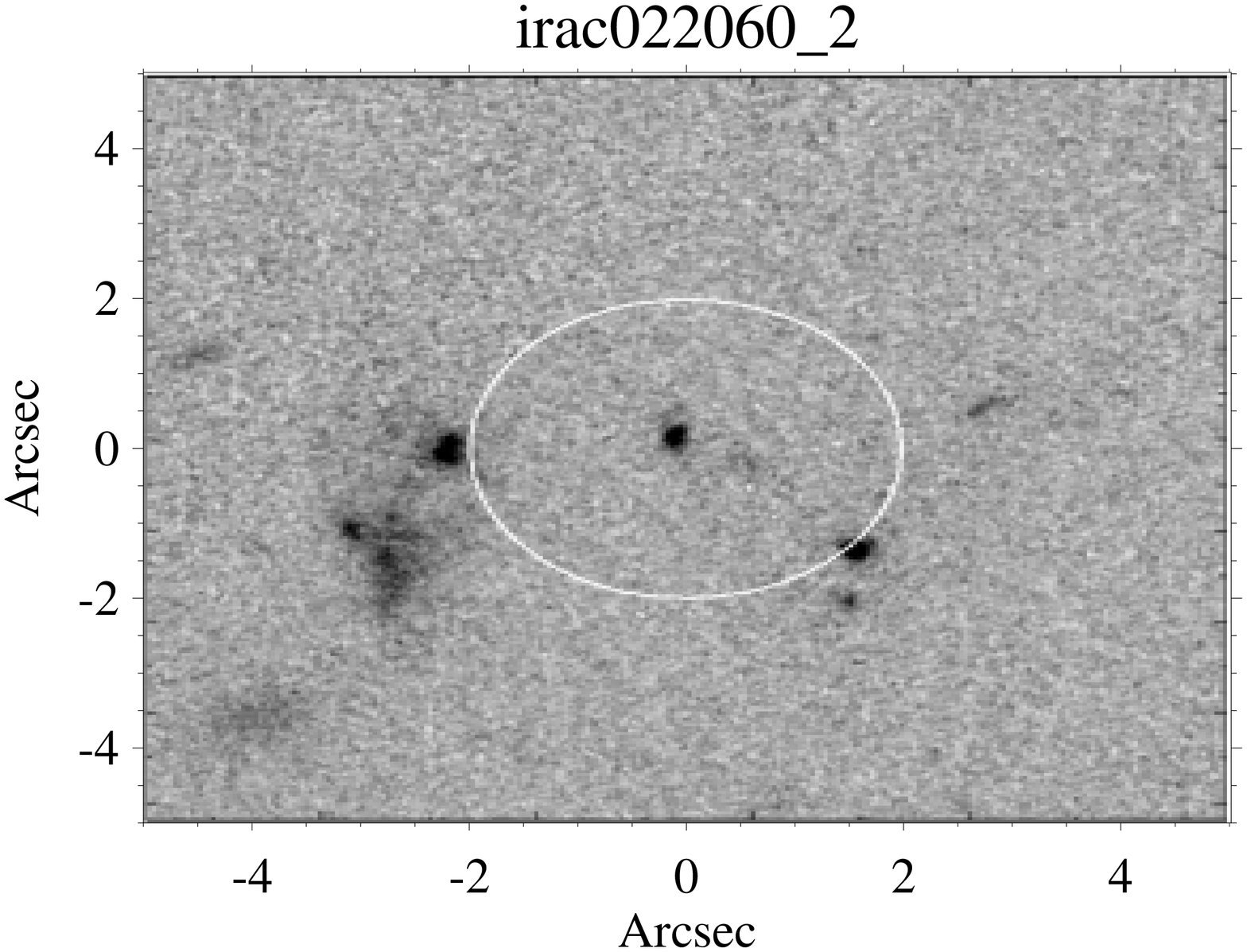}\par}
{\par
\includegraphics[width=4cm,height=4cm,angle=90]{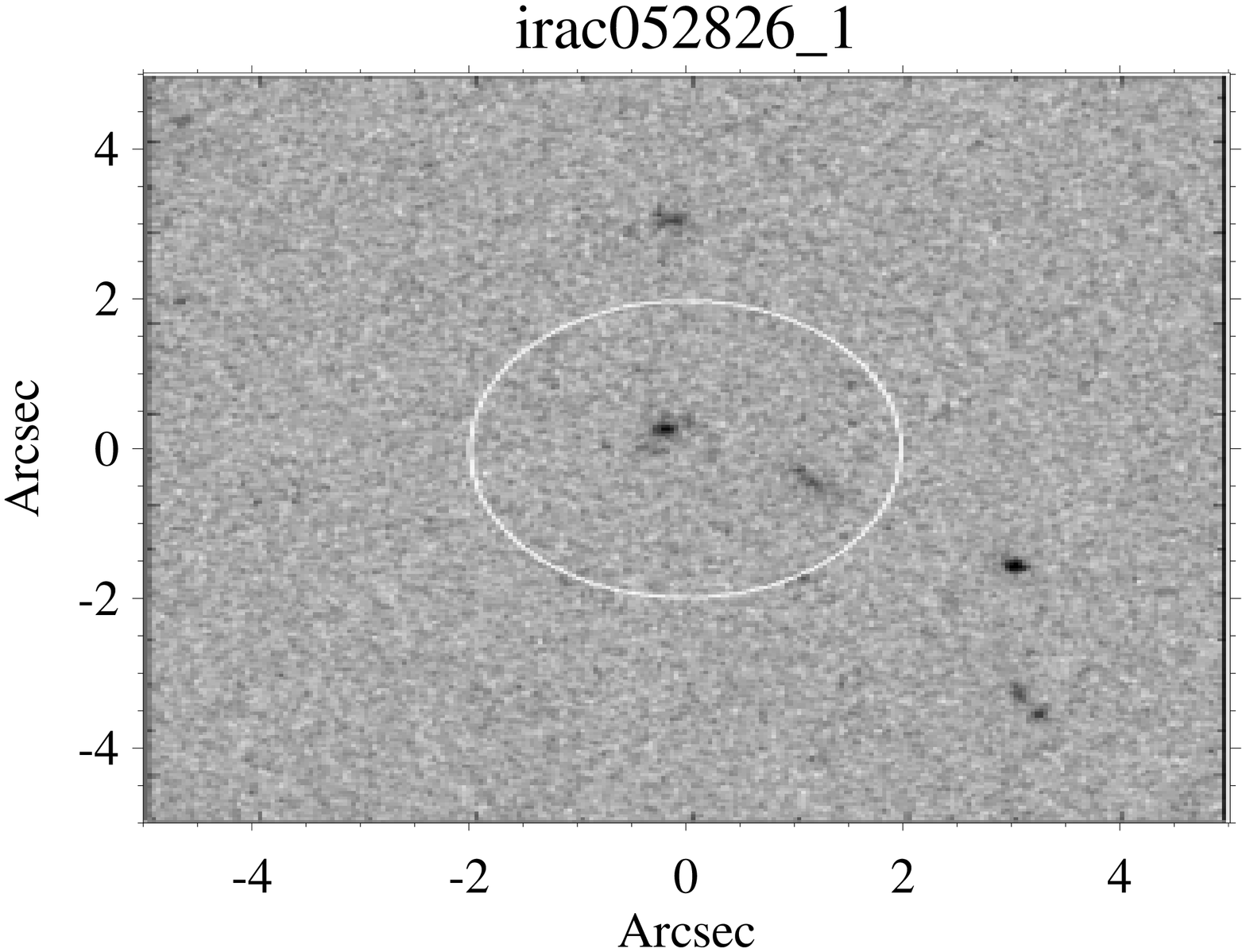}
\includegraphics[width=4cm,height=4cm,angle=90]{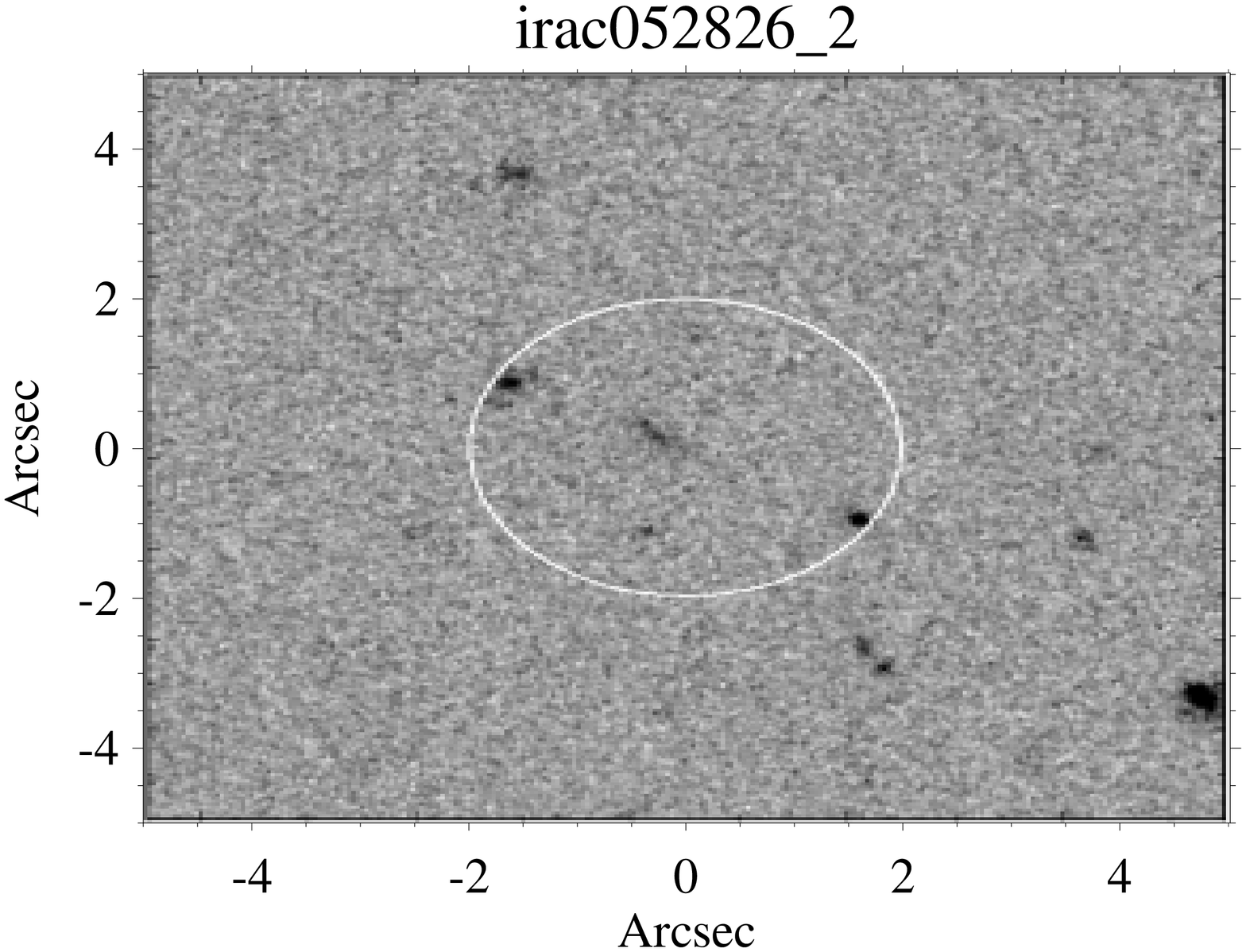}
\includegraphics[width=4cm,height=4cm,angle=90]{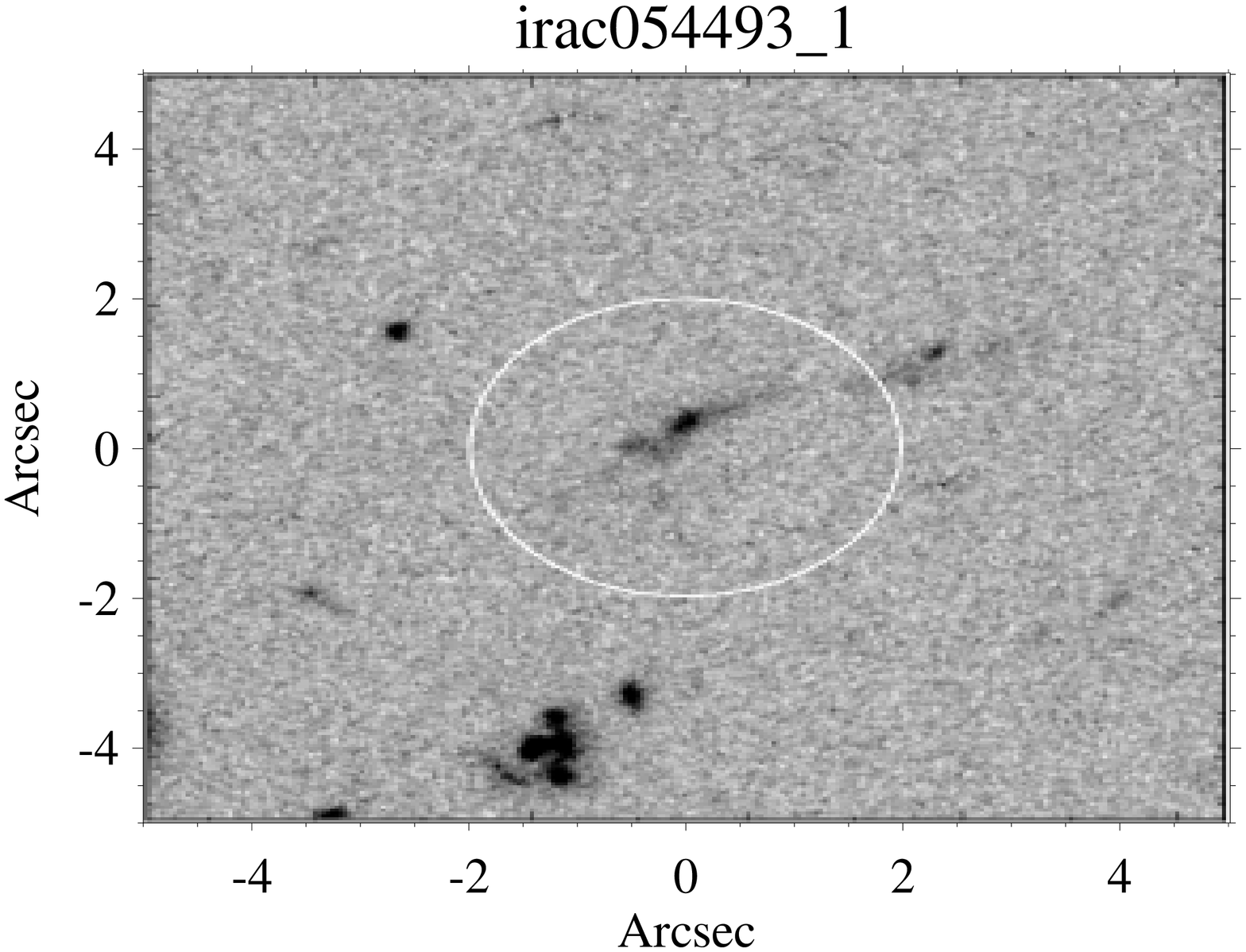}
\includegraphics[width=4cm,height=4cm,angle=90]{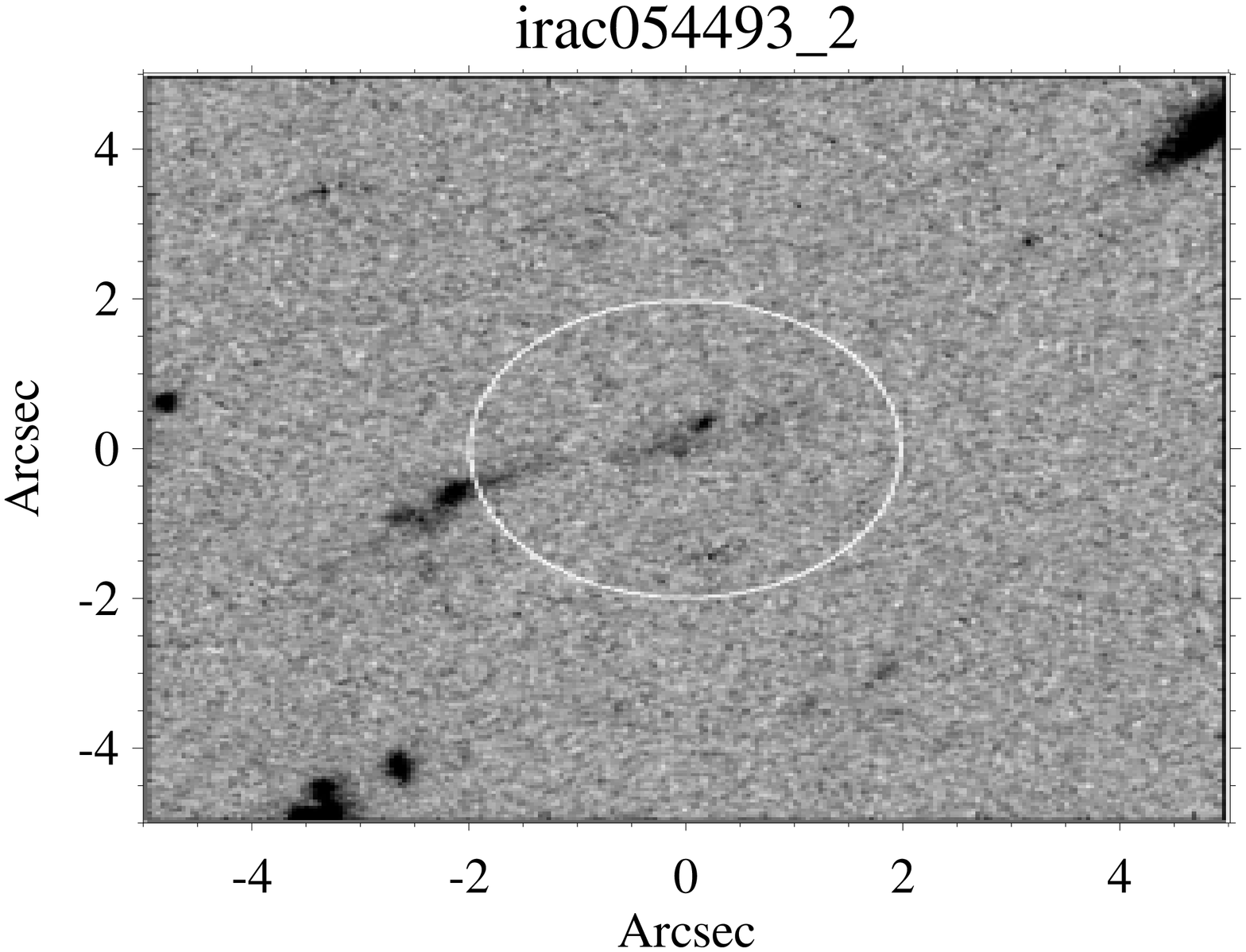}\par}
{\par
\includegraphics[width=4cm,height=4cm,angle=90]{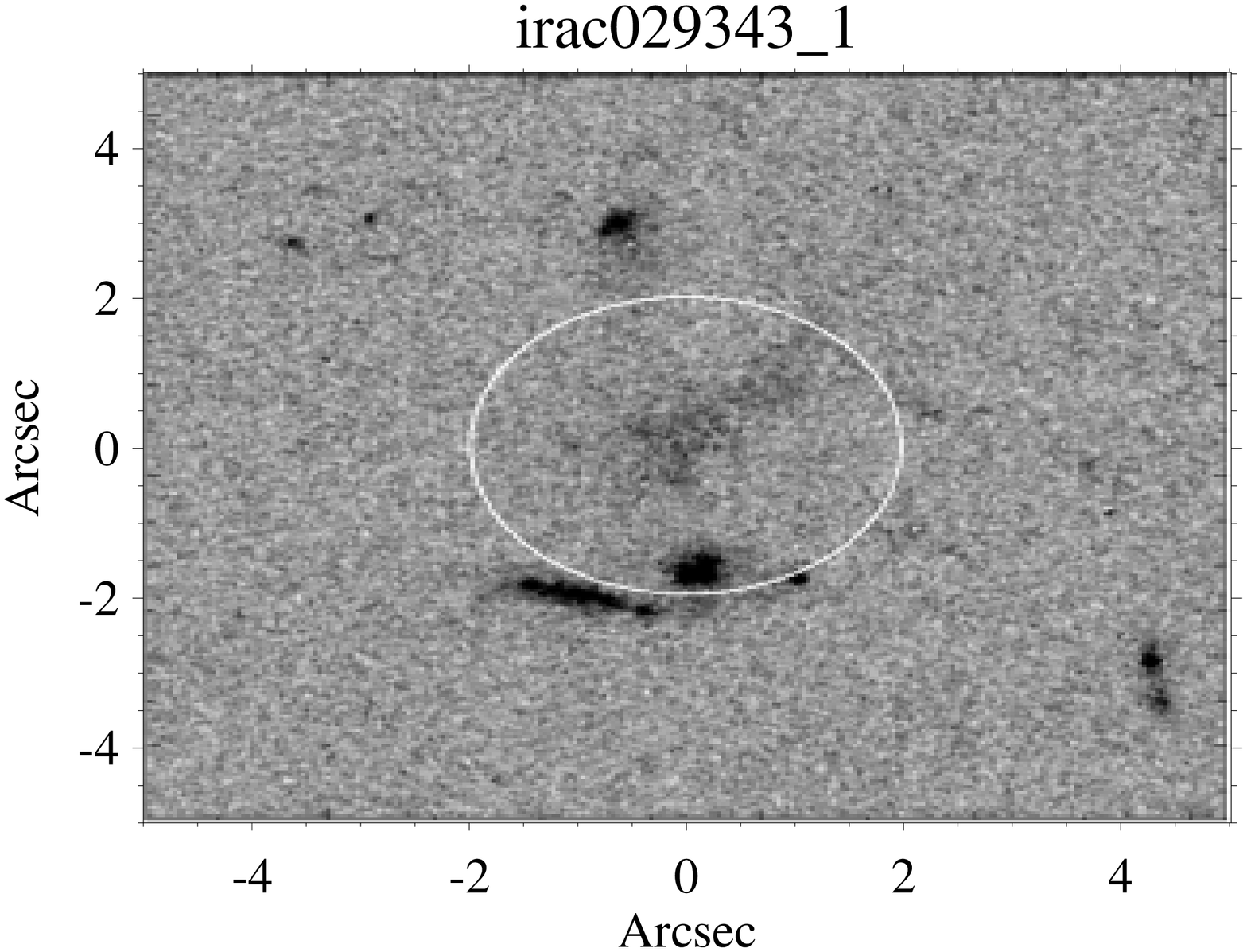}
\includegraphics[width=4cm,height=4cm,angle=90]{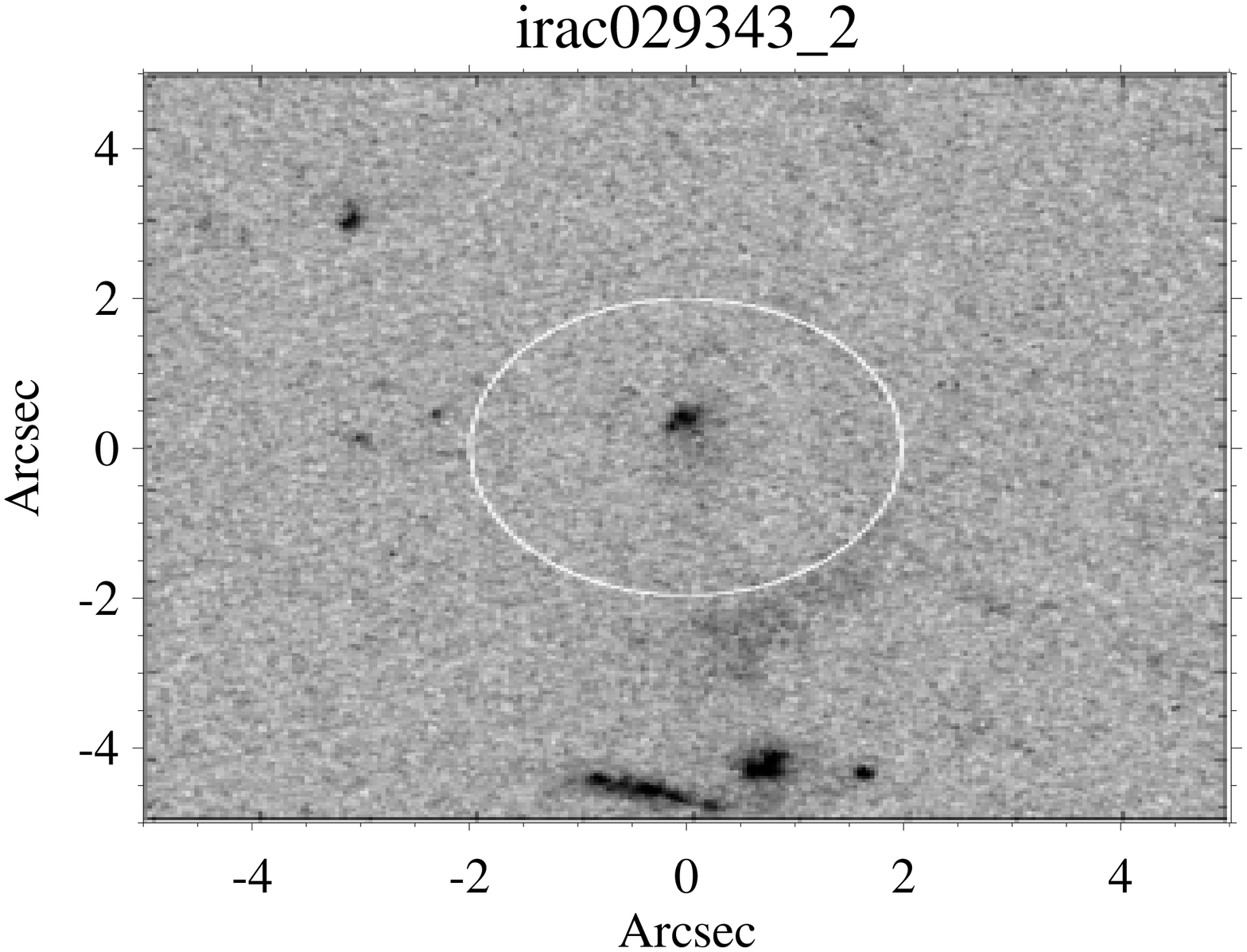}
\includegraphics[width=4cm,height=4cm,angle=90]{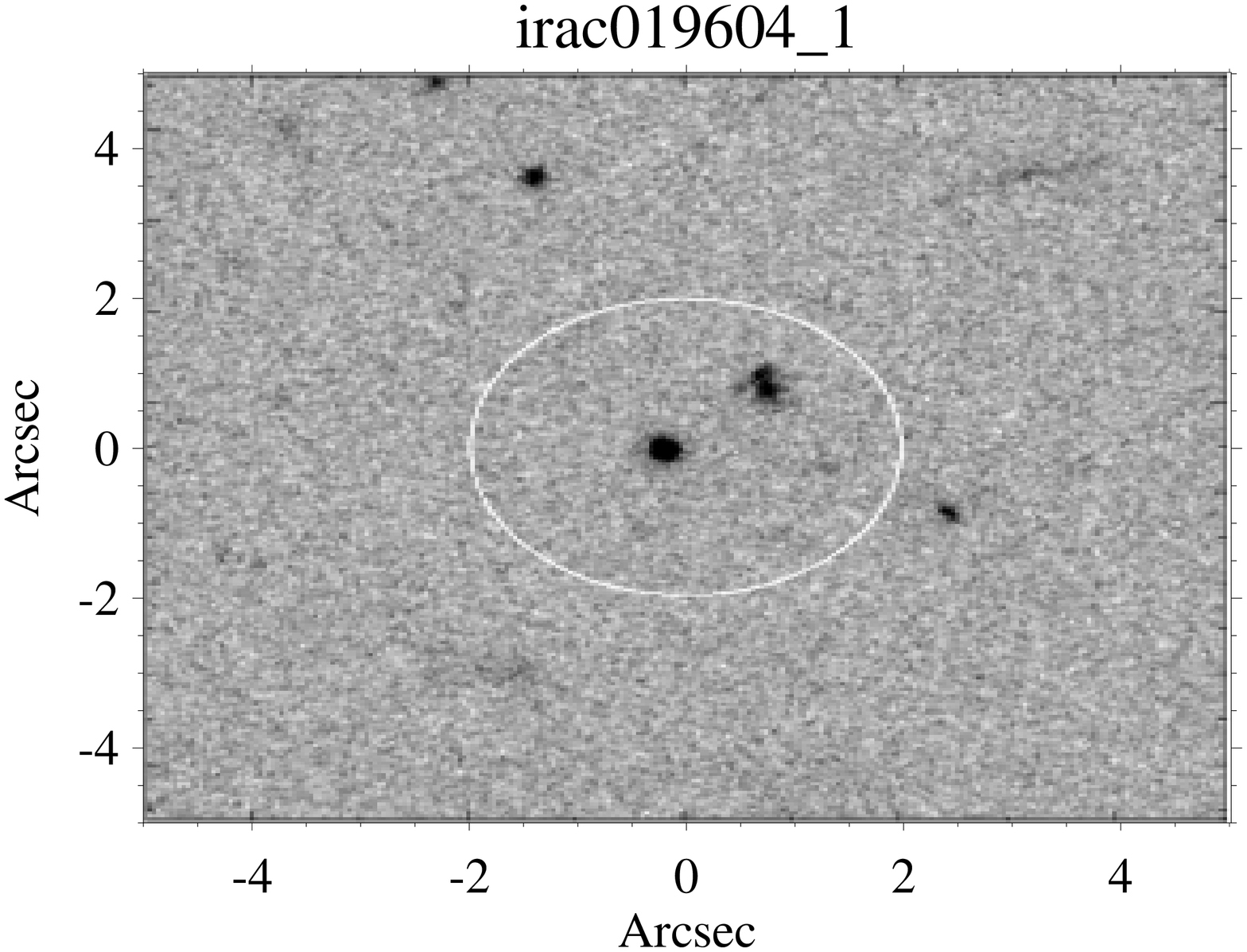}
\includegraphics[width=4cm,height=4cm,angle=90]{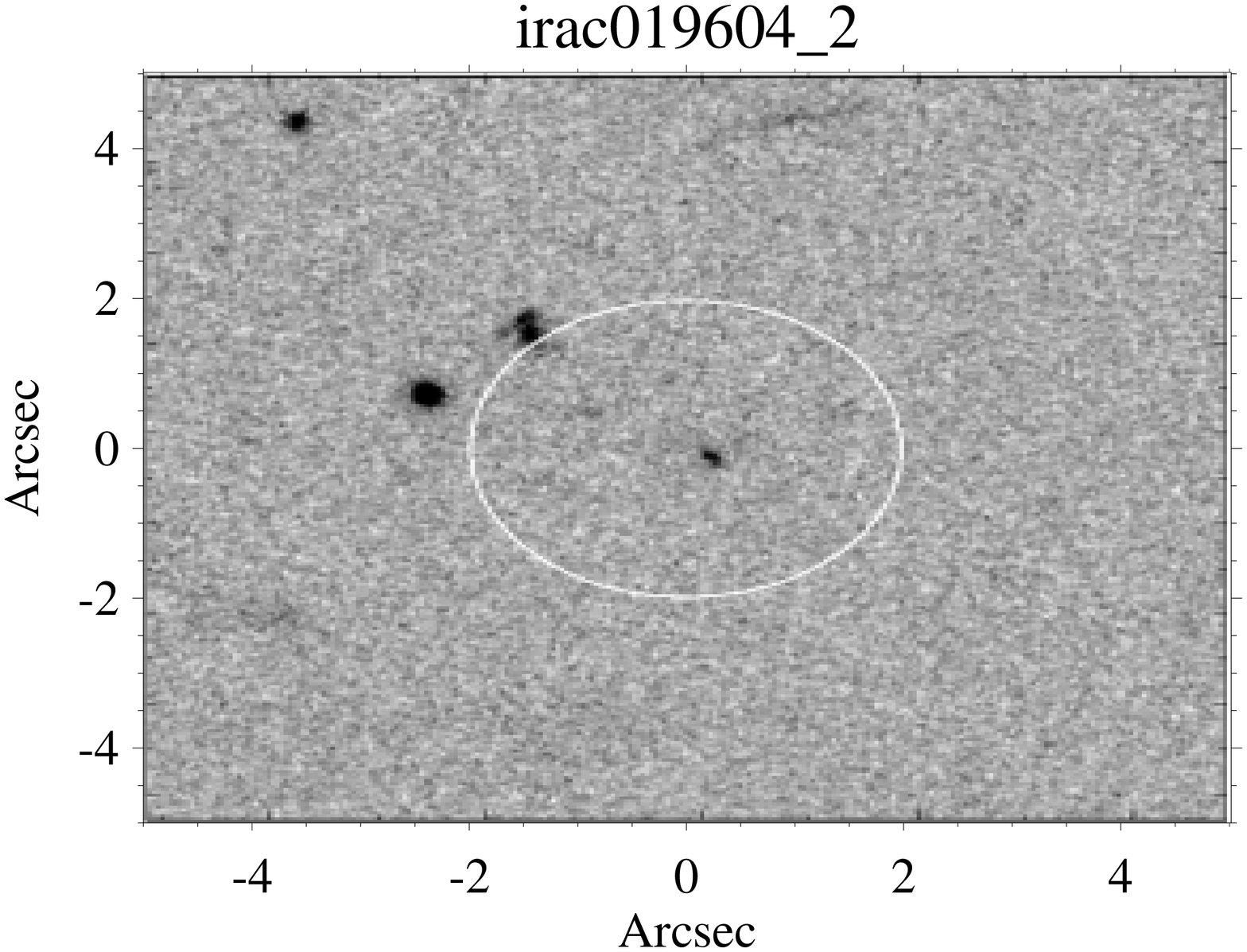}\par}
{\par
\includegraphics[width=4cm,height=4cm,angle=90]{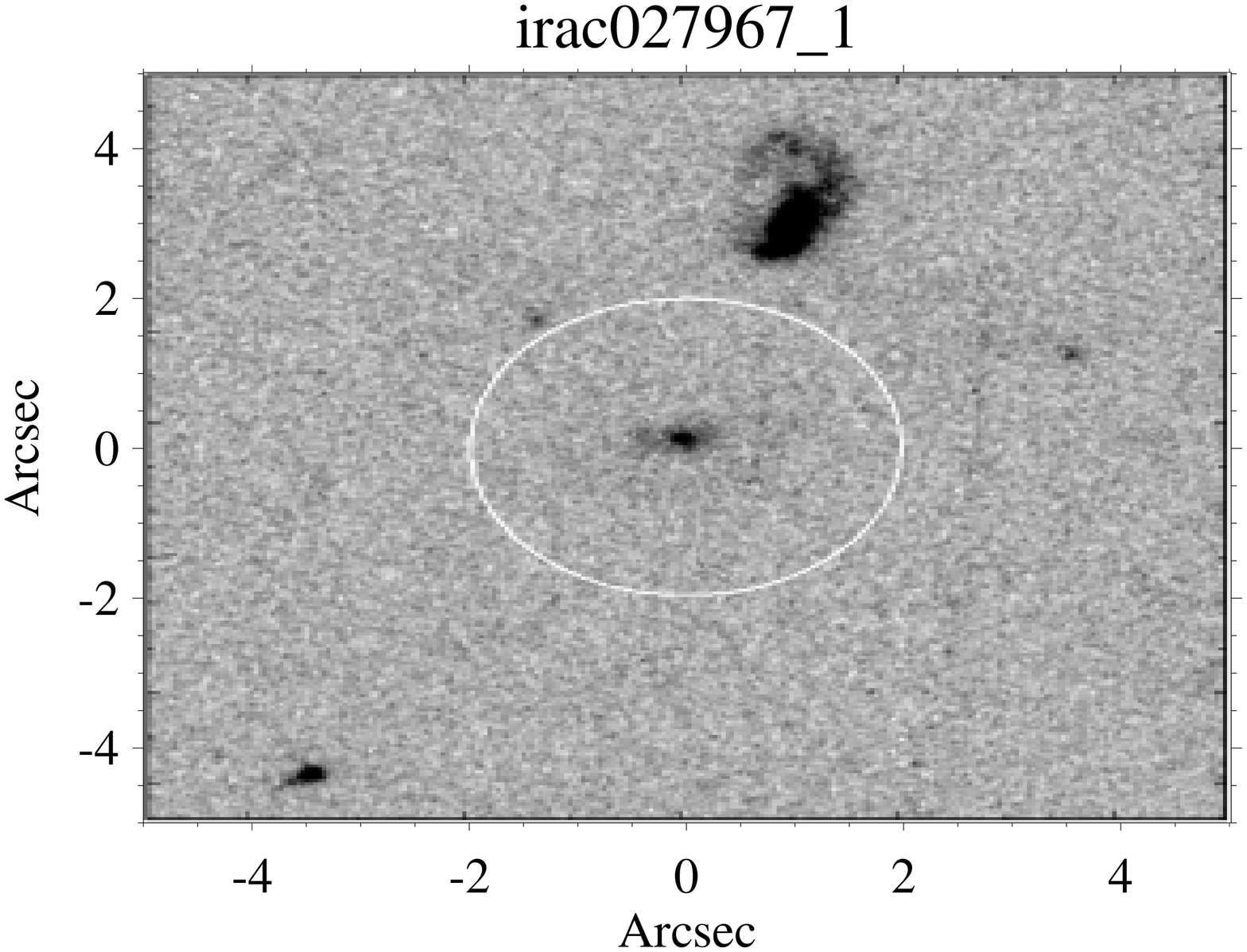}
\includegraphics[width=4cm,height=4cm,angle=90]{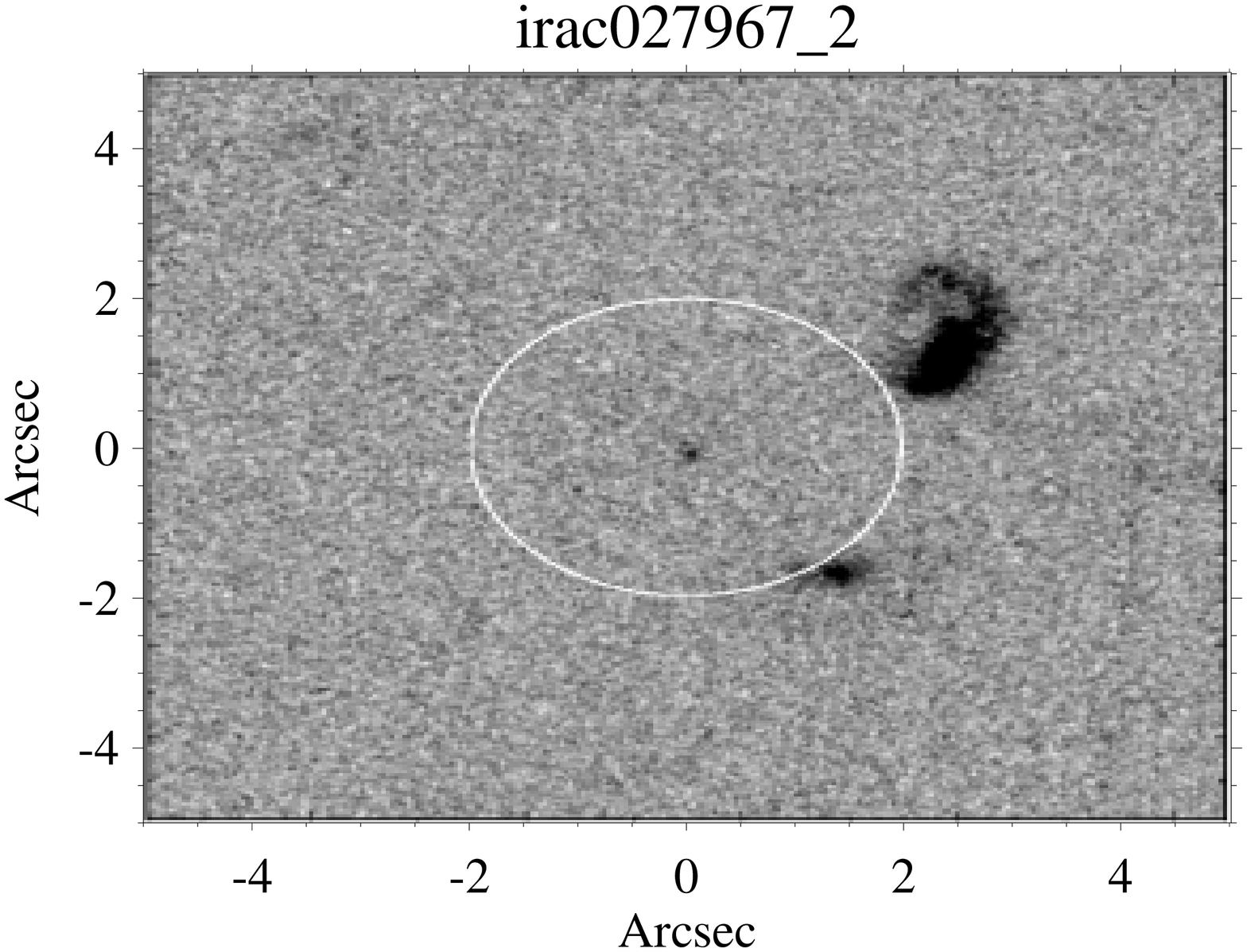}
\includegraphics[width=4cm,height=4cm,angle=90]{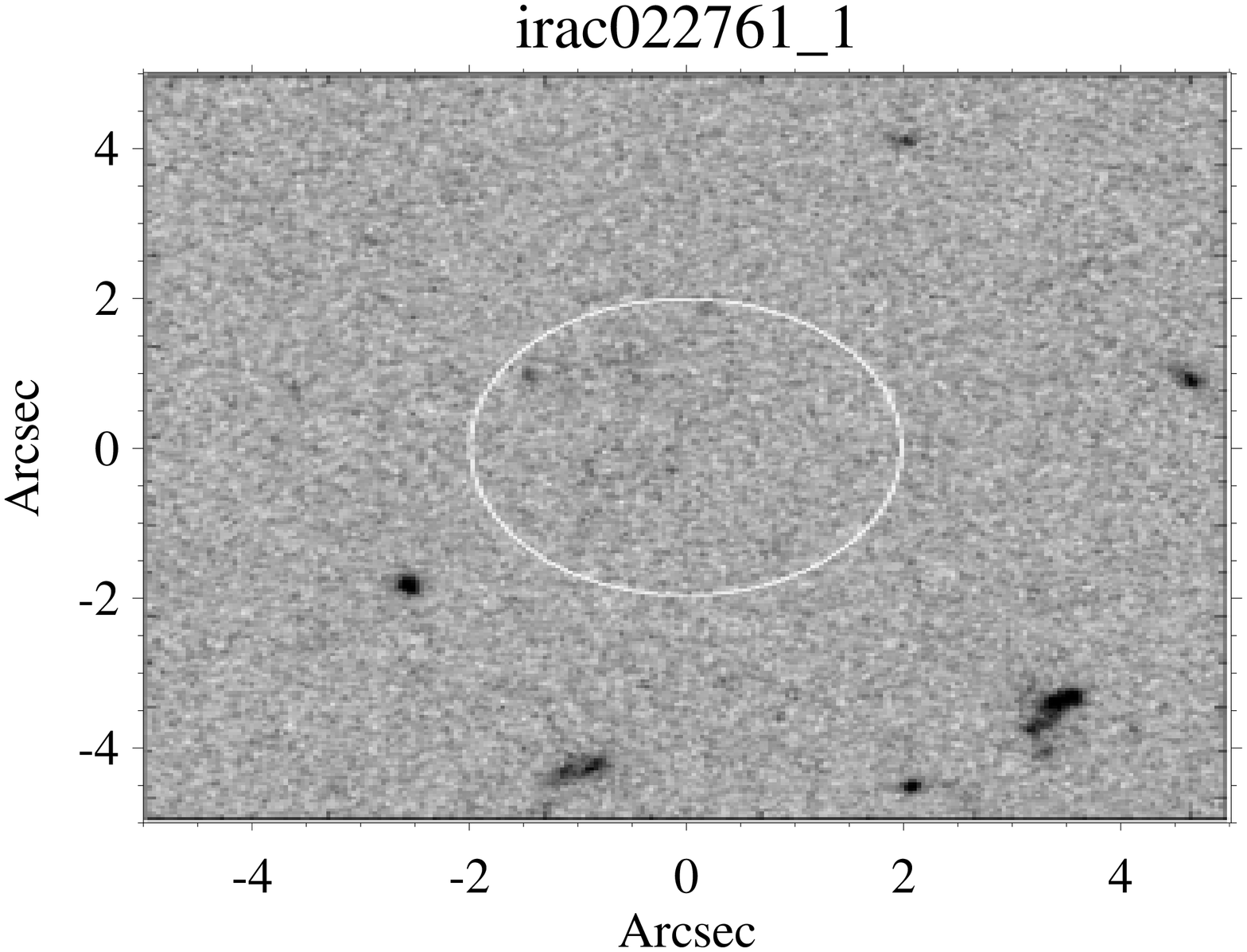}
\includegraphics[width=4cm,height=4cm,angle=90]{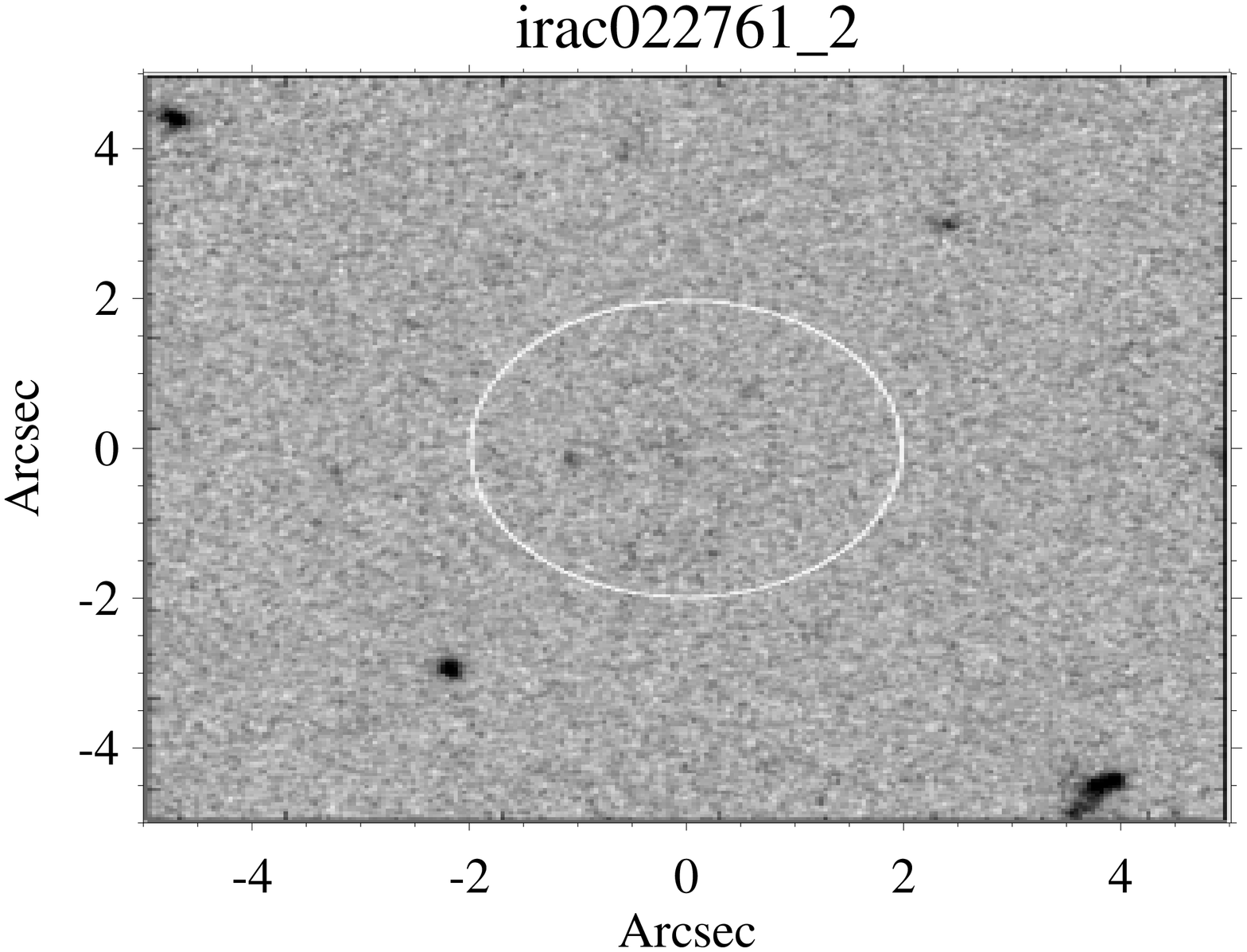}\par}
{\par
\includegraphics[width=4cm,height=4cm,angle=90]{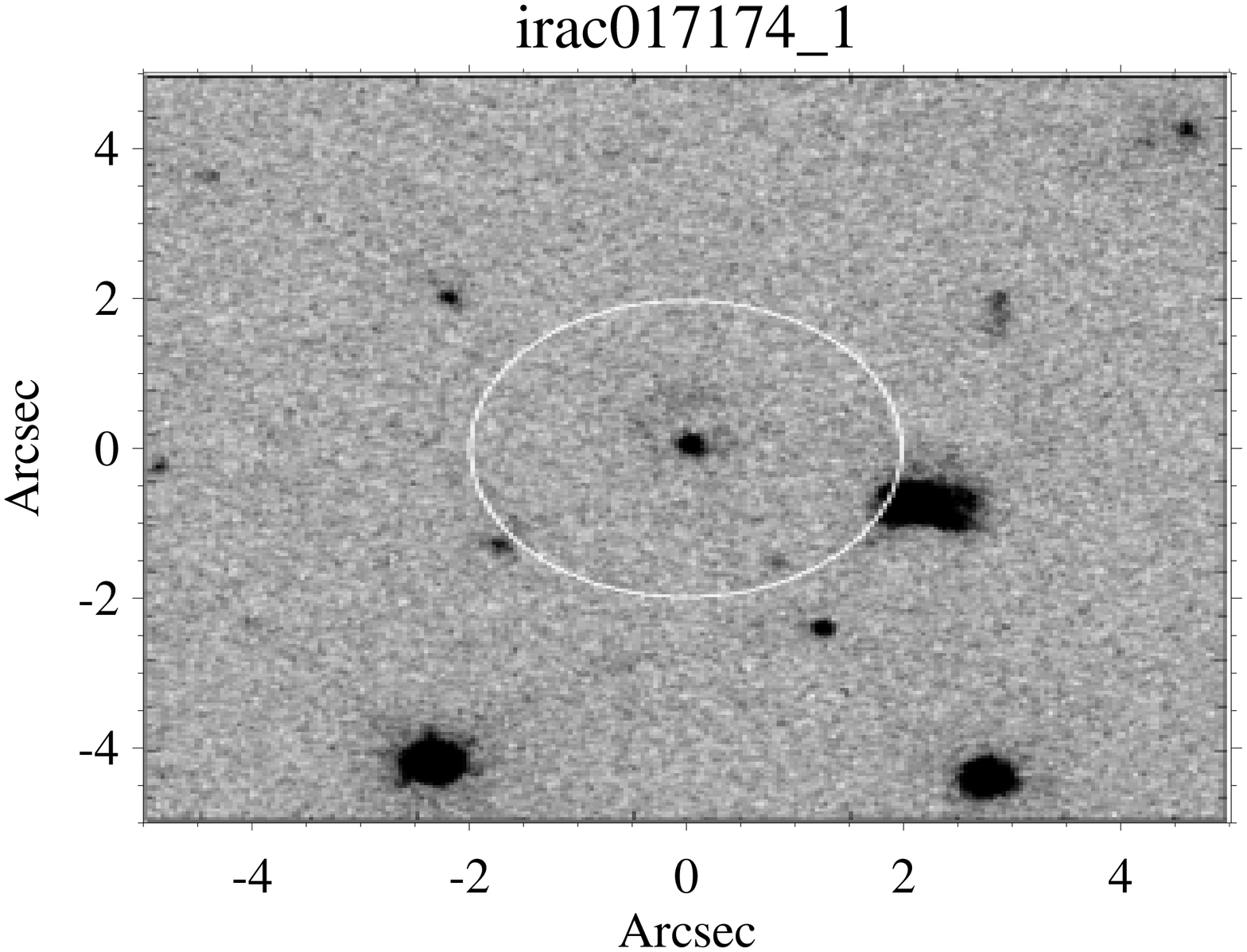}
\includegraphics[width=4cm,height=4cm,angle=90]{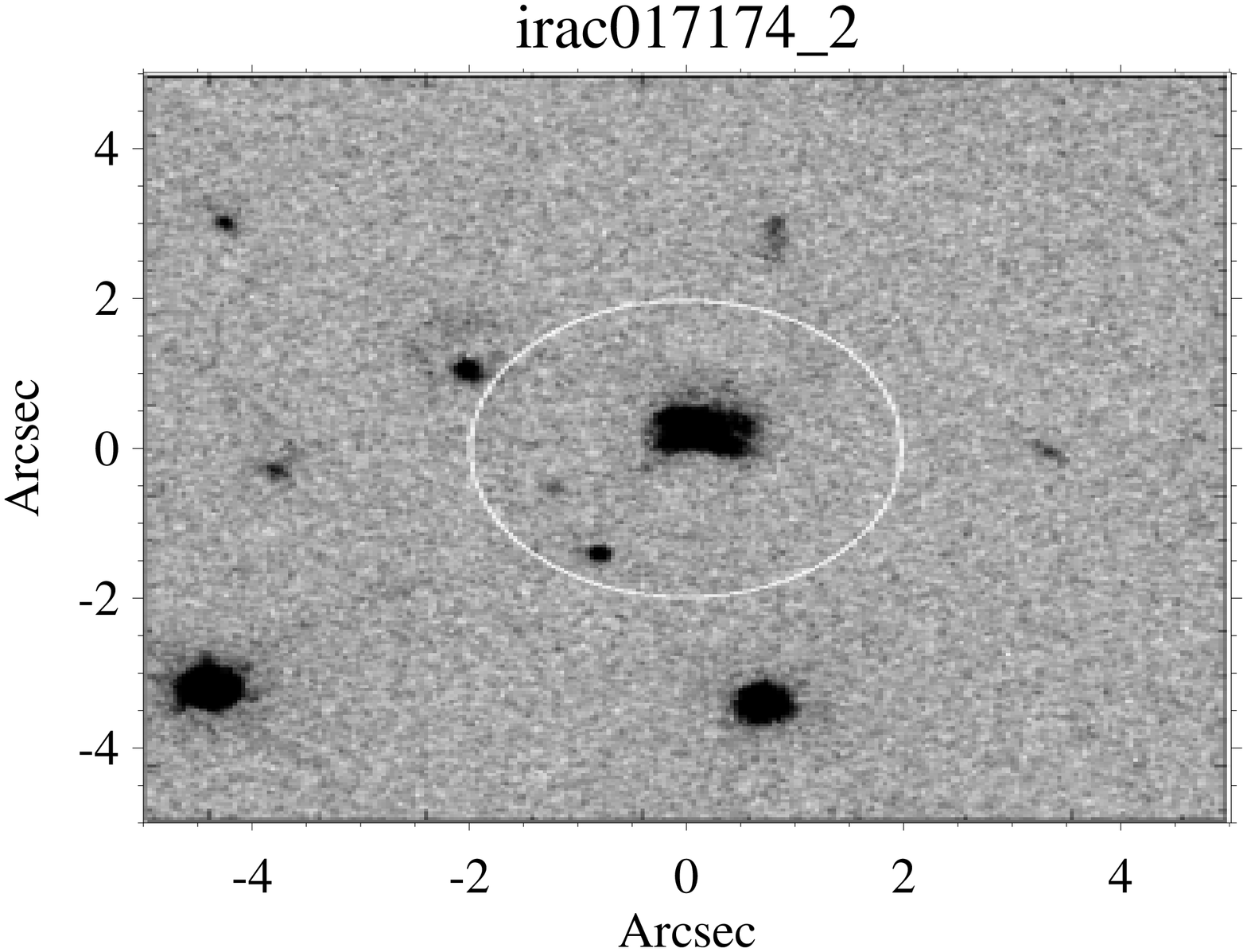}
\includegraphics[width=4cm,height=4cm,angle=90]{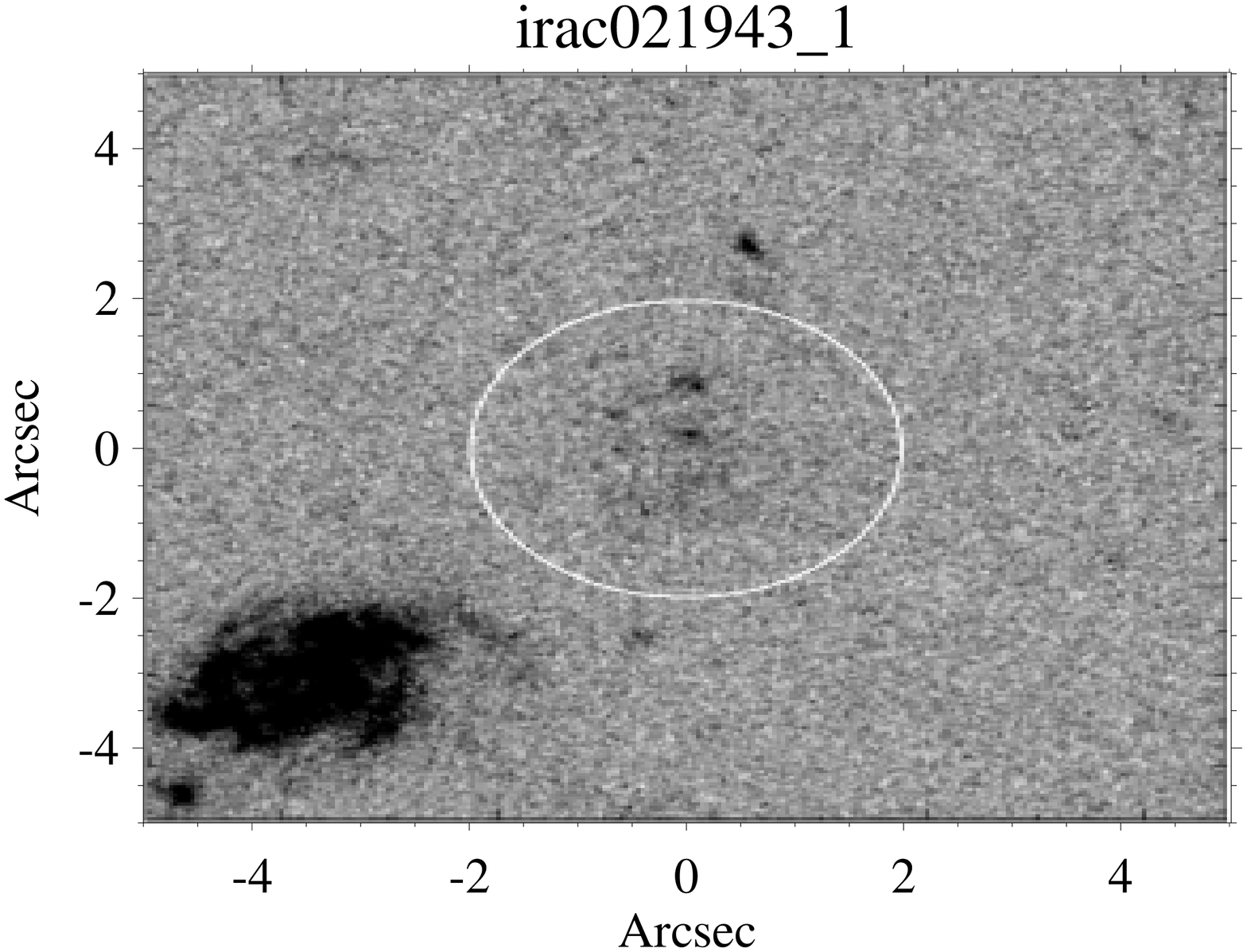}
\includegraphics[width=4cm,height=4cm,angle=90]{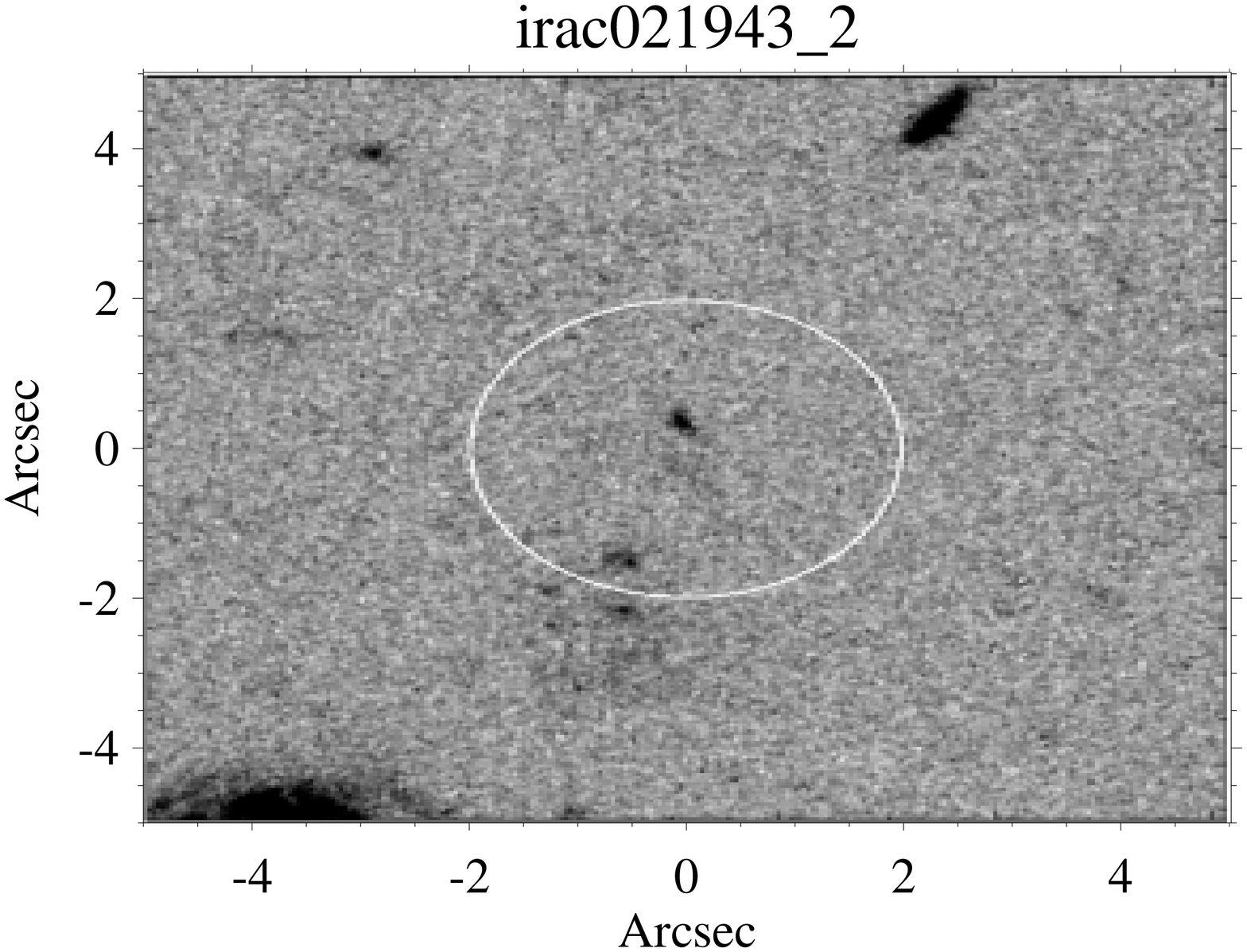}\par}
\figcaption{\footnotesize{Same as in Figure \ref{cromo1} (all the images from ACS V-band/HST), for objects 
irac061881, irac022060, irac052826, irac054493, irac029343, irac019604, irac27967, irac022761, irac017174, 
irac021943.}
\label{cromo2}}
\end{figure}

Twenty out of the 116 objects that comprise our full sample show double detection in the ground-based images, 
thanks to their better spatial resolution. Figs. \ref{cromo1} and \ref{cromo2} show
ACS V-band/HST images of both detections
(indicated with circles) for each pair of galaxies, except for objects irac053271 and irac038708, 
which do not have HST imaging, and for which optical CFHTLS r-band images are shown instead.
The CFHTLS magnitudes have been employed for the calculations with the HyperZ code, although we have 
chosen the
HST images for display purposes, because of their better resolution. These images help us classifying 
morphologically these 20 objects with double
detections as either interacting systems, different star forming regions of the same galaxy, 
or simple source confusion, as described in Table \ref{double}. 

In the case of these objects with double detection, for a single IRAC source, there are two 
counterparts in the ground-based images within the 2.5\arcsec~search radius. As described in Section \ref{sample},
the optical/near-infrared reference image is used to determine the positions of each source. 
The IRAC images are then deconvolved using the IRAC PSFs.
The sources can be resolved for separations $\sim$1\arcsec~from each other, and    
IRAC fluxes are then remeasured by fixing the positions of the objects in each pair, and by scaling the
flux of each object in an aperture of 0.9\arcsec~\citep{Perez05,Perez08}. 
The integrated magnitude is derived applying an aperture correction based on empirical IRAC PSFs (for the 
0.9\arcsec~aperture the factors are 1.01$\pm$0.07, 1.02$\pm$0.08, 1.2$\pm$0.10, and 1.44$\pm$0.14 for the 
channels 3.6, 4.5, 5.8, and 8.0 \micron, respectively). See \citet{Perez08} appendix 
for more details. The flux contamination is found to be smaller than the 10\% 
in most cases, an even smaller for the non-infrared-bright sources.

Once we know the positions of each galaxy in a pair, we can check whether the
mid-infrared emission comes from both, or just from one of the objects in the IRAC and MIPS images. 
In the majority of the cases,
all the mid-infrared flux in a pair of galaxies comes from only one of the objects
(see Table \ref{double}), 
the other probably being a non-active object. Then, we assume that the X-ray emission comes 
from the same mid-infrared emitter, and we calculate photometric redshifts for the active objects only.

In those cases where the mid-infrared emission can not be allocated clearly to one of the objects 
(irac038708, irac056633, and irac046783), photometric
redshifts calculated by HyperZ for both sources in each pair have been obtained and they are reported in Table 
\ref{photoz2} together with their  $\chi_{\nu}^{2}$, probabilities, SED type and A$_{V}$. 
For the other 17 pairs of galaxies, for which the mid-infrared emission comes clearly from only one of the objects,
we calculate photometric redshifts only for the mid-infrared emitter. Spectroscopic redshifts 
from the DEEP database are also given, when available, together with their corresponding reliability flags. 
Unfortunately, this is the case for only four objects, and all of them with low reliability flags (1 or 2, see Table
\ref{photoz2}). Nevertheless, we can assume that the photometric redshifts, obtained as described in 
Section \ref{spectral}, are reasonably good, since we have followed the same methodology as for the 96 sources 
with single detections. 

As reported in Table \ref{double}, irac056633$_{-}$2, irac036704$_{-}$1, irac022060$_{-}$1, irac029343$_{-}$1,
and irac019604$_{-}$1 are interacting systems themselves, as can be seen in the HST images 
(Figs. \ref{cromo1} and 
\ref{cromo2}). These sources must be treated with caution, since their fluxes could be contaminated with extra-emission
coming from their companions. This fact explains the low probabilities of the HyperZ fits for objects 
irac056633$_{-}$2, irac029343$_{-}$1, and irac019604$_{-}$1, reported in Table \ref{photoz2}.

In the same way as we have done for the objects with unique detection in previous sections, 
we distribute here the 23 template fitted-objects with double detection in the same five main categories described before. 
The percentages for each group are: {\it Starburst-dominated AGN} (48 \% of the mid-infrared emitters), 
{\it Starburst-contaminated AGN} (0 \%),
{\it Type-1 AGN} (17 \%), {\it Type-2 AGN} (22 \%), and {\it Normal galaxy hosting AGN} (13 \%).
Note that for this subsample of objects with double detection, almost half of the objects are described by
starburst-type SEDs. If, as in Section 4.1., we split the objects into AGN-dominated and host-dominated 
galaxies, we find that 39\% show AGN-like SEDs while 61\% are host-dominated, a clear overrepresentation. 
This is expected since  
if the pairs of galaxies are interacting objects, the number of starbursts in this subsample of galaxies 
should consequently increase.

\section{Conclusions}

We present a reliable method of classification of hard X-ray and mid-infrared selected AGN, based on 
the fit of well-sampled multiwavelength spectral energy distributions with a complete set of AGN and
starburst galaxy templates. 
The sample studied in this paper consists of 96 AGN with unique detection, 
and 20 AGN with double detection in the EGS. 
The following results were found:

\begin{itemize}

\item{Photometric redshifts have been calculated by
using the HyperZ code. The measured mean discrepancy between our z$_{phot}$'s and a subsample of higly reliable
DEEP spectroscopic redshifts (flag = 3 or 4) is $\Delta$z = -0.03, with $\sigma_{z}$ = 0.11, 
and 3 outliers (8\%). 
We provide more accurate photometric redshifts than the spectroscopic ones for objects with DEEP flag = 1 or 2.}

\item{Five main population groups have been considered according to the set of templates employed. 
For the 96 objects in our sample with unique detection, the following percentages have been found: 
{\it Starburst-dominated AGN} (24 \% of the sample), {\it Starburst-contaminated AGN} (7 \%),
{\it Type-1 AGN} (21 \%), {\it Type-2 AGN} (24 \%), and {\it Normal galaxy hosting AGN} (24 \%). We find that
52\% of the sample has AGN-dominated SEDs and the remaining 48\% host-dominated SEDs.}

\item{58\% of the 96 objects with unique detection in our sample have
z$_{phot}$$<$1, with the rest of the z$_{phot}$ of the sources distributed in a decreasing tail up 
to z$_{phot}$=3. The {\it Starburst-dominated AGN} constitute the high-redshift population of the host-dominated
AGN, whilst the {\it Normal galaxy hosting AGN} are concentrated at low redshifts. In the AGN-dominated group,
{\it Type-1 AGN} are randomly distributed in distance, the {\it Starburst-contaminated AGN} are located
at intermediate values of redshift, and the {\it Type-2 AGN} are the lowest-z objects.}

\item{An evolutionary trend is noticed, in which the {\it Starburst-dominated AGN} would be the progenitors of the 
{\it Type-1 AGN} and {\it Type-2 AGN}, via quenching of the starburst through the AGN feedback.}

\item{Correlations between hard/soft X-ray luminosities and ultraviolet/optical/infrared data are reported for 
such a sample of AGN spanning a wide range of redshift, being in this way represented the behaviour of 
the different AGN types in the various wavelengths considered.}

\item{{\it Type-1 AGN} show the highest values of hard and soft X-ray luminosities of the sample, together with
the {\it Starburst-contaminated AGN}, whilst the {\it Normal galaxy hosting AGN} majority are concentrated
at the lowest values, coinciding with the local cool ULIRGs typical hard X-ray luminosities. 
{\it Starburst-dominated AGN} and {\it Type-2 AGN} present intermediate values of X-ray emission, very similar to
those of warm ULIRGs. This is consistent with a luminosity-dependent evolution of AGN, with low-luminosity AGN
peaking at lower redshifts than luminous active nuclei.}

\item{{\it Type-1 AGN} are all contained in the IRAC color-color diagram region empirically determined by
\citet{Stern05} for spectroscopically selected AGN. There are many {\it Type-2 AGN} and 
{\it Starburst-dominated AGN}
inside this AGN region. These objects have higher mean redshifts than those in the same group but outside the 
pure-AGN area, according with the evolution of the mid-infrared colors with redshift 
for star-forming galaxies described in \citet{Barmby06}.}

\item{Mid-infrared 24 \micron~to optical r band flux ratio versus the
(r-z)$_{AB}$ or the (r-IRAC 3.6 \micron)$_{AB}$ colors show a clear segregation of the different
groups in both diagrams. {\it Starburst-dominated AGN} and {\it Starburst-contaminated AGN}
are displaced towards the highest values of the mid-infrared to optical ratio and display the reddest colors.
{\it Type-1 AGN} and {\it Type-2 AGN} are located at intermediate values, and the {\it Normal galaxy hosting AGN}
have the lowest values of the 24 \micron/r flux ratio and the bluest colors. }

\item{A tentative classification of objects with double detection into the five main population 
groups considered through this paper shows an increase of the {\it Starburst-dominated AGN} of up 
to 48\%, while the others decrease. 
61\% of the fitted objects show AGN-like SEDs, while 39\% is host-dominated.}

\end{itemize}

\acknowledgments

NASA's Chandra X-Ray Observatory was launched in July 1999. The Chandra Data Archive (CDA) is part
of the Chandra X-Ray Center (CXC) which is operated for NASA by the Smithsonian Astrophysical
Observatory.

Based on observations obtained with XMM-Newton, an ESA science mission with instruments and 
contributions directly funded by ESA Member States and NASA.

GALEX (Galaxy Evolution Explorer) is a NASA Small Explorer, launched in April 2003. We gratefully
acknowledge NASA's support for construction, operation, and science analysis of the GALEX mission,
developed in cooperation with the Centre National d'Etudes Spatiales of France and the Korean Ministry
of Science and Technology.)

This work is based in part on observations made with the Spitzer Space Telescope, which is
operated by the Jet Propulsion Laboratory, California, Institute of Technology under a contract with
NASA.

Based on observations obtained with MegaPrime/MegaCam, a joint project of CFHT and CEA/DAPNIA, at
the Canada-France-Hawaii Telescope (CFHT) which is operated by the National Research Council (NRC) of
Canada, the Institut National des Science de l'Univers of the Centre National de la Reserche
Scientifique (CNRS) of France, and the University of Hawaii. This work is based in part on data
products produced at TERAPIX and the Canadian Astronomy data Centre as part of the CFHT Legacy Survey,
a collaborative project of NRC and CNRS.

Based on observations obtained at the Hale Telescope, Palomar Observatory, as part 
of a collaborative agreement between the California Institute
of Technology, its divisions Caltech Optical Observatories and the Jet Propulsion Laboratory 
(operated for NASA), and Cornell University.

Many images of this article are based on observations made with the NASA/ESA Hubble Space Telescope, 
obtained from the data archive at the Space Telescope Science Institute (STScI). STScI is operated by the 
Association of Universities for Research in Astronomy, Inc., under NASA contract NAS5-26555.

This work uses data obtained with support of the National Science Foundation grants 
AST 95-29028 and AST 00-71198.

This work is partially funded by PN AYA2007-67965-C03-01, PN AYA2006-02358 and by the Spanish MEC 
under the Consolider-Ingenio 2010 Program grant CSD2006-00070: First Science with the GTC 
(http://www.iac.es/consolider-ingenio-gtc/). P.G.P.-G. acknowledges support from the Ram\'{o}n y Cajal Program
financed by the Spanish Government and the European Union.
C.R.A., J.R.E., G.B., J.G., and P.G.P.-G. acknowledge Roser Pell\'{o}, Antonio Cabrera Lavers, and Casiana Mu\~{n}oz Tu\~{n}\'{o}n 
for their valuable help. We finally appreciate the very useful report of the anonymous referee.

\clearpage

\begin{deluxetable}{lccccccccccl}
\centering
\tabletypesize{\scriptsize}
\tablecaption{\scriptsize{ID from \citet{Barmby06}, IRAC ID, IRAC 3.6~\micron~ J2000.0 right ascension and declination, 
spectroscopic redshift from DEEP public database with its corresponding reliability between brackets (1-2 = low reliability, 
3-4= high reliability), photometric redshift and its corresponding $\chi_{\nu}^{2}$  and  
probability, optical extinction derived from the \citet{Calzetti00} reddening law, logarithm of $\nu$L$_{\nu}$ in the r band as a reference, in erg~s$^{-1}$,
fitted template, and main group classification.  
Templates: 1,2-Starburst/ULIRG, 3,5-Starburst, 4-Sy1/Starburst/ULIRG, 6-Sy2/Starburst,
13-Sy2/Starburst/ULIRG, 7,12,14-Type-1 QSO, 8-Type-2 QSO, 9-Sy1.8, 10-Sy2, 11-Torus-QSO, 15,16,17-Ellipticals of 2, 5, and 13 Gyr,
18,19,20,21,22,23-Spirals of types S0, Sa, Sb, Sc, Sd, and Sdm.}\label{photoz}}
\tablehead{
\colhead{ID} & \colhead{ID IRAC} & \colhead{RA ($^{o}$)} & \colhead{Dec ($^{o}$)} & \colhead{z$_{spec}$}  &
\colhead{z$_{phot}$} & \colhead{$\chi_{\nu}^{2}$} & \colhead{Prob (\%)} & \colhead{A$_{V}$} &  \colhead{L$_{r}$} &
\colhead{Template} & \colhead{Group}} 
\startdata
1   & 054396 &  213.9870  & 52.2687  &   -	&   0.66    &  0.10  & 100 &   0.3  & 43.80  &   7    &  Type-1 AGN  \\
2   & 067129 &  214.0352  & 52.3547  &   -	&   0.06    &  0.06  &  99 &   0.0  & 41.93  &  15    &    NG hosting AGN   \\     
3   & 045621 &  214.0441  & 52.2727  &   -	&   0.25    &  0.27  &  98 &   0.9  & 43.39  &  19    &    NG hosting AGN   \\     
4   & 068644 &  214.0572  & 52.3766  & 1.701 (2)&   0.05    &  1.15  &  32 &   0.9  & 40.55  &   7    &  Type-1 AGN	\\
5   & 056094 &  214.0591  & 52.3276  & 0.534 (4)&   0.50    &  2.51  &   1 &   0.0  & 43.46  &  22    &    NG hosting AGN   \\
6   & 048319 &  214.0948  & 52.3212  & 1.603 (4)&   0.76    &  1.41  &  18 &   0.0  & 44.21  &   4    &  SB-cont. AGN	  \\
7   & 019994 &  214.0956  & 52.2034  &   -	&   1.37    &  0.48  &  92 &   1.2  & 44.95  &  14    &  Type-1 AGN	\\
8   & 060727 &  214.1236  & 52.3925  &   -	&   0.96    &  0.59  &  81 &   0.3  & 44.40  &  10    &  Type-2 AGN	\\
9   & 053898 &  214.1298  & 52.3695  &   -	&   0.26    &  2.93  &   0 &   0.0  & 42.71  &   9    &  Type-2 AGN	  \\
10  & 040342 &  214.1367  & 52.3171  & 1.028 (4)&   0.95    &  0.83  &  60 &   0.0  & 43.81  &   4    &  SB-cont. AGN	  \\
14  & 052726 &  214.1587  & 52.3857  & 0.417 (4)&   0.35    &  0.40  &  95 &   0.0  & 43.98  &  19    &    NG hosting AGN     \\
16  & 059064 &  214.1765  & 52.4241  &   -	&   0.05    &  2.26  &   1 &   0.3  & 43.89  &  15    &    NG hosting AGN     \\
17  & 029938 &  214.1768  & 52.3034  &   -	&   1.06    &  0.98  &  45 &   0.0  & 45.65  &   7    &  Type-1 AGN	\\
20  & 040860 &  214.1815  & 52.3506  & 0.283 (2)&   0.12    &  0.91  &  53 &   0.0  & 43.20  &  20    &    NG hosting AGN     \\
21  & 045337 &  214.1832  & 52.3720  & 0.510 (4)&   0.45    &  0.38  &  96 &   0.0  & 43.99  &  20    &    NG hosting AGN     \\
22  & 071927 &  214.1891  & 52.4850  & 1.630 (2)&   1.55    &  0.24  &  99 &   0.9  & 44.08  &  14    &  Type-1 AGN	\\
24  & 054089 &  214.2060  & 52.4252  &   -	&   2.35    &  0.49  &  90 &   0.6  & 44.75  &  14    &  Type-1 AGN	\\
25  & 019616 &  214.2065  & 52.2815  & 0.761 (4)&   0.75    &  0.62  &  78 &   0.3  & 43.78  &  22    &    NG hosting AGN  \\
26  & 024423 &  214.2079  & 52.3025  & 0.808 (4)&   0.73    &  0.23  &  99 &   0.0  & 43.87  &   3    &  SB-dom. AGN	     \\
27  & 017652 &  214.2104  & 52.2763  & 0.683 (4)&   0.60    &  0.29  &  98 &   0.9  & 44.16  &   6    &  SB-cont. AGN	   \\
29  & 033772 &  214.2136  & 52.3461  &   -	&   0.85    &  1.67  &   7 &   0.0  & 44.70  &   9    &  Type-2 AGN	\\
30  & 058423 &  214.2163  & 52.4501  &   -	&   0.90    &  0.27  &  98 &   0.3  & 44.32  &  10    &  Type-2 AGN  \\
33  & 042611 &  214.2433  & 52.4036  &   -	&   0.90    &  0.87  &  53 &   0.0  & 42.61  &   4    &  SB-cont. AGN	  \\
35  & 021276 &  214.2529  & 52.3218  &   -	&   0.32    &  0.15  & 100 &   0.3  & 43.61  &   9    &  Type-2 AGN	\\
36  & 041222 &  214.2675  & 52.4149  & 0.281 (4)&   0.25    &  1.13  &  33 &   0.6  & 43.75  &  23    &    NG hosting AGN     \\
38  & 068074 &  214.2737  & 52.5297  & 0.426 (2)&   1.56    &  1.59  &   9 &   0.0  & 44.13  &   2    &  SB-dom. AGN	     \\
41  & 068708 &  214.2850  & 52.5403  &   -	&   1.36    &  0.62  &  80 &   0.3  & 44.16  &   5    &  SB-dom. AGN	     \\
42  & 056274 &  214.2862  & 52.4917  &   -	&   1.26    &  1.17  &  30 &   0.6  & 43.79  &   2    &    SB-dom. AGN       \\
43  & 046787 &  214.2870  & 52.4525  & 0.532 (4)&   0.47    &  0.13  & 100 &   0.0  & 43.90  &  20    &    NG hosting AGN     \\
45  & 050845 &  214.2940  & 52.4747  &   -	&   1.25    &  0.99  &  45 &   0.3  & 44.53  &   1    &    SB-dom. AGN       \\
47  & 039386 &  214.2961  & 52.4280  &   -	&   0.34    &  0.93  &  51 &   0.3  & 44.00  &  19    &    NG hosting AGN     \\
48  & 062600 &  214.2984  & 52.5257  & 0.835 (4)&   0.84    &  0.87  &  57 &   0.0  & 44.13  &  21    &    NG hosting AGN     \\
49  & 016716 &  214.2994  & 52.3366  & 0.433 (4)&   0.45    &  1.37  &  18 &   0.6  & 44.08  &  22    &   NG hosting AGN  \\
50  & 036500 &  214.3096  & 52.4259  &   -	&   0.32    &  0.73  &  60 &   1.2  & 43.59  &  12    &  Type-1 AGN	\\
51  & 071816 &  214.3118  & 52.5720  &   -	&   1.20    &  1.38  &  18 &   1.2  & 44.17  &   2    &    SB-dom. AGN       \\
52  & 026610 &  214.3127  & 52.3869  & 1.271 (3)&   0.48    &  0.27  &  99 &   0.0  & 43.41  &  14    &  Type-1 AGN	\\
53  & 041138 &  214.3134  & 52.4474  & 0.723 (4)&   0.67    &  1.08  &  38 &   0.0  & 44.14  &   9    &  Type-2 AGN	\\
55  & 041987 &  214.3290  & 52.4623  & 1.211 (3)&   1.23    &  0.27  &  98 &   0.0  & 44.10  &   1    &    SB-dom. AGN       \\
56  & 042538 &  214.3303  & 52.4655  & 1.208 (3)&   1.19    &  1.35  &  19 &   0.0  & 43.71  &  10    &  Type-2 AGN	   \\	
57  & 030161 &  214.3335  & 52.4168  &   -	&   0.88    &  1.03  &  42 &   0.3  & 44.19  &   8    &  Type-2 AGN	\\
59  & 055009 &  214.3456  & 52.5288  & 0.465 (4)&   0.46    &  0.42  &  94 &   0.0  & 43.52  &  21    &     NG hosting AGN     \\
60  & 055370 &  214.3475  & 52.5316  & 0.484 (4)&   0.50    &  0.29  &  98 &   0.0  & 43.64  &   9    &  Type-2 AGN    \\ 
61  & 031265 &  214.3483  & 52.4320  &   -	&   1.19    &  1.01  &  43 &   1.2  & 44.42  &   1    &    SB-dom. AGN      \\ 
62  & 057218 &  214.3510  & 52.5416  & 0.902 (4)&   0.83    &  1.39  &  17 &   0.0  & 43.59  &   3    &    SB-dom. AGN      \\ 
63  & 048619 &  214.3525  & 52.5069  & 0.482 (4)&   0.54    &  1.50  &  12 &   0.3  & 44.49  &  14    &  Type-1 AGN    \\ 
64  & 069965 &  214.3553  & 52.5956  &   -	&   1.16    &  0.18  &  97 &   0.0  & 43.22  &   4    &  SB-cont. AGN	  \\
66  & 051055 &  214.3637  & 52.5254  &   -	&   1.57    &  1.38  &  18 &   1.2  & 43.79  &   2    &    SB-dom. AGN      \\ 
67  & 068063 &  214.3704  & 52.5984  &   -	&   1.43    &  0.85  &  58 &   0.3  & 44.40  &   3    &    SB-dom. AGN      \\ 
69  & 034221 &  214.3748  & 52.4633  &   -	&   0.91    &  0.38  &  92 &   0.6  & 44.50  &  10    &   Type-2 AGN	 \\
72  & 035715 &  214.3784  & 52.4718  &   -	&   1.42    &  1.19  &  29 &   1.2  & 43.79  &   2    &    SB-dom. AGN       \\
73  & 049420 &  214.3859  & 52.5342  & 0.986 (4)&   0.91    &  0.38  &  96 &   0.6  & 43.87  &  16    &    NG hosting AGN    \\
74  & 055653 &  214.3909  & 52.5637  & 0.551 (4)&   0.52    &  0.39  &  95 &   0.3  & 43.90  &  22    &    NG hosting AGN	\\
75  & 019988 &  214.3911  & 52.4155  &   -	&   0.92    &  1.14  &  33 &   0.0  & 43.72  &   3    &    SB-dom. AGN       \\
76  & 044463 &  214.3932  & 52.5186  & 0.271 (4)&   0.29    &  0.42  &  95 &   0.6  & 43.86  &  16    &    NG hosting AGN   \\
77  & 032243 &  214.3952  & 52.4696  &   -	&   1.73    &  1.52  &  11 &   0.0  & 45.29  &   1    &     SB-dom. AGN    \\
78  & 040934 &  214.3998  & 52.5083  &   -	&   2.42    &  0.61  &  82 &   0.3  & 45.61  &  14    &  Type-1 AGN	\\
79  & 039818 &  214.4012  & 52.5047  &   -	&   0.93    &  1.31  &  21 &   0.3  & 42.85  &   2    &    SB-dom. AGN       \\
80  & 061825 &  214.4014  & 52.5957  & 0.197 (1)&   2.32    &  1.56  &  10 &   0.0  & 44.92  &   1    &      SB-dom. AGN    \\
81  & 016037 &  214.4037  & 52.4084  &   -	&  2.33  &    1.37  &  19  &   0.3  & 44.79  &   14   & Type-1 AGN   \\
82  & 062180 &  214.4043  & 52.5994  &   -	&  0.25  &    1.22  &  27  &   0.0  & 44.79  &   22   &    NG hosting AGN	 \\
83  & 035272 &  214.4056  & 52.4893  &   -	&  1.19  &    0.36  &  87  &   0.9  & 43.83  &    9   & Type-2 AGN    \\
84  & 053837 &  214.4112  & 52.5706  &   -	&  1.11  &    1.17  &  30  &   0.3  & 43.65  &    2   &   SB-dom. AGN  \\
86  & 031503 &  214.4127  & 52.4789  &   -	&  0.90  &    0.74  &  69  &   1.2  & 43.68  &    9   & Type-2 AGN   \\
87  & 031796 &  214.4137  & 52.4806  &   -	&  1.02  &    0.53  &  88  &   0.0  & 44.32  &   10   & Type-2 AGN   \\
89  & 057956 &  214.4228  & 52.5959  &   -	&  1.39  &    0.37  &  95  &   0.0  & 44.33  &   10   & Type-2 AGN   \\
90  & 028146 &  214.4244  & 52.4732  & 1.148 (4)&  1.15  &    2.43  &	1  &   0.6  & 44.53  &    4   & SB-cont. AGN   \\
91  & 031444 &  214.4393  & 52.4976  & 0.873 (4)&  0.87  &    1.27  &  24  &   0.9  & 43.76  &    1   & SB-dom. AGN   \\
92  & 024070 &  214.4401  & 52.4672  & 0.224 (2)&  0.23  &    0.61  &  82  &   0.0  & 43.03  &   22   &  NG hosting AGN   \\
93  & 033761 &  214.4415  & 52.5091  & 0.985 (3)&  0.97  &    1.95  &	3  &   0.6  & 44.43  &    8   &  Type-2 AGN  \\
95  & 027043 &  214.4445  & 52.4829  &   -	&  1.78  &    0.93  &  50  &   0.9  & 43.90  &    2   &   SB-dom. AGN  \\
97  & 024055 &  214.4460  & 52.4713  &   -	&  2.58  &    0.27  &  90  &   1.2  & 43.83  &   12   &  Type-1 AGN   \\
98  & 051437 &  214.4472  & 52.5862  & 1.547 (3)&  2.37  &    0.92  &  51  &   0.3  & 43.81  &    7   & Type-1 AGN   \\
99  & 021585 &  214.4550  & 52.4676  & 0.996 (4)&  1.00  &    0.50  &  87  &   0.3  & 44.65  &   17   &    NG hosting AGN      \\
101 & 035904 &  214.4575  & 52.5290  &   -	&  2.65  &    0.47  &  88  &   0.9  & 44.30  &    2   &   SB-dom. AGN  \\
105 & 030608 &  214.4657  & 52.5129  &   -	&  0.87  &    0.58  &  82  &   0.3  & 43.69  &   10   & Type-2 AGN   \\
106 & 022680 &  214.4684  & 52.4814  &   -	&  1.00  &    0.57  &  84  &   0.6  & 43.92  &    9   & Type-2 AGN   \\
107 & 021273 &  214.4707  & 52.4775  & 0.671 (3)&  0.60  &    0.40  &  95  &   0.3  & 43.96  &   22   &    NG hosting AGN     \\
108 & 045400 &  214.4737  & 52.5795  & 0.719 (4)&  0.65  &    0.51  &  88  &   0.6  & 43.77  &    9   & Type-2 AGN   \\
109 & 028312 &  214.4748  & 52.5095  &   -	&  1.34  &    1.06  &  39  &   0.3  & 44.43  &    8   & Type-2 AGN   \\
110 & 031338 &  214.4760  & 52.5232  &   -	&  0.65  &    0.81  &  61  &   1.2  & 43.45  &   10   & Type-2 AGN   \\
111 & 044246 &  214.4775  & 52.5774  & 0.948 (3)&  0.85  &    0.61  &  81  &   0.3  & 43.99  &   21   &    NG hosting AGN	\\
112 & 047305 &  214.4803  & 52.5924  &   -	&  2.75  &    0.34  &  96  &   0.3  & 43.91  &   12   & Type-1 AGN   \\
113 & 029613 &  214.4868  & 52.5235  &   -	&  0.50  &    0.52  &  88  &   1.2  & 42.79  &    7   & Type-1 AGN   \\
116 & 027980 &  214.4893  & 52.5186  &   -	&  0.16  &    0.70  &  67  &   0.9  & 42.79  &    7   & Type-1 AGN   \\
118 & 029054 &  214.4956  & 52.5275  &   -	&  0.63  &    0.15  & 100  &   0.6  & 43.69  &    9   & Type-2 AGN   \\
119 & 046309 &  214.5015  & 52.6030  &   -	&  2.45  &    0.31  &  93  &   0.0  & 43.69  &   13   &  SB-cont. AGN	  \\
124 & 041429 &  214.5082  & 52.5875  &   -	&  1.33  &    0.19  & 100  &   0.6  & 45.13  &   14   & Type-1 AGN   \\
125 & 042989 &  214.5119  & 52.5965  &   -	&  2.05  &    1.14  &  33  &   0.3  & 43.45  &    1   &   SB-dom. AGN  \\
126 & 044785 &  214.5191  & 52.6092  & 0.387 (1)&  1.61  &    1.88  &	4  &   1.2  & 44.49  &    7   & Type-1 AGN   \\
127 & 032921 &  214.5270  & 52.5662  &   -	&  1.30  &    0.70  &  71  &   0.0  & 45.05  &   10   & Type-2 AGN   \\
128 & 018428 &  214.5305  & 52.5083  &   -	&  0.90  &    0.57  &  84  &   0.0  & 44.35  &   20   &    NG hosting AGN	\\
133 & 024215 &  214.5679  & 52.5586  &   -	&  2.38  &    2.28  &	1  &   0.6  & 44.54  &    2   &   SB-dom. AGN  \\
134 & 016978 &  214.5751  & 52.5340  &   -	&  1.32  &    2.04  &	3  &   0.3  & 43.89  &    3   &   SB-dom. AGN  \\
135 & 022888 &  214.5841  & 52.5647  &   -	&  0.25  &    1.36  &  19  &   0.6  & 42.35  &    7   & Type-1 AGN   \\
136 & 018192 &  214.5888  & 52.5485  & 0.036 (1)&  0.05  &    0.59  &  81  &   1.2  & 40.70  &    7   & Type-1 AGN   \\
137 & 030219 &  214.5939  & 52.6020  &   -	&  0.84  &    0.92  &  50  &   1.2  & 43.51  &    9   & Type-2 AGN   \\
\enddata
\end{deluxetable}

\clearpage

\begin{table}[ !ht ]
\scriptsize
\centering
\begin{tabular}{lcccccccccccc}
\hline
\hline
Band &\multicolumn{2}{c}{Total fit} &\multicolumn{2}{c}{SB-dom. AGN} &\multicolumn{2}{c}{SB-cont. AGN} 
&\multicolumn{2}{c}{Type-1 AGN} & \multicolumn{2}{c}{Type-2 AGN} 
&\multicolumn{2}{c}{NG hosting AGN} \\
& $\alpha$ & r& $\alpha$ & r &  $\alpha$ & r & $\alpha$ & r & $\alpha$ & r &  $\alpha$ & r\\
\hline
FUV    	   &  1.22 &  0.84 & 1.05 & 0.81 & 1.37 & 0.94 &   1.29    &  0.95   &    -    &      -   &    1.12	&     0.71	    \\
NUV    	   &  0.83 &  0.73 & 1.08 & 0.83 & 0.87 & 0.87 &   0.99    &  0.85   &    -    &      -   &    0.72	&     0.52	    \\
u    	   &  0.75 &  0.73 & 0.90 & 0.77 &  -   &  -   &   1.01    &  0.89   &    -    &      -   &	 -	&	-	    \\
g          &  0.70 &  0.72 & 0.79 & 0.72 & 1.12 & 0.80 &   0.99    &  0.91   &   0.76  &    0.58  &      -      &	-	    \\  
r    	   &  0.67 &  0.71 & 0.71 & 0.66 &  -   &  -   &   0.97    &  0.92   &   0.71  &    0.57  &      -      &	-	    \\  
i    	   &  0.66 &  0.73 & 0.71 & 0.66 &  -	&  -   &   0.96    &  0.93   &   0.64  &    0.55  &    0.47     &     0.58	    \\
z    	   &  0.67 &  0.73 & 0.78 & 0.74 &  -	&  -   &   0.97    &  0.93   &   0.64  &    0.59  &    0.48     &     0.57	    \\
J    	   &  0.68 &  0.75 & 0.82 & 0.79 &  -	&  -   &   0.97    &  0.94   &   0.75  &    0.62  &    0.48     &     0.58	    \\
K    	   &  0.72 &  0.78 & 0.78 & 0.82 & 1.25 & 0.87 &   0.91    &  0.92   &   0.66  &    0.51  &    0.62     &     0.67	    \\
IRAC3.6    &  0.81 &  0.80 & 1.01 & 0.83 & 1.40 & 0.88 &   0.90    &  0.88   &   0.67  &    0.49  &    0.72     &     0.66          \\
IRAC4.5    &  0.87 &  0.80 & 1.07 & 0.82 &  -   &  -   &   0.90    &  0.86   &   0.77  &    0.51  &    0.82     &     0.67	    \\
IRAC5.8    &  0.91 &  0.79 & 0.96 & 0.73 & 1.38 & 0.84 &   0.90    &  0.85   &   0.81  &    0.53  &    0.88     &     0.65  	    \\
IRAC8.0    &  0.93 &  0.80 & 0.69 & 0.64 & 1.40 & 0.85 &   0.92    &  0.87   &   0.82  &    0.56  &    0.98     &     0.69  	    \\
MIPS24     &  1.02 &  0.70 &   -  &   -  &  -   &  -   &   0.89    &  0.84   &   0.87  &    0.50  &    1.23     &     0.67  	    \\ 
\hline
\hline	     											    		    				    		        										        																																																																																																		      
Band &\multicolumn{2}{c}{Total fit} &\multicolumn{2}{c}{SB-dom. AGN}&\multicolumn{2}{c}{SB-cont. AGN}
&\multicolumn{2}{c}{Type-1 AGN} 
&\multicolumn{2}{c}{Type-2 AGN}&\multicolumn{2}{c}{NG hosting AGN}  \\
& $\alpha$ & r& $\alpha$ & r& $\alpha$ & r &  $\alpha$ & r & $\alpha$ & r &  $\alpha$ & r\\
\hline
FUV    	   &  1.07    &   0.83    &  0.92   &  0.80  &   0.92    &  0.95 &  1.17   &  0.95 &  -   &  -  &   0.92	&  0.63	   \\
NUV    	   &  0.75    &   0.74    &  0.87   &  0.85  &    -	 &    -	 &  0.87   &  0.86 &  -   &  -  &   0.68	&  0.54	   \\
u    	   &  0.66    &   0.73    &  0.70   &  0.78  &    -	 &    -	 &  0.87   &  0.89 &  -   &  -  &    -  	&    -     \\
g          &  0.61    &   0.71    &  0.64   &  0.72  &    -	 &    -	 &  0.85   &  0.90 &  -   &  -  &    -  	&    -     \\  
r    	   &  0.58    &   0.70    &   -     &   -    &    -	 &    -	 &  0.84   &  0.90 &  -   &  -  &    - 	        &    -	   \\  
i    	   &  0.57    &   0.70    &   -     &   -    &    -	 &    -	 &  0.83   &  0.91 &  -   &  -  &   0.44	&  0.58	   \\    
z    	   &  0.57    &   0.70    &   -     &   -    &    -	 &    -	 &  0.83   &  0.91 &  -   &  -  &   0.45	&  0.59	   \\    
J  	   &  0.59    &   0.72    &  0.52   &  0.65  &    -	 &    -	 &  0.82   &  0.91 &  -   &  -  &   0.45	&  0.59	   \\    
K    	   &  0.62    &   0.76    &  0.49   &  0.67  &    -	 &    -	 &  0.79   &  0.91 &  -   &  -  &   0.55	&  0.65	   \\
IRAC3.6    &  0.70    &   0.79    &  0.64   &  0.71  &    -	 &    -	 &  0.79   &  0.88 &  -   &  -  &   0.64	&  0.65	   \\
IRAC4.5    &  0.76    &   0.79    &  0.73   &  0.74  &    -	 &    -	 &  0.80   &  0.87 &  -   &  -  &   0.71	&  0.65	   \\    
IRAC5.8    &  0.79    &   0.77    &  0.72   &  0.72  &    -	 &    -	 &  0.81   &  0.87 &  -   &  -  &   0.76	&  0.62	   \\    
IRAC8.0    &  0.80    &   0.77    &  0.58   &  0.67  &    -	 &    -	 &  0.81   &  0.88 &  -   &  -  &   0.81	&  0.63	   \\    
MIPS24     &  0.87    &   0.68    &  0.66   &  0.61  &    -	 &    -	 &  0.77   &  0.84 &  -   &  -  &   1.02	&  0.61	   \\    
\hline	     											    		    				    		        										        																																																																																																		      
\end{tabular}										
\caption{Photometric band considered, fit slope and correlation coefficient $r$ of the scatter diagrams between the 
ultraviolet/optical/infrared bands and the hard X-ray (top) and soft X-ray luminosities (bottom) for the total
sample and for the {\it Starbust-dominated AGN}, 
{\it Starburst-contaminated AGN}, {\it Type-1 AGN}, {\it Type-2 AGN} , and  {\it Normal galaxy hosting AGN}
groups. Spearman's rank correlation tests have been performed for all scatter 
diagrams. The values reported in Table \ref{types} are indeed significant (p$<$0.01).} 
\label{types}
\end{table}

\clearpage

\begin{table}[ !ht ]
\begin{tabular}{lccccc}
\hline
\hline
X-ray range & {SB-dom. AGN} &{SB-cont. AGN} &{Type-1 AGN} & {Type-2 AGN} &{NG hosting AGN} \\
\hline
Hard & [42,44] & [43,44] & [40,45] & [42,43] & [40,43] \\
Soft & [41,44] & [42,44] & [39,45] & [41,43] & [40,43] \\
\hline
\end{tabular}										
\caption{Hard and soft X-ray luminosity ranges (log) for the {\it Starburst-dominated AGN}, 
{\it Starburst-contaminated AGN}, {\it Type-1 AGN}, {\it Type-2 AGN} , and  
{\it Normal galaxy hosting AGN} groups. Luminosities are not corrected for absorption.} 
\label{luminosity}
\end{table}

\clearpage

\begin{table}[ !ht ]
\centering
\scriptsize
\begin{tabular}{lcccc}
\hline
\hline                                    
ID & ID IRAC & s1 IR emission & s2 IR emission & comments \\
\hline
11  &  053271 & Dominant & Negligible & s2 diffuse region, s2 probably not AGN \\
13  &  038708 & Dominant & Dominant   & Interacting system	 \\
23  &  056633 & Dominant & Dominant   & s1 diffuse region with stellar knots, s2 interacting system itself \\
31  &  034779 & Dominant & Negligible & Interacting system \\
32  &  031799 & Dominant & Negligible & s1 QSO-like, s2 probably not AGN \\
34  &  036704 & Dominant & Negligible & s1 interacting system itseft, s2 probably not AGN  \\
37  &  028084 & Dominant & Negligible & s1 face-on disky galaxy, s2 probably not AGN \\
39  &  071060 & Dominant & Negligible & Interacting system spectroscopically confirmed, s2 probably not AGN \\
58  &  046783 & Dominant & Dominant   & s1 and s2 QSO-like   \\
65  &  042079 & Dominant & Negligible & s2 probably not AGN, s2 diffuse region with stellar knots \\
88  &  061881 & Dominant & Negligible & s2 probably not AGN, s2 diffuse region with stellar knots \\
100 &  022060 & Dominant & Negligible & s1 interacting system itself , s2 probably not AGN \\
102 &  052826 & Dominant & Negligible & s2 probably not AGN, s2 diffuse region with stellar knots \\
103 &  054493 & Dominant & Negligible & Interacting system, s2 probably not AGN  \\
117 &  029343 & Dominant & Negligible & s1 diffuse region and interacting system itseft, s2 probably not AGN \\
121 &  019604 & Dominant & Negligible & s1 interacting system itself, s2 probably not AGN \\
122 &  027967 & Dominant & Negligible & s2 probably not AGN \\
129 &  022761 & Dominant & Negligible & s1 and s2 optical dropouts, s2 probably not AGN \\
131 &  017174 & Dominant & Negligible & Minicluster, s2 probably not AGN \\
132 &  021943 & Dominant & Negligible & s1 diffuse regions with stellar knots, s2 probably not AGN \\
\hline
\end{tabular}
\caption{\footnotesize{ID from \citet{Barmby06}, IRAC ID, mid-infrared emission of sources 1 and 2 in each pair of galaxies, and comments 
based on visual inspection of the objects. Classification of objects as "probably not AGN" is based on their mid-infrared emission.}
\label{double}}
\end{table}

\clearpage

\begin{deluxetable}{lccccccccccl}
\centering
\tabletypesize{\scriptsize}
\tablecaption{\footnotesize{ID from \citet{Barmby06}, IRAC ID, IRAC 3.6~\micron~ J2000.0 right ascension and declination, spectroscopic 
redshift from the DEEP public database with its corresponding reliability between brackets (1-2 = low reliability, 3-4= high reliability),
photometric redshifts 
for both blended galaxies when mid-infrared emission comes from both (13,23 and 58) or for the mid-infrared 
emitter in rest of the cases, and their 
corresponding $\chi_{\nu}^{2}$, probabilities, optical extinctions derived from the \citet{Calzetti00}
reddening law, logarithm of $\nu$L$_{\nu}$ in the r band as a reference, in erg~s$^{-1}$,
fitted templates, and general classification. Templates are the same described in Table \ref{photoz}.}
\label{photoz2}}
\tablehead{
\colhead{ID} & \colhead{ID IRAC} & \colhead{RA ($^{o}$)} & \colhead{Dec ($^{o}$)} & \colhead{z$_{spec}$}  &
\colhead{z$_{phot}$} & \colhead{$\chi_{\nu}^{2}$} & \colhead{Prob (\%)} & \colhead{A$_{V}$} &  \colhead{L$_{r}$} &
\colhead{Template} & \colhead{Group}} 
\startdata
 11  &  053271$_{-}$1  &   214.1439  & 52.3775 &  2.089 (2) & 2.44  &  1.29 &  23  &  1.20  &	45.21  &  14  &   Type-1 AGN  \\
 13  &  038708$_{-}$1  &   214.1499  & 52.3200 &    -	    & 1.09  &  0.94 &  49  &  0.00  &	44.17  &   1  &   SB-dom. AGN \\
 13  &  038708$_{-}$2  &   214.1499  & 52.3200 &    -	    & 1.10  &  1.21 &  28  &  0.00  &	44.11  &   3  &   SB-dom. AGN \\
 23  &  056633$_{-}$1  &   214.2032  & 52.4330 &    -	    & 1.21  &  0.35 &  95  &  0.00  &	44.30  &   5  &   SB-dom. AGN	  \\
 23  &  056633$_{-}$2  &   214.2032  & 52.4330 &    -	    & 0.26  &  1.73 &	6  &  0.30  &	43.18  &  22  &   NG hosting AGN    \\
 31  &  034779$_{-}$1  &   214.2239  & 52.3567 &    -	    & 0.22  &  0.91 &  53  &  0.90  &	42.76  &  10  &  Type-2 AGN \\
 32  &  031799$_{-}$1  &   214.2246  & 52.3453 &    -	    & 2.67  &  0.67 &  76  &  0.30  &	44.41  &  12  &  Type-1 AGN \\
 34  &  036704$_{-}$1  &   214.2506  & 52.3845 &    -	    & 1.41  &  0.42 &  92  &  0.30  &	44.47  &  21  &   NG hosting AGN   \\
 37  &  028084$_{-}$1  &   214.2677  & 52.3611 &    -	    & 0.44  &  0.89 &  55  &  0.90  &	43.48  &  10  &  Type-2 AGN \\
 39  &  071060$_{-}$1  &   214.2738  & 52.5418 &  0.170 (2) & 2.34  &  0.80 &  64  &  0.30  &	44.90  &   7  &  Type-1 AGN   \\
 58  &  046783$_{-}$1  &   214.3350  & 52.4867 &    -	    & 1.23  &  1.85 &	5  &  0.60  &	43.73  &   2  &   SB-dom. AGN \\
 58  &  046783$_{-}$2  &   214.3350  & 52.4867 &    -	    & 0.92  &  3.14 &	0  &  0.00  &	42.67  &   2  &  SB-dom. AGN \\
 65  &  042079$_{-}$1  &   214.3627  & 52.4867 &    -	    & 1.16  &  2.30 &	1  &  0.00  &	43.69  &   8  &  Type-2 AGN \\
 88  &  061881$_{-}$1  &   214.4148  & 52.6053 &    -	    & 0.26  &  1.72 &	8  &  0.00  &	42.22  &   9  &  Type-2 AGN   \\
100  &  022060$_{-}$1  &   214.4551  & 52.4699 &  0.998 (2) & 0.98  &  0.75 &  68  &  0.00  &	44.13  &  10  &  Type-2 AGN  \\
102  &  052826$_{-}$1  &   214.4590  & 52.6005 &    -	    & 1.26  &  0.33 &  95  &  0.90  &	43.61  &   2  &   SB-dom. AGN \\
103  &  054493$_{-}$1  &   214.4620  & 52.6093 &    -	    & 2.43  &  1.79 &	6  &  0.60  &	44.43  &  12  &   Type-1 AGN  \\
117  &  029343$_{-}$1  &   214.4923  & 52.5262 &    -	    & 1.12  &  2.09 &	2  &  0.00  &	43.71  &   2  &   SB-dom. AGN \\
121  &  019604$_{-}$1  &   214.5047  & 52.4950 &  0.623 (1) & 0.14  &  1.90 &	4  &  0.00  &	41.48  &   1  &   SB-dom. AGN	  \\
122  &  027967$_{-}$1  &   214.5062  & 52.5305 &    -	    & 0.89  &  1.31 &  22  &  0.30  &	43.85  &  21  &   NG hosting AGN     \\
129  &  022761$_{-}$1  &   214.5373  & 52.5309 &    -	    & 1.06  &  0.77 &  61  &  1.20  &	43.19  &   1  &   SB-dom. AGN \\
131  &  017174$_{-}$1  &   214.5545  & 52.5202 &    -	    & 1.15  &  1.04 &  41  &  0.30  &	44.07  &   3  &   SB-dom. AGN \\
132  &  021943$_{-}$1  &   214.5641  & 52.5466 &    -	    & 0.98  &  0.63 &  77  &  0.30  &	43.44  &   2  &   SB-dom. AGN \\
\enddata
\end{deluxetable}

\end{document}